%% file: main.tex
\documentclass[%
 reprint,
superscriptaddress,
nofootinbib,
amsmath,amssymb,
prb,
floatfix,
]{revtex4-2}
\usepackage[normalem]{ulem}

\usepackage{graphicx}
\graphicspath{{./figures/}}

\usepackage{dcolumn}

\usepackage{bm}

\usepackage{hyperref}
\hypersetup{
    colorlinks,
    linkcolor={blue},
    citecolor={blue},
    urlcolor={blue}
}

\usepackage{booktabs} 

\usepackage{multirow} 

\usepackage{tabularx} 

\usepackage{braket} 

\usepackage[caption=false, justification=justified]{subfig} 

\usepackage{makecell} 

\usepackage{lipsum}

\usepackage{amsthm}
\newtheorem{lemma}{Lemma}

\usepackage{xcolor}

\definecolor{ultramarine}{RGB}{0,150,96}

\definecolor{darkblue}{RGB}{0,0,127}

\definecolor{olive}{RGB}{128, 128, 0}

\begin{document}

\title{%
No Strings Attached: 
Boundaries and Defects in the Cubic Code} 

\author{Cory T. Aitchison}
 \email{cory.aitchison@sydney.edu.au}
\affiliation{Centre for Engineered Quantum Systems, School of Physics, University of Sydney, Sydney, NSW 2006, Australia}

\author{Daniel Bulmash}
\affiliation{Department of Physics and Center for Theory of Quantum Matter, University of Colorado Boulder, Boulder, Colorado 80309, USA}

\author{Arpit Dua}
\affiliation{Department of Physics, California Institute of Technology, Pasadena, CA 91125, USA}
\affiliation{Institute for Quantum Information and Matter, California Institute of Technology, Pasadena, California 91125, USA}

\author{Andrew C. Doherty}
\affiliation{Centre for Engineered Quantum Systems, School of Physics, University of Sydney, Sydney, NSW 2006, Australia}

\author{Dominic J. Williamson}
\affiliation{Centre for Engineered Quantum Systems, School of Physics, University of Sydney, Sydney, NSW 2006, Australia}

\date{July 31, 2023}

\begin{abstract}
\noindent
Haah's cubic code is the prototypical type-II fracton topological order. 
It instantiates the no string-like operator property that underlies the favorable scaling of its code distance and logical energy barrier. 
Previously, the cubic code was only explored in translation-invariant systems on infinite and periodic lattices. 
In these settings, the code distance scales superlinearly with the linear system size, while the number of logical qubits within the degenerate ground space exhibits a complicated functional dependence that undergoes large fluctuations within a linear envelope. 
Here, we extend the cubic code to systems with open boundary conditions and crystal lattice defects. 
We characterize the condensation of topological excitations in the vicinity of these boundaries and defects, finding that their inclusion can introduce local string-like operators and enhance the mobility of otherwise fractonic excitations.
Despite this, we use these boundaries and defects to define new encodings where the number of logical qubits scales linearly without fluctuations, and the code distance scales superlinearly, with the linear system size. 
These include a subsystem encoding with open boundary conditions and a subspace encoding using lattice defects. 
\end{abstract}
\maketitle

\tableofcontents

\section{\label{sec:intro}Introduction}
\input{intro.tex}

\section{\label{sec:bg}Background}
\input{background.tex}

\section{\label{sec:bdries}Boundaries}
\input{boundaries/boundaries.tex}

\section{\label{sec:superbdries}Superlinear-Distance Boundary Codes}
\input{boundaries/bdries_super.tex}

\section{\label{sec:defects}Defects}
\input{defects/defects.tex}

\section{\label{sec:superdefects}Superlinear-Distance Defect Codes}
\input{defects/defects_super.tex}

\section{\label{sec:conc}Conclusion}
\input{conclusion.tex}

\bibliography{ref}

\appendix 
\section{\label{sec:otherbdries} Additional Boundary Codes}
\input{appendix/app_bdries.tex}

\section{\label{sec:otherdefects} Additional Defect Codes}
\input{appendix/app_defects.tex}

\section{\label{sec:appendix}Additional Material}
\input{appendix/appendix.tex}

\end{document}

%% file: intro.tex
Quantum computers are required to operate effectively
in the presence of errors and noisy operations~\cite{shorSchemeReducingDecoherence1995, steaneErrorCorrectingCodes1996,shorFaulttolerantQuantumComputation1997, preskillFaultTolerantQuantumComputation1998}. 
A primitive component of a quantum computer is the 
quantum hard drive: a system capable of safely storing quantum information for long periods of time.
In comparison to leading approaches such as the surface code~\cite{kitaevFaulttolerantQuantumComputation2003,bravyiQuantumCodesLattice1998}, which require \emph{active}
procedures to continually detect and correct for errors~\cite{dennisTopologicalQuantumMemory2002}, such a hard drive should be \emph{passively}
self-correcting. To this end, one can envision a system where quantum
information is encoded in an energetic ground state and errors that corrupt
this information are suppressed by macroscopic energy barriers~\cite{brownQuantumMemoriesFinite2016}. Unfortunately, this behavior is
impossible to achieve in many cases - such as the surface code - due to 
no-go theorems that prohibit self-correction in $(2+1)$D systems~\cite{alickiThermalizationKitaev2D2009,
bravyiNogoTheoremTwodimensional2009,
bravyiTradeoffsReliableQuantum2010,
alickiThermalStabilityTopological2010,
yoshidaFeasibilitySelfcorrectingQuantum2011,
haahLogicaloperatorTradeoffLocal2012,
landon-cardinalLocalTopologicalOrder2013,
brownQuantumMemoriesFinite2016}.

Fortunately, these theorems do not apply to higher spatial dimensions.
Already in $(3+1)$D there are topological codes with no \emph{string-like logical operators} that have significantly better energy barriers than any $(2+1)$D code~\cite{haahLocalStabilizerCodes2011,haahLatticeQuantumCodes2013}. 
The earliest such example is Haah's \emph{cubic code}, which was found via a computational search~\cite{haahLocalStabilizerCodes2011}. 
The cubic code model is part of a larger classification of unconventional topological phases of matter, known as
\emph{fracton topological orders}~\cite{chamonQuantumGlassinessStrongly2005,bravyiTopologicalOrderExactly2011,haahLocalStabilizerCodes2011,castelnovoTopologicalQuantumGlassiness2012,kim20123d,haahLatticeQuantumCodes2013,yoshidaExoticTopologicalOrder2013,vijayNewKindTopological2015,vijayFractonTopologicalOrder2016,williamsonFractalSymmetriesUngauging2016}.
In this classification, topological codes with no string-like operators are called type-II fracton phases~\cite{vijayFractonTopologicalOrder2016}.
As a type-II fracton phase, the cubic code only supports topological excitations that are completely immobile~\cite{haahLocalStabilizerCodes2011}. 
When used as an error-correcting code, this immobility results in a code distance that scales superlinearly with the linear system size i.e., $d \sim \mathcal O(L^\alpha)$ for $\alpha > 1$, where $L$ is the number of lattice sites along one axis. 
Moreover, the minimum energy required to map between degenerate ground states via local operations - also known as the energy barrier - scales as $\mathcal O(\log(L))$~\cite{bravyiEnergyLandscape3D2011}. 
This energy barrier enables the cubic code to be partially self-correcting:
its quantum memory time increases with the system size only up to a finite threshold that decreases with temperature~\cite{bravyiEnergyLandscape3D2011}. For comparison, the surface code in $(2+1)$D has a memory time independent of system size. This property makes the cubic code a leading candidate for creating a quantum hard drive in $(3+1)$D.

There are, however, additional features of the cubic code that are undesirable for applications to quantum error correction (QEC). 
The number of encoded qubits, $k$, varies sporadically with the system size~\cite{haahLatticeQuantumCodes2013} (see Fig.~\ref{fig:cubic_k}). 
In an actual implementation, achieving a large $k$, therefore, 
requires the system to be from a family of carefully chosen system sizes that have large jumps between them. 
Moreover, the model was formulated in a translation-invariant setting with periodic boundary conditions, arranging the physical qubits on a $3$-torus. This topology is not feasible in a strictly local architecture. 

Since the discovery of the cubic code, there has not yet been an examination into how the model's topology or geometry may be modified, and how these modifications may affect its core characteristics, such as its no-string property or fracton topological order. 
A significant question is whether open boundary conditions or the inclusion of lattice defects affect the error-correcting properties of the code, either by improving or worsening them. 
In other codes, including such modifications has the potential to increase the number of encoded qubits or enable additional fault-tolerant quantum gates~\cite{bravyiQuantumCodesLattice1998, bombinTopologicalOrderTwist2010, brownPokingHolesCutting2017,bulmashBraidingGappedBoundaries2019}. However, in our setting, open boundary conditions and defects have the potential to introduce string-like operators and reduce the energy barrier for logical errors.

In this work, we characterize the properties of boundaries and defects in the cubic code, including their interactions with quasiparticle excitations. We investigate whether modifying the cubic code by introducing boundaries and defects can affect the scaling of the number of qubits
while maintaining the superlinear code distance and favorable energy barrier of
the periodic model. We approach this with a combination of analytic arguments, visualizations, and numerical computations. 
For the latter, we simulate lattices with up to approximately
$20^3$ qubits, and assume a consistent extrapolation for larger systems.

\subsection{Summary of Results}

\begin{table*}
  \centering
  \begin{tabular}{ccccc}
    \toprule
    Diagram & Notation & Name & Encoded Qubits ($k$) & Logical Weight \\
    \cmidrule(r){1-1} \cmidrule(lr){2-2} \cmidrule(lr){3-3} \cmidrule(lr){4-4}
    \cmidrule(l){5-5}
    \raisebox{-0.5\totalheight}{
  \includegraphics[width=0.19\linewidth]{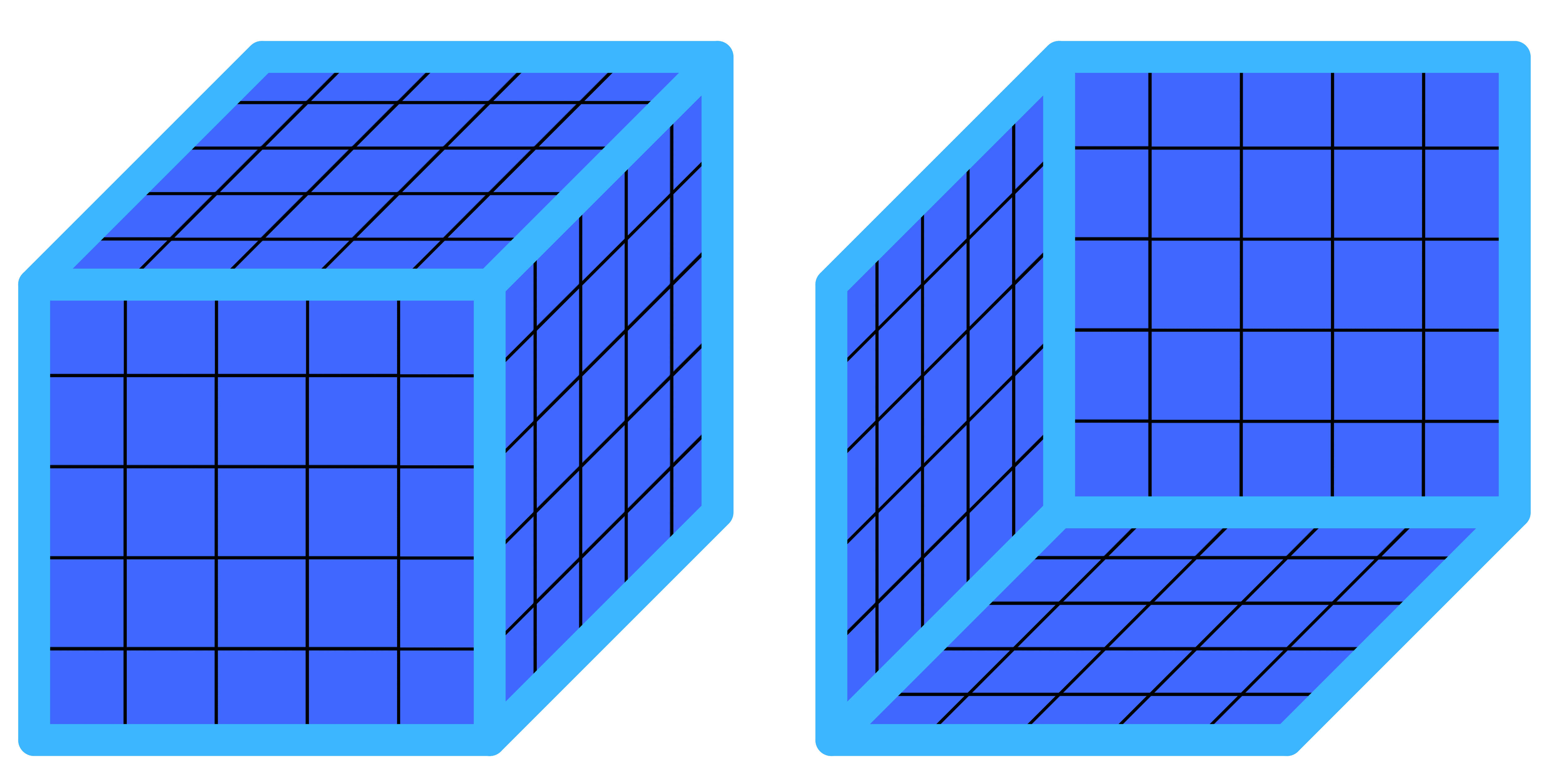}}&
  $(eee;eee)$ & 
    Only $e$ & $0$ & - \\ 
    \raisebox{-0.5\totalheight}{\includegraphics[width=0.19\linewidth]{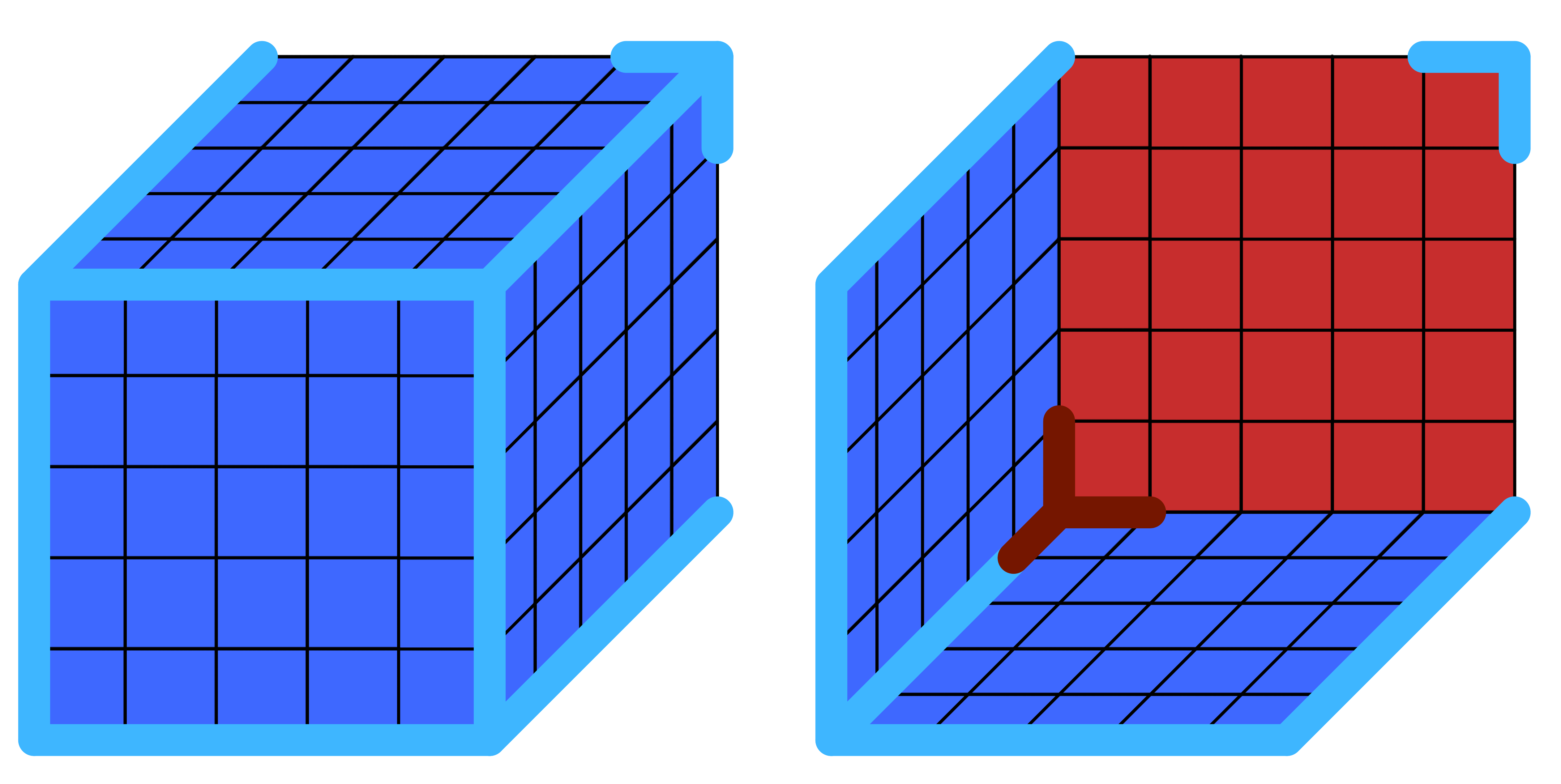}}& 
    $(eee;mee)$ & 
      One $(m)$ & $0$ & - \\
    \raisebox{-0.5\totalheight}{\includegraphics[width=0.19\linewidth]{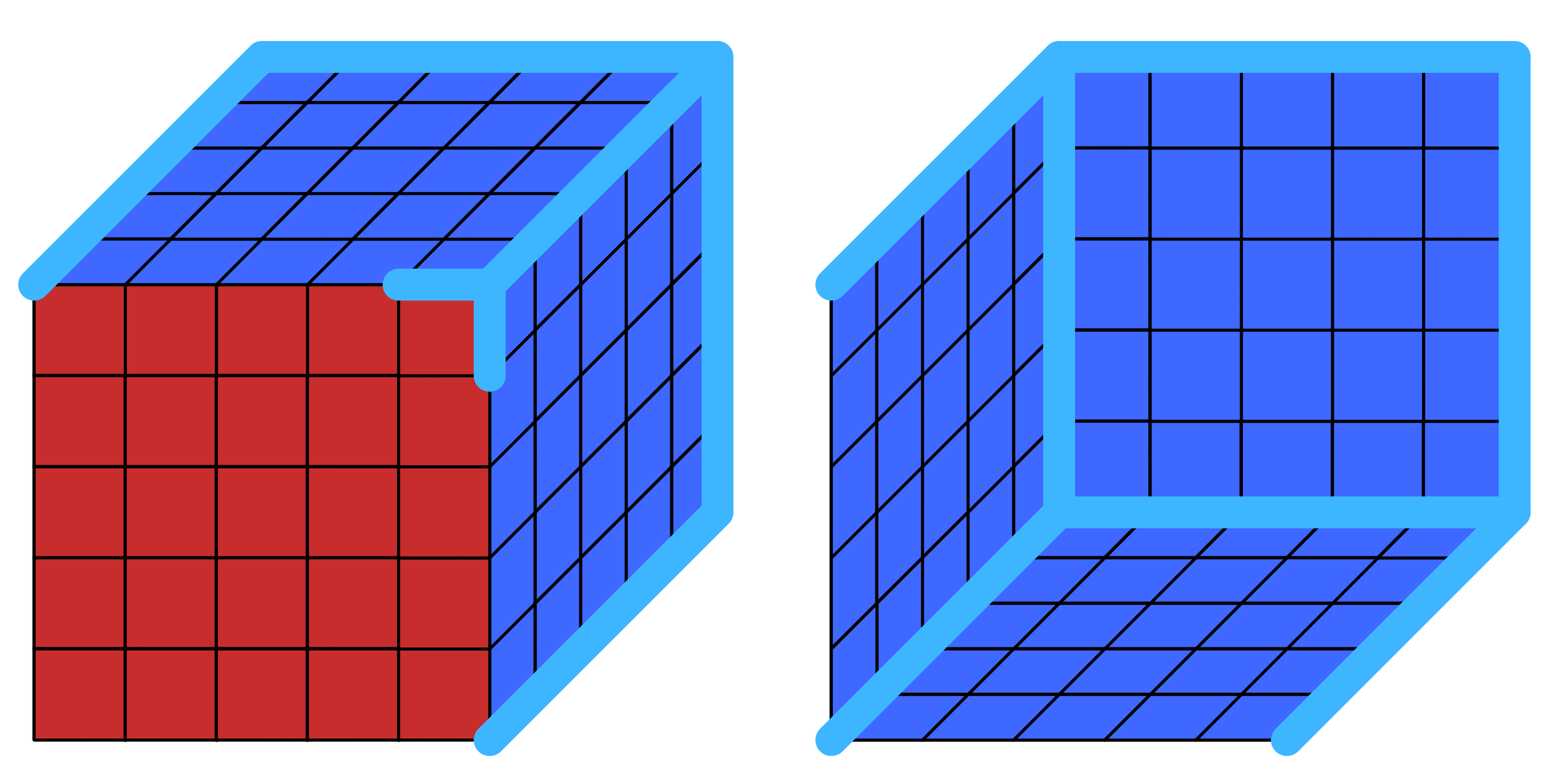}}& 
    $(mee;eee)$ & 
      One $(m_{ABC})$ & $0$ & - \\
    \raisebox{-0.5\totalheight}{\includegraphics[width=0.19\linewidth]{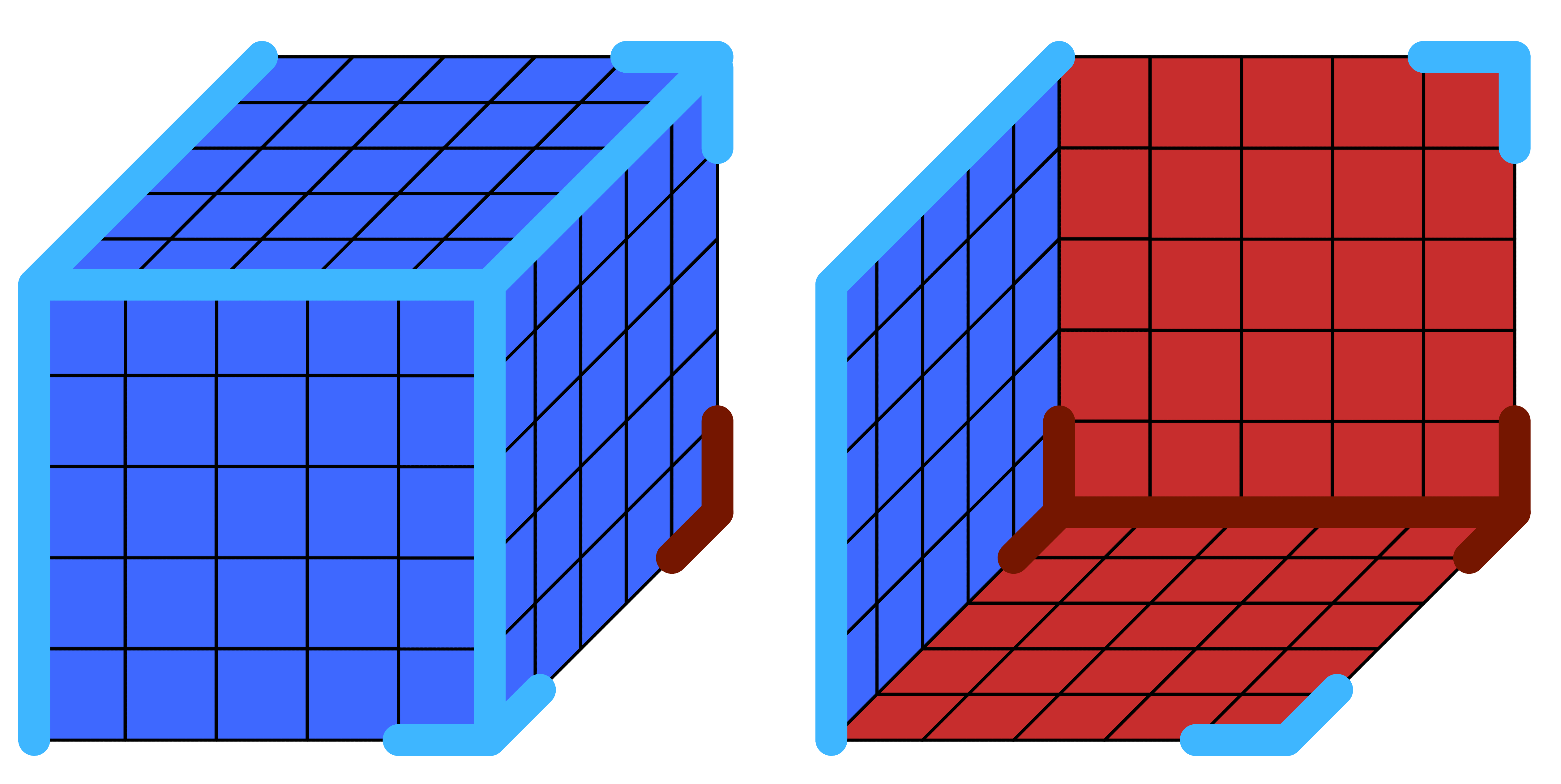}}& 
    $(eee;mem)$ & Two $(m)$ & $2\min\{L_x,L_z\} - 6$ & \makecell{Constant,\\Superlinear}
    \\
    \raisebox{-0.5\totalheight}{\includegraphics[width=0.19\linewidth]{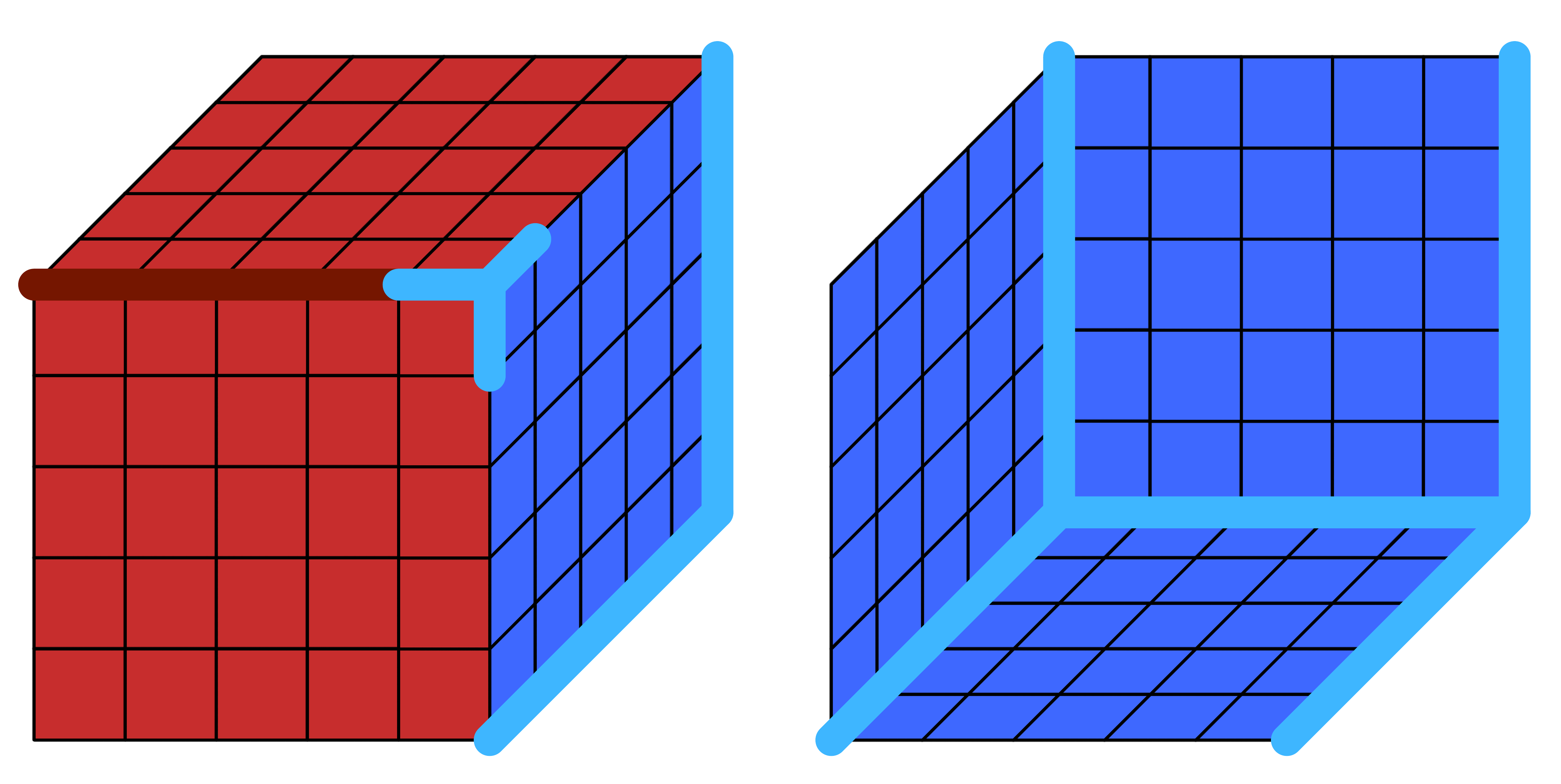}}& 
    $(mem;eee)$ & Two $(m_{ABC})$ & $0$ & - \\
    \raisebox{-0.5\totalheight}{\includegraphics[width=0.19\linewidth]{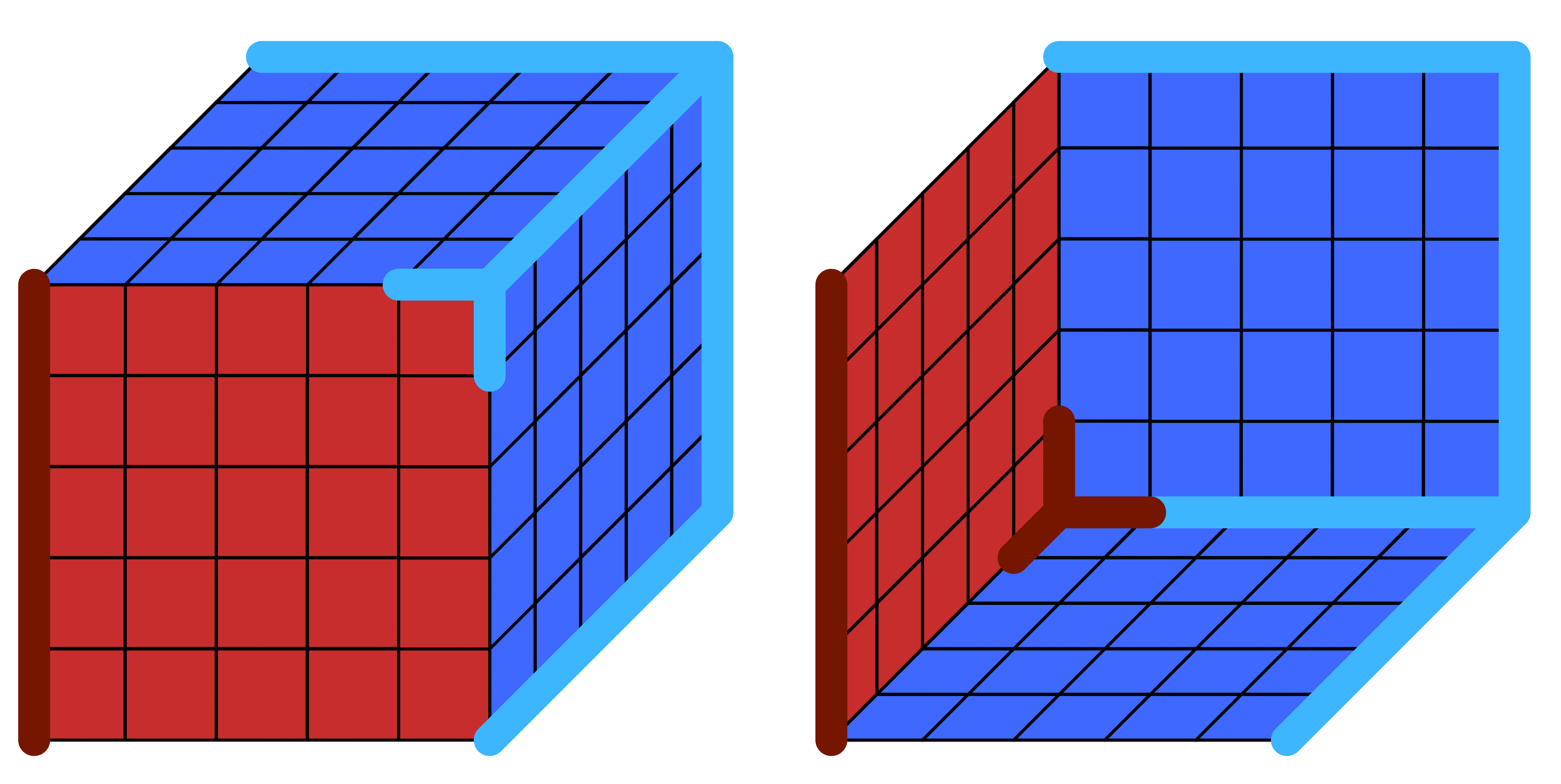}}& 
    $(mee;eme)$ & $(m)$ \& $(m_{ABC})$ & $0$ & - \\ 
    \raisebox{-0.5\totalheight}{\includegraphics[width=0.19\linewidth]{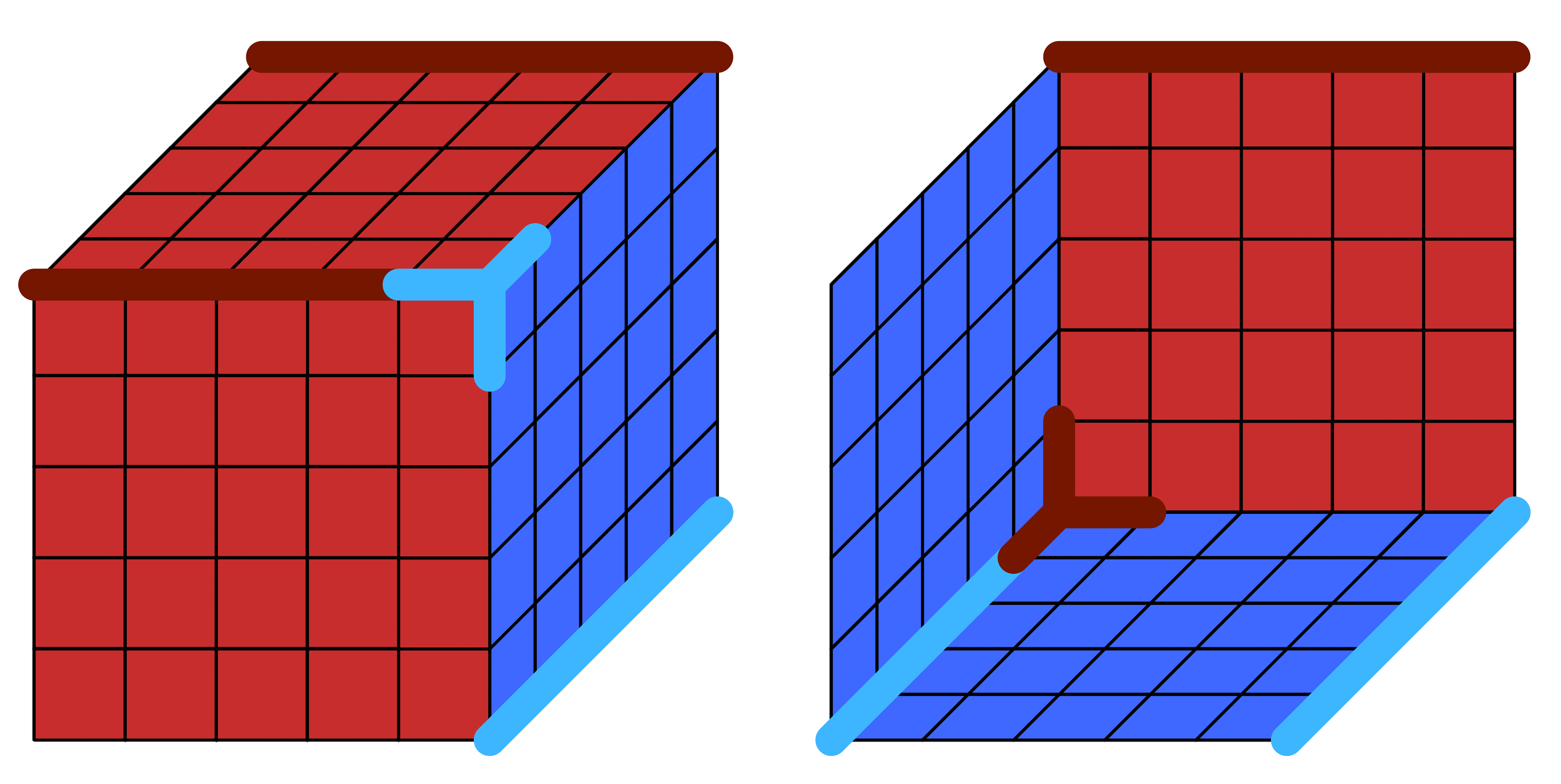}}&
    $(mem;mee)$ & 
      \emph{Tennis ball 1} & $2L_z$ & Linear, Superlinear \\
    \raisebox{-0.5\totalheight}{\includegraphics[width=0.19\linewidth]{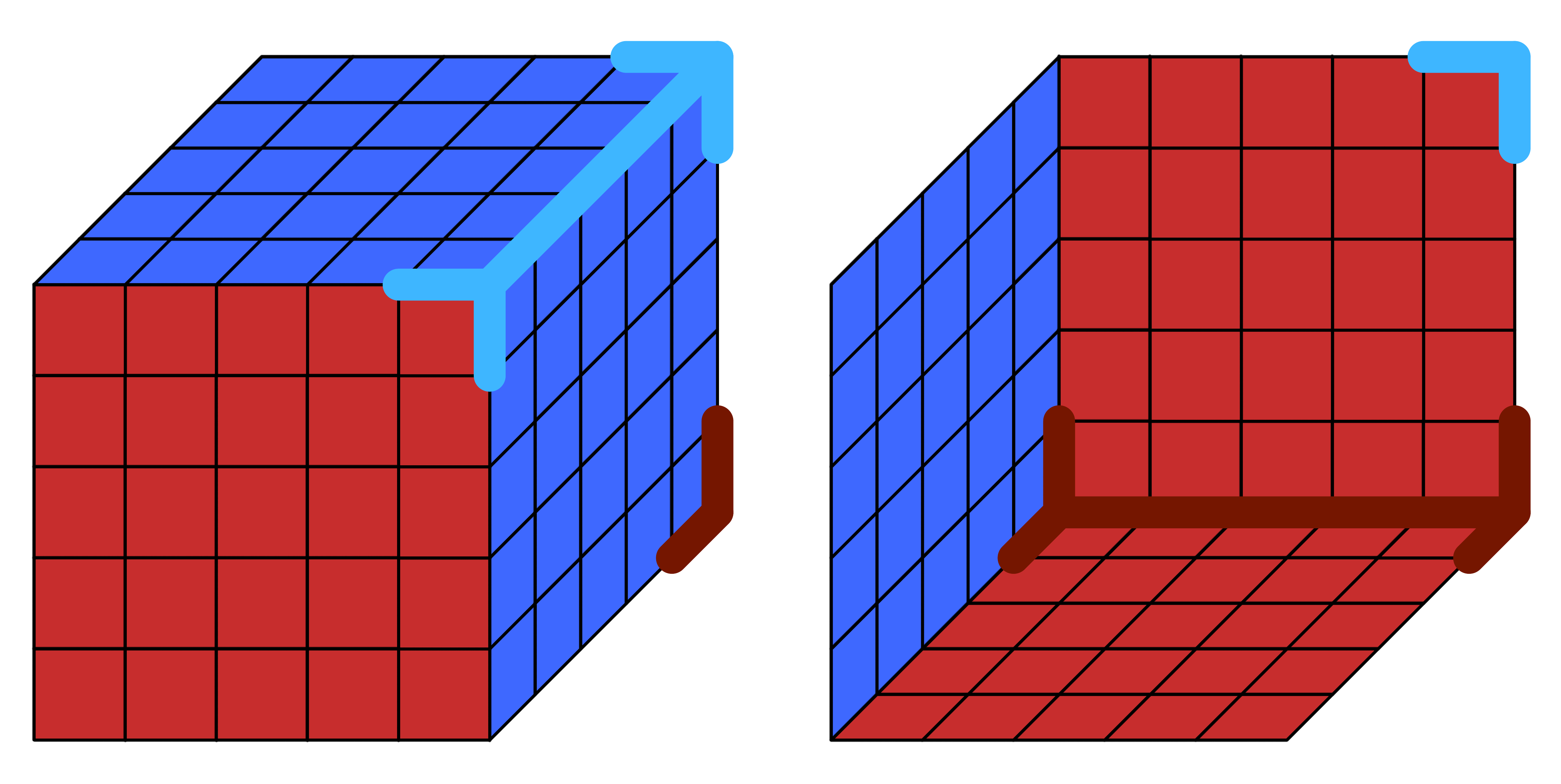}}& 
    $(mee;mem)$ & 
      \emph{Tennis ball 2} & $2L_z - 6$ & \makecell{Constant,
      \\Superlinear} \\ 
    \raisebox{-0.5\totalheight}{\includegraphics[width=0.19\linewidth]{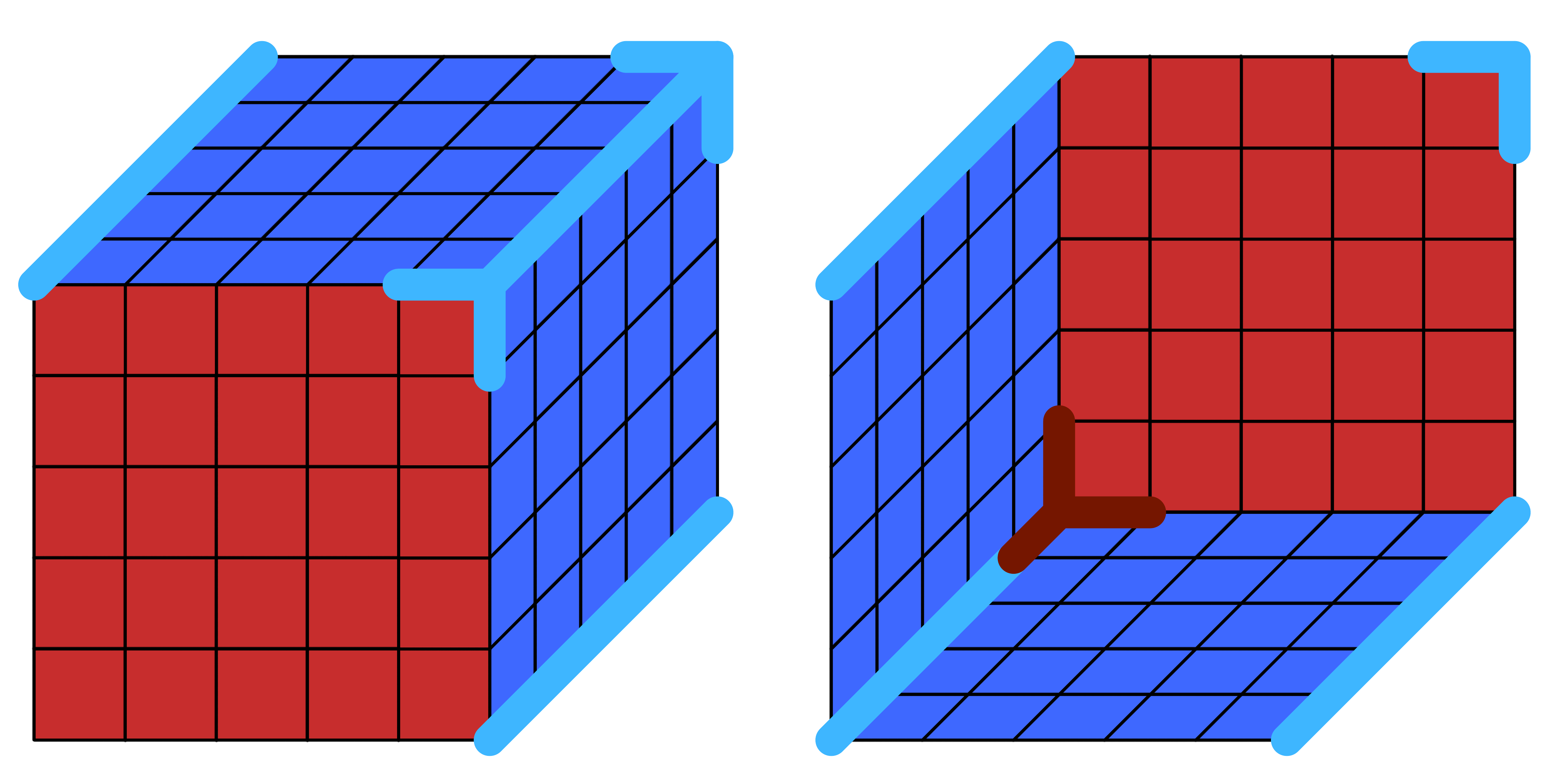}}& 
    \hspace{24pt}$(mee;mee)$\hspace{24pt} & 
    \emph{Tube} & $2(L_y+L_z-L_x)-3$ & \makecell{Constant, Linear, \\ Superlinear} \\
    \raisebox{-0.5\totalheight}{\includegraphics[width=0.19\linewidth]{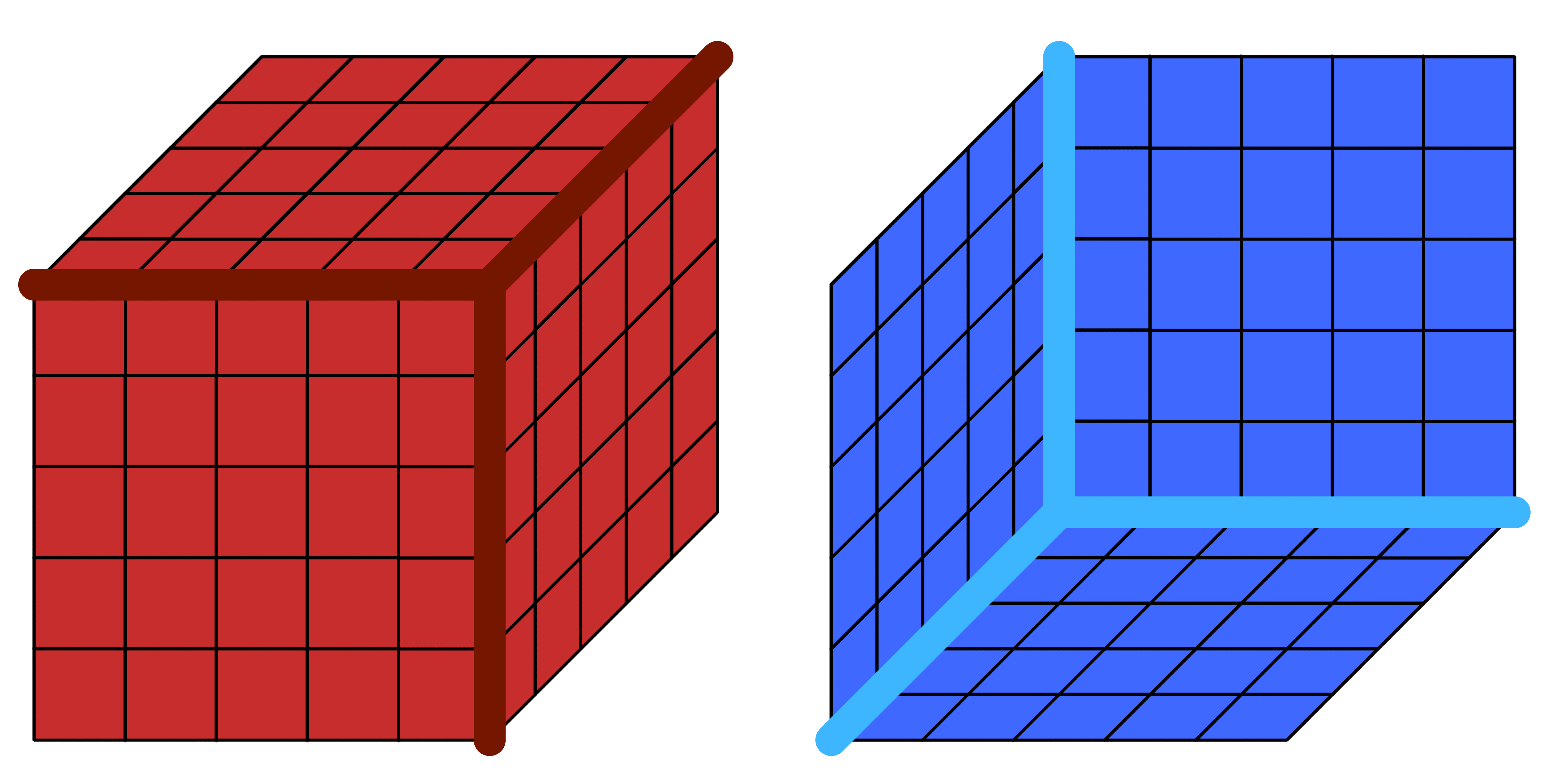}}&
    $(mmm;eee)$ & 
      \emph{Half-half 1} & $0$ & - \\ 
    \raisebox{-0.5\totalheight}{\includegraphics[width=0.19\linewidth]{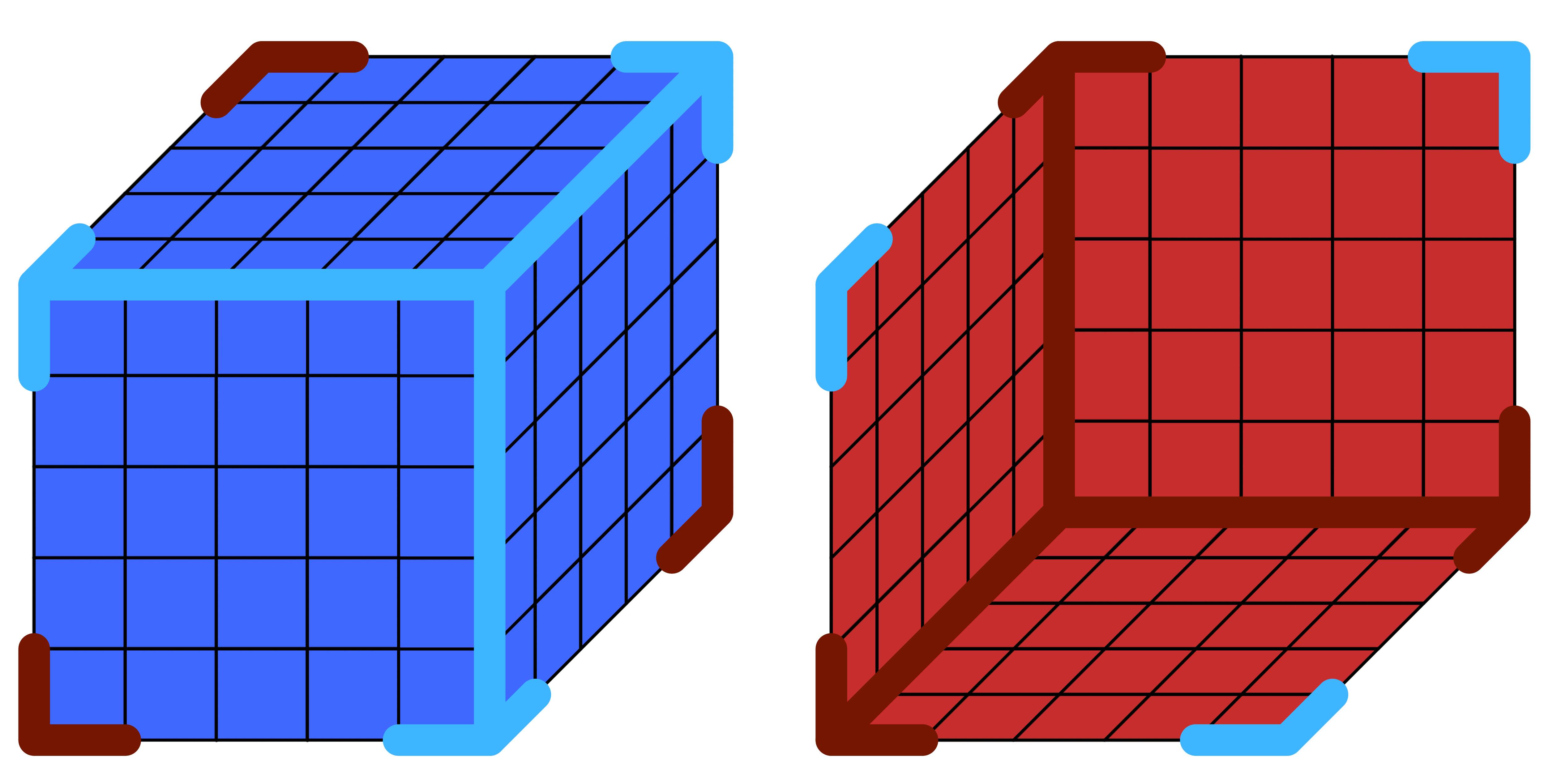}}&
    $(eee;mmm)$ & 
      \emph{Half-half 2} & $4\min\{L_x,L_y,L_z\} -12$ & \makecell{Constant,
      \\Superlinear} \\
    \raisebox{-0.5\totalheight}{\includegraphics[width=0.19\linewidth]{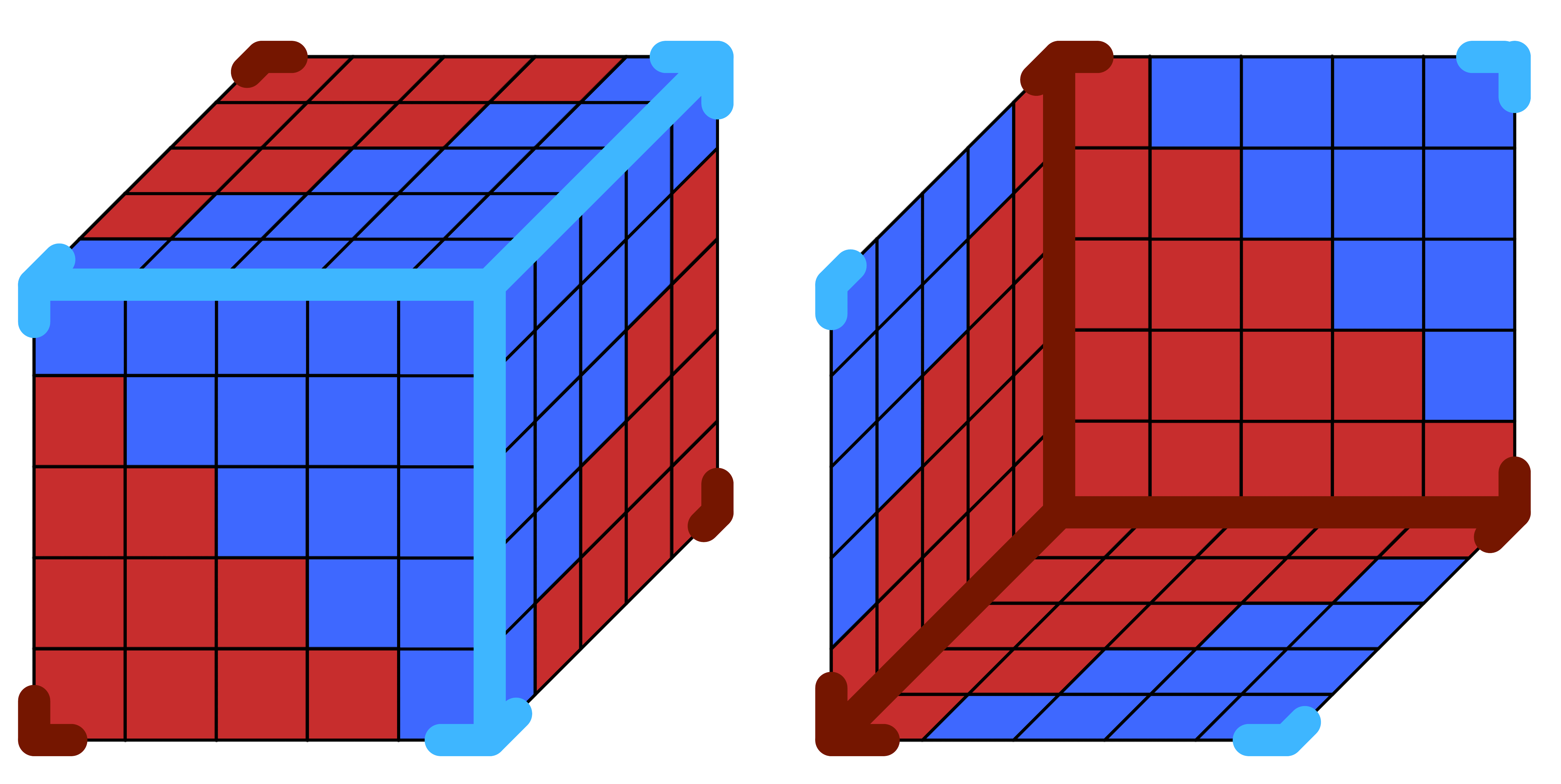}}& 
    - & 
    \emph{Triangular} & $\begin{aligned}4L-&4\\
\text{\small$(L=L_x=$}&\text{\small$L_y=L_z)$}\end{aligned}$ & Linear, Superlinear \\ 
    \bottomrule
  \end{tabular}
  \caption{Summary of the properties for codes with open boundaries on a
    lattice with linear system size $(L_x, L_y, L_z)$, showing the positive faces on the
    left and the negative faces on the right diagram. Red and dark red (blue and
  light blue) indicate $X$ ($Z$) stabilizers. Vertex stabilizers are implied
when three edges of the same color meet.
The half-length edges at the vertices of the \emph{triangular} configuration have vertex stabilizers but no edge stabilizers. 
    The notation column corresponds to the $(xyz;\bar x\bar y\bar z)$ faces of the lattice, where $p$ indicates a
    periodic boundary and $e$ and $m$ are open boundaries that interact with $e$ and $m$ excitations respectively. The final column identifies the presence of at least one logical operator ($\bar X$ or $\bar Z$) with a minimum weight that scales with the indicated function of $L_x, L_y,$ and/or $L_z$. 
}
  \label{table:boundary_codes}
\end{table*}

In this paper, we consider the construction of $X$- or $Z$-type open boundaries normal to a crystallographic axis. These boundaries are gapped using \emph{plaquette} stabilizers formed from truncated bulk $X$ and $Z$ cube stabilizers respectively. Both cases exhibit two topologically distinct interactions with the fractonic excitations: $X$-type boundaries on the negative-oriented faces of the lattice (and $Z$-type on the positive faces) condense single fractons. Conversely, $X$-type positive faces (and $Z$-type negative faces) cause their corresponding fractons to gain a $(2+1)$D mobility within diagonal subsystems along the surface. These boundary layers have a direct correspondence to the 6-6-6 color code~\cite{bombinTopologicalQuantumDistillation2006}.

Our results demonstrate that the no-string property of the closed cubic code model is not readily retained in the presence of open boundary conditions.
Because of this, it is nontrivial to construct QEC codes with open boundary conditions that have the desired superlinear code distance. Table \ref{table:boundary_codes} summarizes the different configurations of open boundary conditions considered in this paper; none contain only logical operators with superlinear weight. Notably, however, the scaling of $k$ in all nontrivial cases is now a simple function of the linear system size $L$ that does not exhibit large fluctuations.

It is nevertheless possible to create a cubic code with open boundary conditions that has a superlinear distance. For this, we use the \emph{tennis ball 1} configuration from Table \ref{table:boundary_codes}. In the corresponding code, logical $\bar X$ and $\bar Z$ operators stretch between two boundaries in the $\hat x$ and $\hat y$ lattice directions respectively. There exist logical $\bar X$ operators supported solely near the $+\hat z$ face, and $\bar Z$ operators near the $-\hat z$ face, that have linear weights. However, those further in the bulk have weight superlinear in $L$. A subsystem code~\cite{Bacon2005a,Poulin2005} can be used to gauge out the logical qubits with either an $\bar X$ or $\bar Z$ that can be supported near a boundary face - thus producing a code with superlinear distance. 

Alternatively, with boundary conditions that are periodic in the $\hat z$ direction only, we are able to construct stabilizer codes with simple linear scaling of $k$ and a superlinear code distance without resorting to subsystem codes. This result, along with other periodic boundary condition codes, is summarized in Table \ref{table:periodic_codes}.

In addition to open boundary conditions, we study the inclusion of crystal lattice defects including vacancies, edge dislocations, and screw dislocations. Similar to the open boundaries, we focus on configurations that are aligned with the crystallographic axes. While modified stabilizer terms are provided for vacancies and edge dislocations, screw dislocations do not admit additional deformed stabilizers. We explore how condensation of fractons on defects affects the fracton mobility in the vicinity of these features. We propose several encodings using defects; configurations such as a pair of edge dislocations or multiple vacancies wrapped around a periodic boundary can form stabilizer code families with superlinear distances and a simple linear scaling of $k$. 
These results are discussed in detail in Section \ref{sec:superdefects}. 

To the best of our knowledge, this work constitutes the first exploration of boundaries and defects in a type-II fracton topological order. 
Our results demonstrate that introducing defects and boundaries into the cubic code leads to encodings with new features that could prove advantageous over encodings based on periodic boundary conditions. 
This includes encodings with a number of logical qubits that scales linearly with the linear system size, without fluctuations. 
This work is a first step towards a general theory of translation symmetry enrichment in type-II fracton topological orders. 

\begin{table}
  \centering
  \begin{tabular}{cccc}
    \toprule
    Notation & Encoded Qubits ($k$) & Code Distance \\
    \cmidrule(r){1-1} \cmidrule(lr){2-2} \cmidrule(l){3-3} 
    
    $(ppp;ppp)$ & Eq.~(\ref{eq:cubic_k}) & Superlinear \\
     
    $(ppe;ppe)$ & $\frac12 k_{(ppp;ppp)} + 2\tau(L;\; \infty)$ & Linear \\

    $(ppm;ppe)$ & $4\min\{\tau(L_x;\;L_z),\, \tau(L_y;\;L_z)\}$ &
    Linear \\

    $(ppe;ppm)$ & 0 & - \\

    $(pmm;pem)$ & $2\tau(L_x;\;L_z)$ & Linear  \\

    $(pem;pem)$ & $2L_x$ & Superlinear$^*$ \\

    $(pem;pme)$ & 0 & - \\

    $(pem;pmm)$ & 0 & - \\ 

    $(pmm;pmm)$ & 0 & -  \\

    $(pmm;pee)$ & 0 & - \\

    $(pee;pmm)$ & 0 & - \\
     
    \bottomrule
  \end{tabular}
  \caption{Summary of the properties for boundary codes on a lattice with
    linear system size $(L_x, L_y, L_z)$, where some direction is periodic.
    In cases where the behavior of anisotropic systems is unclear, we take
    $L_x=L_y=L_z \equiv L$. The notation column corresponds to the $(xyz;\bar x\bar y\bar z)$ faces of the lattice, where $p$ indicates a
    periodic boundary and $e$ and $m$ are open boundaries that interact with $e$ and $m$ excitations respectively. $\tau(L_1;\;L_2)$ is a fractal-like function defined in Section \ref{sec:periodic} that encapsulates the number of string-like operators that can form logical operators by wrapping around a periodic boundary. 
    \hspace{\textwidth} $\,^*$Superlinearity is only ensured when
    $3\!\nmid\!\!L_x$, otherwise there are linear-weight operators.}
  \label{table:periodic_codes}
\end{table}

\subsection{Outline of Paper}

This paper is organized as follows: In Section \ref{sec:bg} we present 
background on quantum error-correcting codes and outline the key properties of the
cubic code. In Section \ref{sec:bdries} we
characterize the open boundary conditions of the cubic code. 
In Section \ref{sec:superbdries} we discuss constructions of superlinear-distance codes. 
In Section \ref{sec:defects} we characterize the inclusion of defects - vacancies, edge dislocations, and screw dislocations. 
In Section \ref{sec:superdefects} we discuss the use of defects to construct superlinear-distance codes. 
In Section \ref{sec:conc} we present our conclusions. 
The appendices include a discussion of further open and periodic boundary codes (Appendix \ref{sec:otherbdries}) and defect codes (Appendix \ref{sec:otherdefects}) considered in this study. 
A summary of all the potential boundary encodings is provided in Tables \ref{table:boundary_codes} and \ref{table:periodic_codes}.

%% file: background.tex
We begin with a brief review of quantum error-correcting codes and self-correcting
quantum memories, before discussing the cubic code in particular.

\subsection{\label{sec:qec}Review of Quantum Error-Correcting Codes}

A quantum error-correcting (QEC) code is a scheme to encode one or more quantum
states within a higher-dimensional Hilbert space in order to provide the
ability to detect and correct a class of errors. Arbitrary errors are generated by the algebra of single-qubit Pauli errors, spanned by the Pauli $X$ and $Z$ operators
\begin{equation}
  X := \begin{pmatrix} 0 & 1 \\ 1 & 0 \end{pmatrix},\qquad 
  Z := \begin{pmatrix} 1 & 0 \\ 0 & -1 \end{pmatrix},
\end{equation}
written in the computational basis $\{\ket0,\ket1\}$ corresponding to the two states of a physical qubit. One approach for
constructing QEC codes is to employ the \emph{stabilizer formalism}~\cite{calderbankQuantumErrorCorrection1997,gottesmanHeisenbergRepresentationQuantum1998}: Consider a system of $n$ physical
qubits. We select a commuting collection of (tensor) products of Pauli operators and consider the stabilizer group $\mathcal{S}$ that they generate. We require that $-I \notin \mathcal{S}$, where $I$ is the identity operator. Quantum
information is then encoded in the eigenvectors of the degenerate
$+1$-eigenspace common to all elements of $\mathcal S$. That is, any physical
measurement of the encoded state using $S_i \in \mathcal S$ returns a $+1$ value.
Importantly, single-qubit $X$ and $Z$ operators anti-commute with some $S_i$,
thus mapping any encoded state out of the $+1$-eigenspace and producing a change
in the measurement outcomes. This can be detected and corrected with appropriate QEC codes. 

For a system with $s$ independent stabilizer generators and $n$ physical qubits, the degeneracy of the
$+1$-eigenspace is $2^{n-s}$. Equivalently, the number of encoded \emph{logical
qubits} that are protected from errors is $k = n-s$. 

An effective QEC code should have a large number of encoded qubits but also
should make it difficult for errors to affect the encoded information. A logical
operator, denoted as $\bar X$ or $\bar Z$, is an operator that 
commutes with all stabilizers, yet is not
itself in the stabilizer group. In this way, logical operators act on the encoded
states within the $+1$-subspace, changing the state of the logical qubit while not being detectable using stabilizer measurements. Effective QEC codes must
therefore make it difficult for errors to create a logical operator (logical
error). We quantify this difficulty in two ways: code distance and energy barriers. 

\paragraph{Code Distance}
Operators that act on encoded quantum states are only uniquely defined modulo
multiplication by stabilizers. The weight of a logical operator - the number of
single-qubit Pauli operators required to construct it - is therefore variable.
The code distance is defined as the minimum weight operator that can
create a logical error on the code, taking into account this multiplication by
stabilizers.

\paragraph{\label{sec:selfcorrection}Energy Barriers}

Additionally, we can consider the physical qubits in a QEC code as forming a quantum condensed matter system, evolving under a Hamiltonian 
\begin{equation}
  H = -\sum_{i=1}^s S_i
\end{equation}
where $\{S_i\}_{i=1}^s$ are spatially-local operators that generate the stabilizer group $\mathcal S$. Since the encoded states belong to the $+1$-eigenspace of all $S_i$, they also
correspond to the ground state (minimum energy state) of this system. Pauli errors then map an encoded state into the $-1$-eigenspace of some $S_i$, thus increasing the energy. These \emph{flipped} or \emph{excited} stabilizers can be interpreted as the location of
excitations or quasiparticles, with emergent behavior such as mobility, charge,
and even braiding statistics. The ability of a code to correct against local errors is equivalent to the condition of \emph{topological order}: the state of the system cannot be determined solely by local operations~\cite{wenTopologicalOrdersRigid1990,dennisTopologicalQuantumMemory2002, bravyiTopologicalQuantumOrder2010, wenColloquiumZooQuantumtopological2017}.

\begin{figure}[t]
  \centering
  \includegraphics[width=0.95\linewidth]{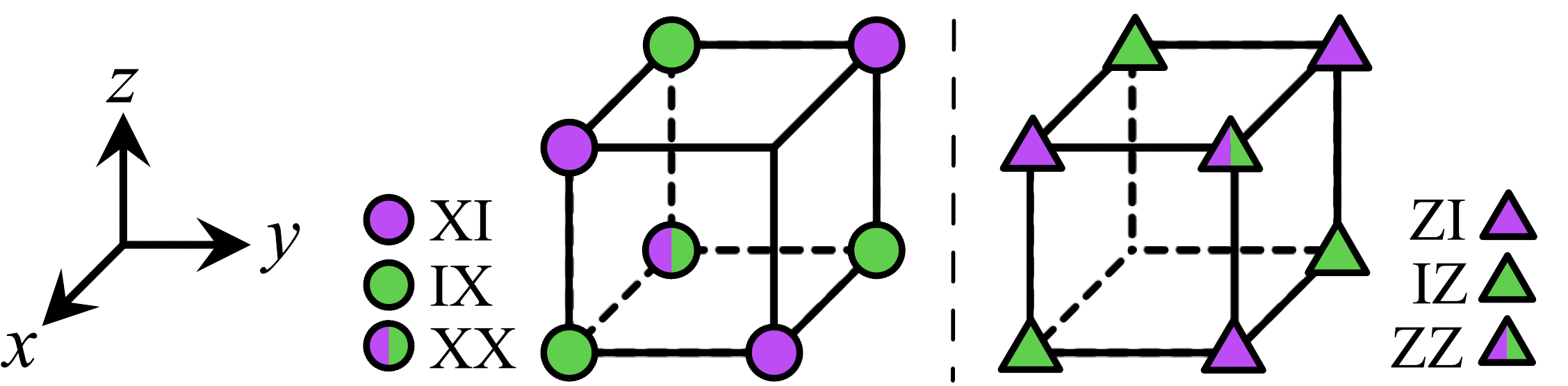} 
  \caption{Generators of the cubic code stabilizers, comprised of single-qubit
    Pauli operators. {(Left)} The $C_X$ operator. {(Right)} The $C_Z$
  operator.}
  \label{fig:cubic_stabilisers}
\end{figure} 

The energy barrier of this system is then defined as the minimum energy that
must be surpassed in order to create a logical error by sequentially applying
single-qubit Pauli errors. Importantly, larger energy barriers will cause the evolution 
of the system to naturally suppress the creation of logical errors when at nonzero temperatures. 
That is, for an energy barrier $E$, entropy change
$S$, temperature $T$ and Boltzmann constant $k_B$, the time a system can remain
in its encoded state (quantum memory time) scales approximately via the Arrhenius
Law~\cite{robertsSymmetryProtectedSelfCorrectingQuantum2020} 
\begin{equation}
  \tau_\text{lifetime} \sim \exp\left(\frac{E-TS}{k_BT}\right)
  \label{eq:arrhenius}
\end{equation}
A \emph{self-correcting quantum memory} at finite temperature is then defined as a QEC code where the
lifetime grows without bound in the number of physical qubits, at sufficiently
small nonzero temperature~\cite{bombinSelfcorrectingQuantumComputers2013}. A
necessary condition for this, therefore, is that the energy barrier must grow
with the system size. However, this behavior is impossible in all stabilizer
codes formed by arranging the qubits in $(2+1)$D and demanding the stabilizers be spatially local~\cite{alickiThermalizationKitaev2D2009,
bravyiNogoTheoremTwodimensional2009,
bravyiTradeoffsReliableQuantum2010,
alickiThermalStabilityTopological2010,
yoshidaFeasibilitySelfcorrectingQuantum2011,
haahLogicaloperatorTradeoffLocal2012,
landon-cardinalLocalTopologicalOrder2013,
brownQuantumMemoriesFinite2016}.
These no-go theorems do not necessarily apply to higher dimensions. In
particular, self-correction is readily possible in $(4+1)$D~\cite{dennisTopologicalQuantumMemory2002,alickiThermalStabilityTopological2010}. In $(3+1)$D, there have been
several attempts at creating such behavior~\cite{chesiSelfcorrectingQuantumMemory2010, yoshidaFeasibilitySelfcorrectingQuantum2011, bombinSelfcorrectingQuantumComputers2013, 
beckerDynamicGenerationTopologically2013, brellProposalSelfcorrectingStabilizer2016, brownQuantumMemoriesFinite2016, robertsSymmetryProtectedSelfCorrectingQuantum2020}. One of the more
promising candidates, and one of the only exactly-solvable candidates, is known as the \emph{cubic code}.

\begin{figure}[t]
  \centering 
  \includegraphics[height=78pt]{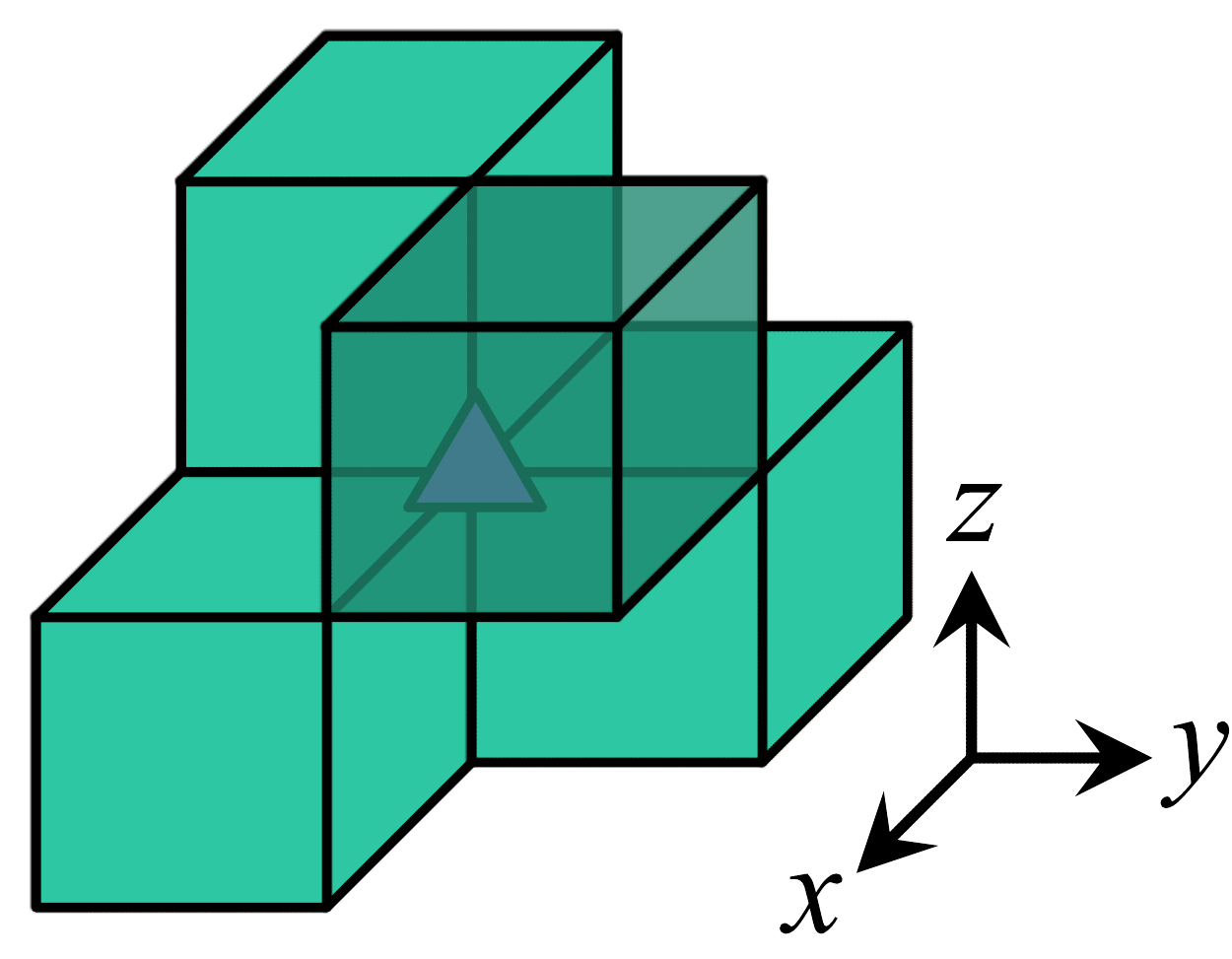}
  \includegraphics[height=78pt]{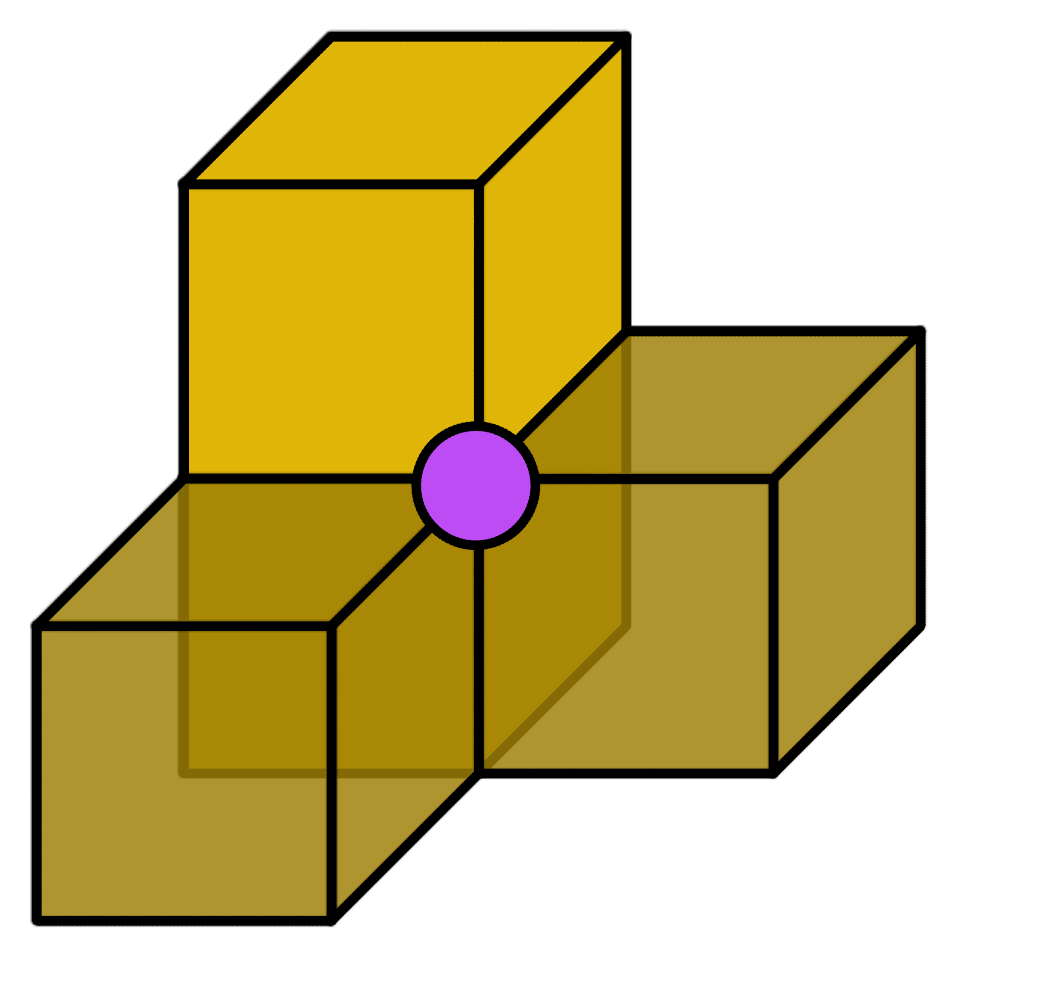}

  \includegraphics[height=78pt]{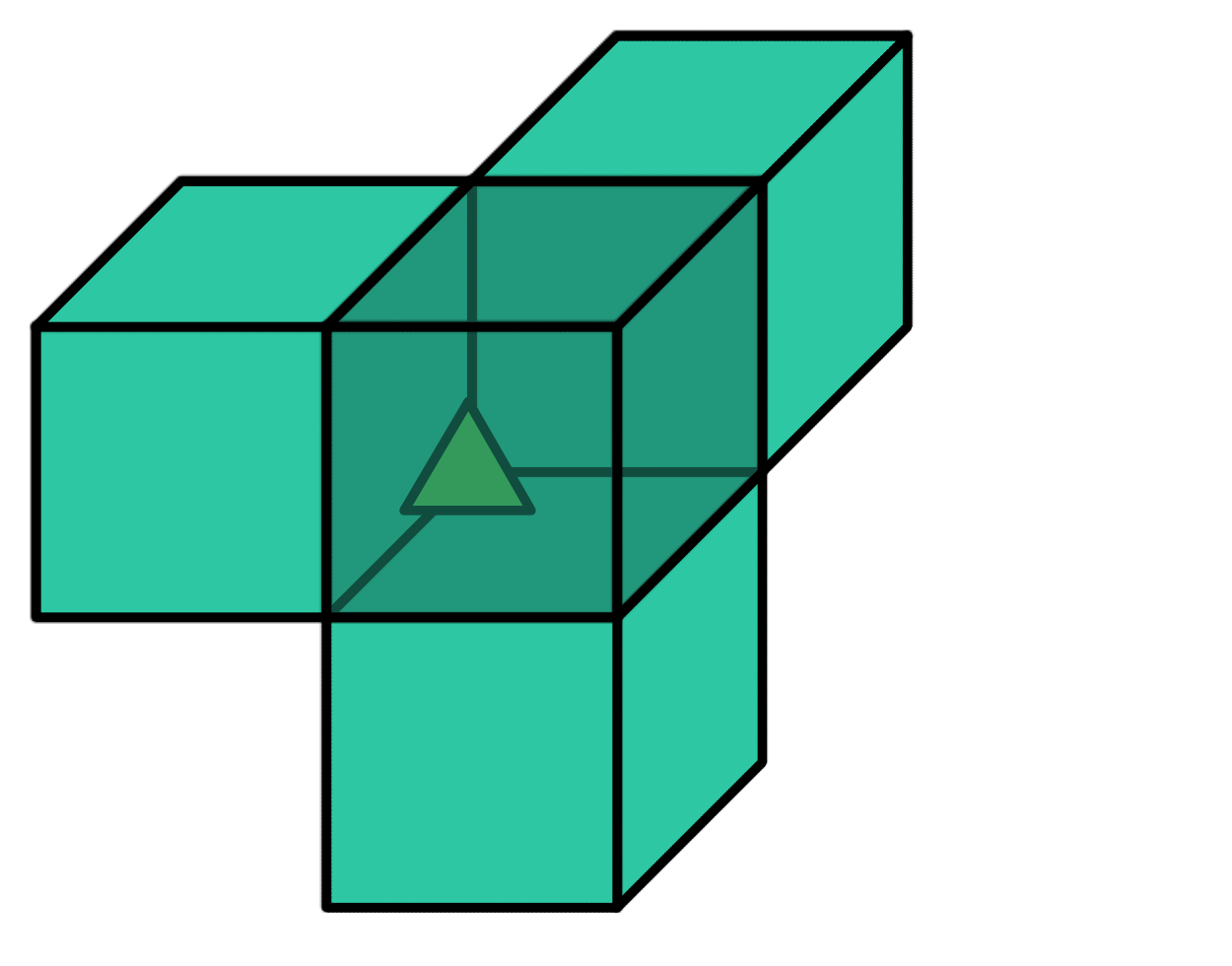}
  \includegraphics[height=78pt]{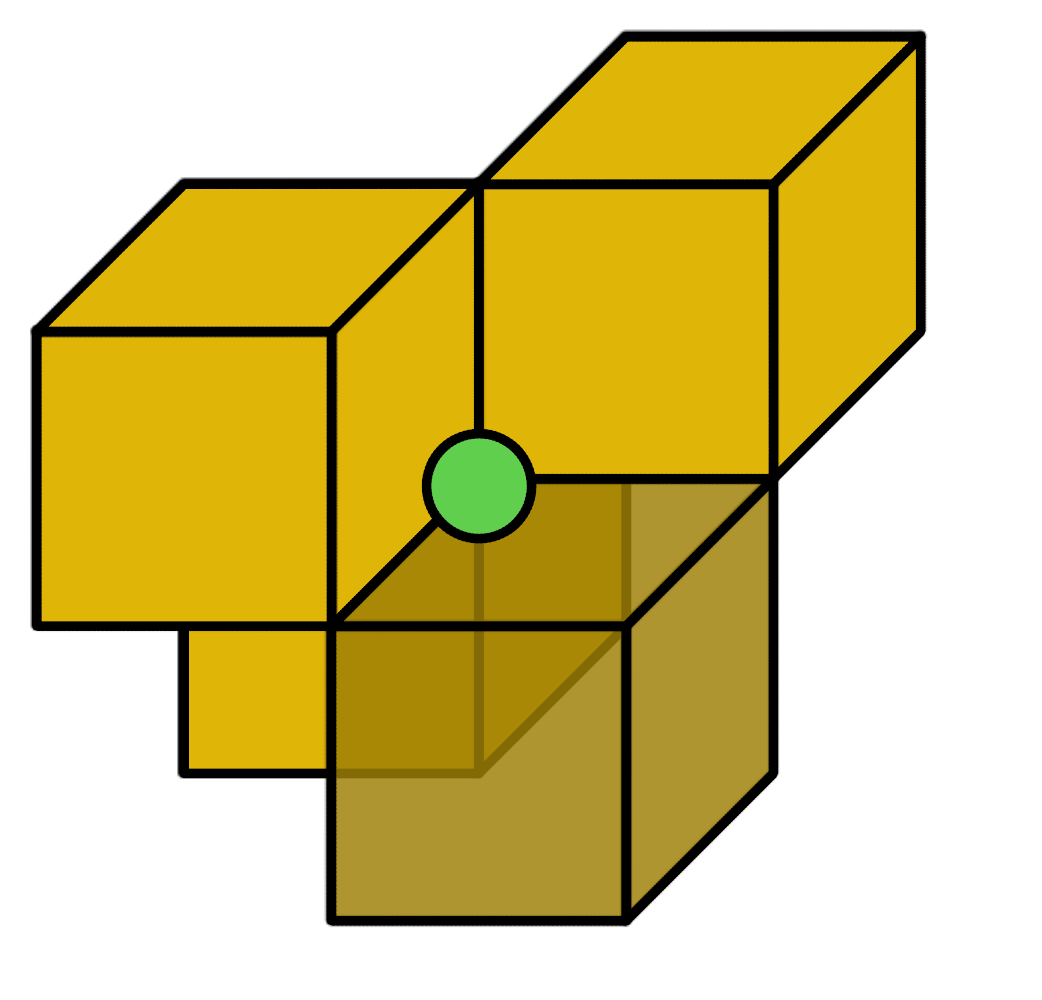}
  \caption{As per
    Table \ref{table:fracton_charges}, each single-qubit Pauli operator
    creates $4$ excitations of cubic code stabilizers, arranged in a tetrahedral shape. Transparency is used to show cubes hidden by
  the 3D perspective.}
  \label{fig:cubic_excitations}
\end{figure}

\begin{table}[t]
  \centering
    \begin{tabular}{cccc}
      \toprule
      Charge & Created By & Excited Stabilizer & Color
      \\ 
      \cmidrule(r){1-1} \cmidrule(lr){2-2} \cmidrule(lr){3-3} \cmidrule(l){4-4}
      $e$ & \raisebox{-0.5\totalheight}{\includegraphics[height=32pt]{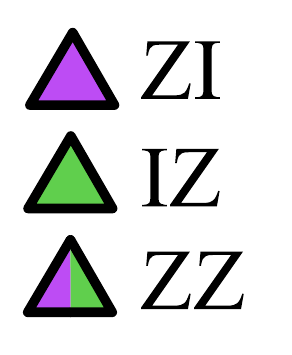}} & $C_X$ &
      \raisebox{-0.2\totalheight}{\includegraphics[height=12pt]{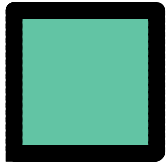}}  \\[16pt]
      $m$ & \raisebox{-0.5\totalheight}{\includegraphics[height=32pt]{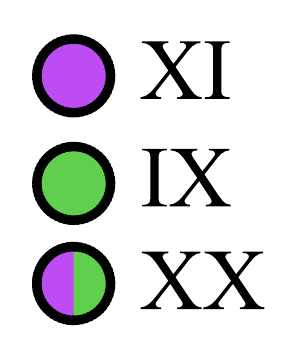}} & $C_Z$ & \raisebox{-0.2\totalheight}{\includegraphics[height=12pt]{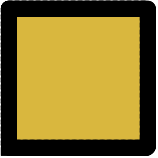}} \\ 
    \bottomrule
  \end{tabular}
  \caption{Elementary charges of the cubic code model. Filled triangles and circles are used throughout this manuscript to refer to the Pauli operators. Color refers to the
    convention for indicating created charges.}
  \label{table:fracton_charges}
\end{table}

\subsection{\label{sec:cubic}The Cubic Code}

Initially proposed in Ref.~\cite{haahLocalStabilizerCodes2011}, the cubic code is defined on a $(3+1)$D simple cubic
lattice with two qubits at each lattice site and periodically identified boundaries in all directions, forming a $3$-torus. 
We use the notation $XI$ to denote the operator $X\otimes I$ acting on the two
qubits at a particular lattice site, with the identity on all other qubits.
The model has two kinds of stabilizer generators, both with support on a subset of the 16
qubits at the 8 vertices of a unit cube. These generators, referred to here
as $C_X$ and $C_Z$, are shown in Fig.~\ref{fig:cubic_stabilisers}.

Pauli operators create tetrahedral excitation patterns in the neighboring
stabilizers, highlighted in Fig.~\ref{fig:cubic_excitations}. Following the
convention with the surface code, we denote these two types of excitations as
$e$ and $m$ as in Table \ref{table:fracton_charges}.

\subsubsection{\label{sec:symmetries} Lattice Symmetries}

Noting the form of the stabilizers and their excitation patterns, there are three lattice symmetries of the cubic code relevant for discussions in this paper:
\begin{enumerate}
  \item $3$-fold rotation about $\hat x + \hat y + \hat z$.
  \item Mirror symmetry about the plane normal to $\hat x - \hat y$.
  \item The map $\{IX\leftrightarrow ZI, IZ \leftrightarrow
    XI\}$ combined with spatial inversion.
\end{enumerate}
These symmetries are used to relate boundaries and defects with different orientations in later sections.

\subsubsection{\label{sec:cubic_properties}Encoding Properties}
The defining property of the cubic code is that all its nontrivial topological quasiparticle excitations\footnote{A nontrivial topological excitation is one that cannot be created locally.} are fundamentally immobile \emph{fractons}. This is equivalent to a property called ``no
string-like operators''~\cite{haahLocalStabilizerCodes2011}. A string-like operator is any operator that creates two
constant-sized regions of nontrivial topological excitations, separated by an arbitrary distance $\Delta$, with a weight that
scales linearly with $\Delta$. In doing so, these operators incur a maximum energy cost that 
is independent of $\Delta$. 
Since such string-like operators do not exist in the cubic code, fractons cannot be moved large distances with constant energy. This immobility property is discussed further in Section \ref{sec:cubic_movement}. 

Importantly, this behavior results in a code distance that scales
superlinearly with the linear system size $L$ (the number of sites in each lattice direction), and an energy barrier that scales
logarithmically with $L$. Although seemingly promising for use in self-correction, we also need to consider the effects of entropy. The entropy of a system is typically extensive, scaling polynomially with system size~\cite{bravyiEnergyLandscape3D2011}. As per
Eq.~(\ref{eq:arrhenius}), at finite temperature there will thus exist a given system size where increasing $n$ further results in a decrease to the lifetime, as entropic contributions overwhelm the system. This energy barrier is
therefore enough to ensure only \emph{partial self-correction} of the system~\cite{bravyiAnalyticNumericalDemonstration2013}.
Nevertheless, it remains one of the only stabilizer codes to achieve even this behavior.

\begin{figure}[t]
  \centering 
  \includegraphics[width=\linewidth]{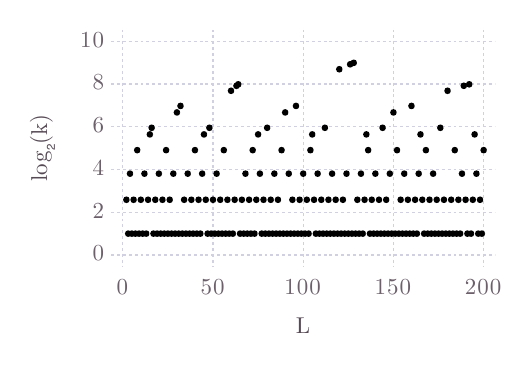}
  \vspace{-10mm}
  \caption{Number of encoded qubits, $k$, in the periodic cubic code model with
    linear system size $(L,L,L)$ in $x,y,z$, as per Eq.~(\ref{eq:cubic_k}). The value is
    bounded by $2\leq k \leq 4L-2$.}
  \label{fig:cubic_k}
\end{figure}

A key theorem for self-correcting quantum memories is that for there to be no
string-like operators, the ground state degeneracy - or equivalently, the number of encoded qubits $k$ - in a $(3+1)$D translation-invariant stabilizer code must depend on the system size~\cite{yoshidaFeasibilitySelfcorrectingQuantum2011}. In the case of the
cubic code, $k$ fluctuates significantly, bounded by $2 \leq k \leq
4L-2$, where we take a geometry with equal linear system size $L$ in $x,y,$ and $z$. For notation,
we define 
\begin{equation}
  q_n(L) = \begin{cases} 1 & n | L \\ 
    0 & \text{otherwise} 
  \end{cases}
  \label{eq:qn}
\end{equation}
and 
\begin{equation}
  \zeta(L) = \begin{cases} 
    \operatorname{max}\left\{ 2^z \, : \, 2^z | L,\, z\in\mathbb Z
    \right\} & L \in \mathbb Z \\ 
    0 & \text{otherwise}
  \end{cases}
  \label{eq:zeta}
\end{equation}
Using this, the exact empirical formula for $2 \leq L \leq 200$ is given by (see
Ref.~\cite{haahLocalStabilizerCodes2011})
\begin{equation}
  k = 2\left[ 1 -2q_2 + 2\zeta(L) \left(q_2 + 12q_{15} + 60q_{63} \right)\right]
  \label{eq:cubic_k}
\end{equation}
where we have written $q_2 = q_2(L)$ etc. for readability. This relationship is plotted
in Fig.~\ref{fig:cubic_k}. A number-theoretic exact formula for $k(L)$ is known for all $L$ but the formula is cumbersome to write down and not enlightening for our purposes~\cite{haahCommutingPauliHamiltonians2013}.  Importantly, there is a strong dependence on the exact divisibility of $L$; changing $L \mapsto L+1$ can cause $k$ to
fluctuate by several orders of magnitude. 
A guiding question for this work is
whether the scaling can become a more consistent - ideally linear -
function of $L$ by introducing open boundaries and lattice defects.

\subsubsection{\label{sec:cubic_movement}Fracton Mobility}

As noted previously, there are no string-like operators through the bulk of the
cubic code model. Consequently, the topological quasiparticle excitations are strictly immobile and cannot
be moved through the lattice without incurring an additional energy penalty that
scales with the distance, $\Delta$. There
are three sufficient behaviors, presented below, that describe this generalized motion and form the basis of our arguments for code distance and energy barriers in later sections:

\begin{figure}[t]
  \centering 
  \includegraphics[width=\linewidth]{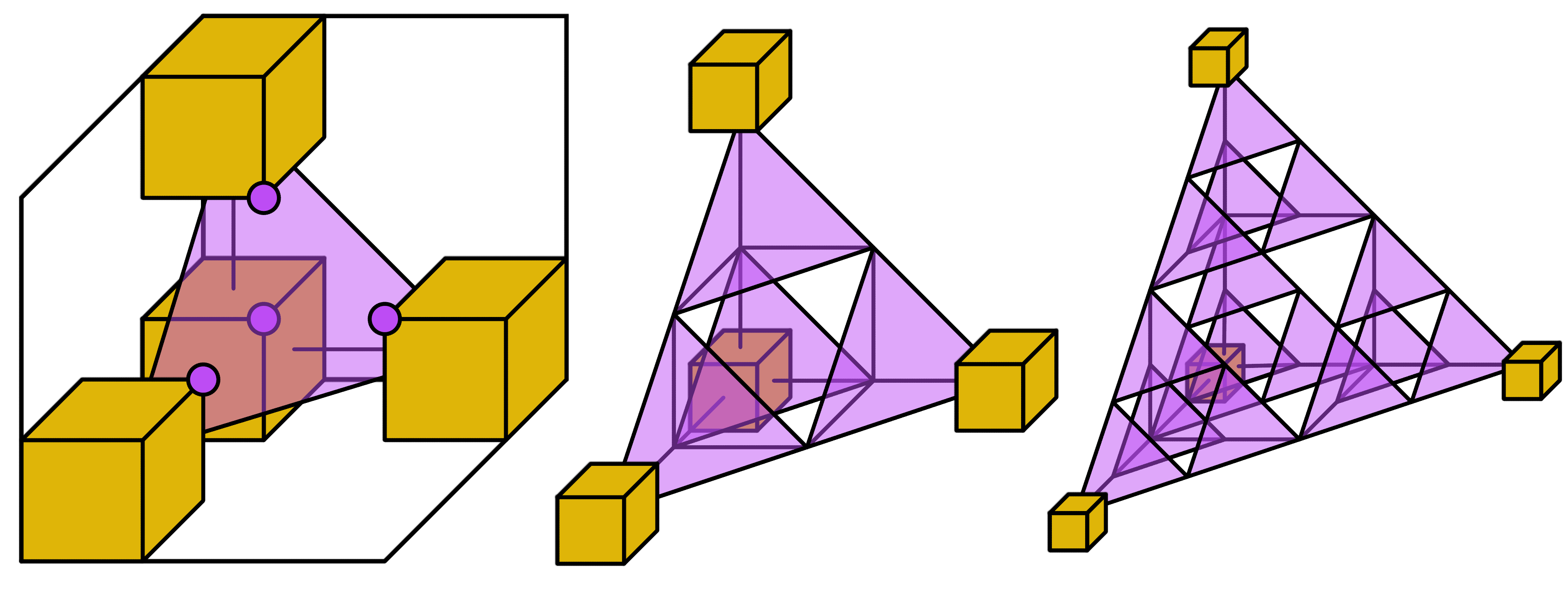}
  \caption{Fractal pattern used to expand the separation between a set of $4$
  $m$ excitations in the bulk, using repeated $XI$ Pauli operators (purple
circles). The shaded purple regions are used to highlight repeated applications
of the leftmost operator, showing similarities to a Sierpinski pyramid.}
  \label{fig:fractal_tetra}
\end{figure}

\paragraph{\label{sec:fractal}Fractal Operators}
Firstly, fractons can be moved through the bulk of the model using operators arranged in the shape of \emph{fractal} tetrahedra. Originally described in
Ref.~\cite{haahLocalStabilizerCodes2011}, the excitation patterns in
Fig.~\ref{fig:cubic_excitations} can be repeated in a fractal pattern to create
increasingly larger separations of charge, as in Fig.~\ref{fig:fractal_tetra}.
Notably, doing so requires multiple excitations to move outwards and it
involves intermediary high-energy states. It was from this fractal
behavior that the logarithmic energy barrier of the periodic cubic code 
was derived~\cite{bravyiEnergyLandscape3D2011}. Due to this fractal nature,
this process can create excitations separated by a distance $\Delta$ where
$\Delta = 2^j$ for $j=0,1,2,\ldots$. This dependence on powers of $2$
contributes towards the sporadic scaling of logical qubits in the
periodic cubic code: if the width of the lattice is not a power of $2$, multiple
smaller tetrahedra need to be combined in a nontrivial way, wrapping around the
periodic boundary to annihilate all charges. This motivates the form of
Eq.~(\ref{eq:zeta}).

\begin{figure}[t]
  \centering 
  \includegraphics[width=\linewidth]{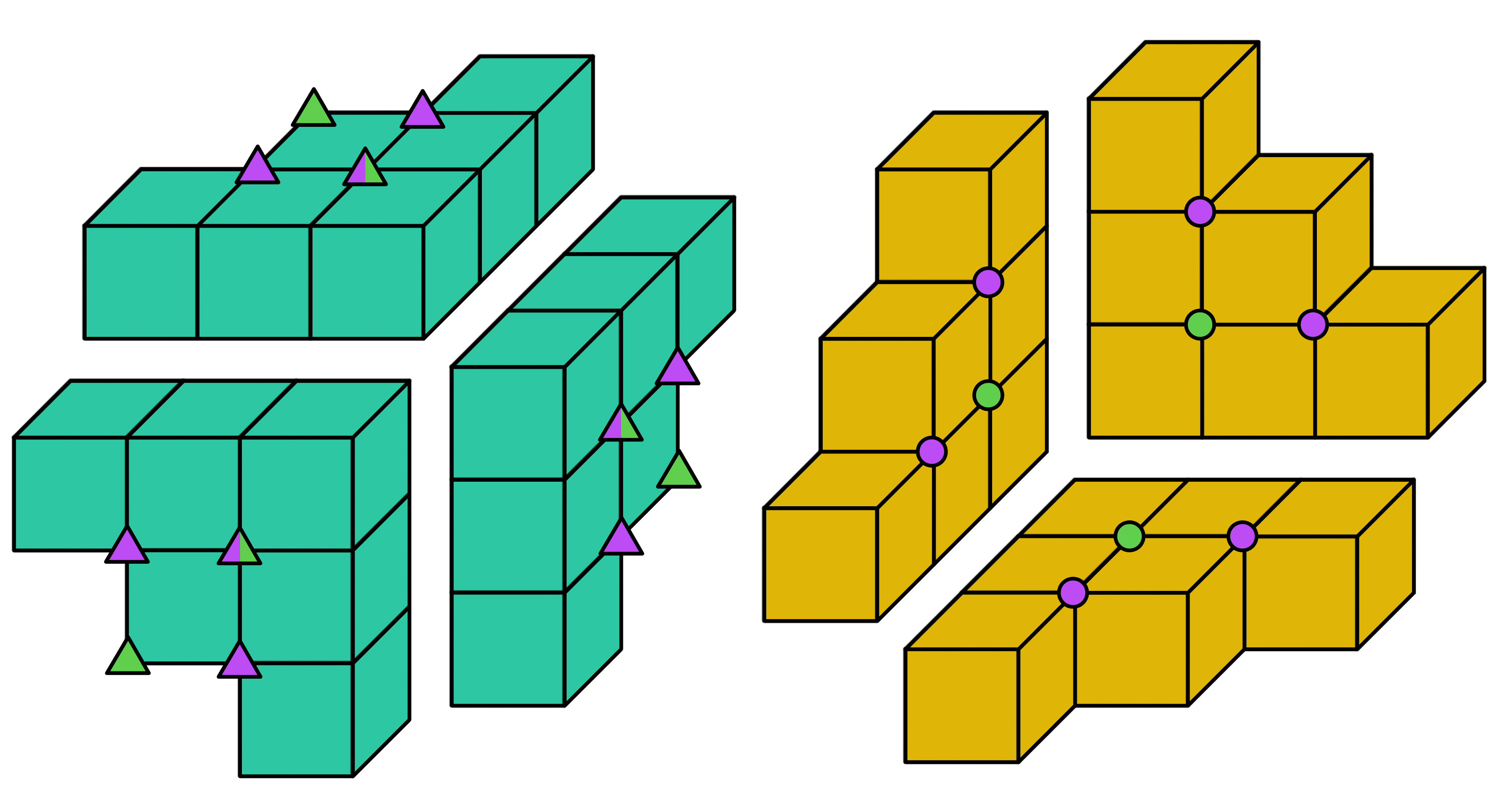}
  \caption{The $F$ operators that can cascade excitations through the lattice,
    as in Fig.~\ref{fig:fractal_move}.
  {(Left)} $F_e^{xy}$, $F_e^{xz}$, and $F_e^{yz}$ operators. The labels
indicate the plane and directions along which the excitations can move.
{(Right)} $F_m^{\bar x\bar y}, F_m^{\bar x \bar z}$, and $F_m^{\bar y \bar
z}$ operators, which move excitations in the negative lattice directions.}
\label{fig:f_operators}
\end{figure}

\begin{figure*}
  \centering 
  \subfloat[\label{fig:fractal_move0}$\Delta = 0$]{
  \includegraphics[height=7cm]{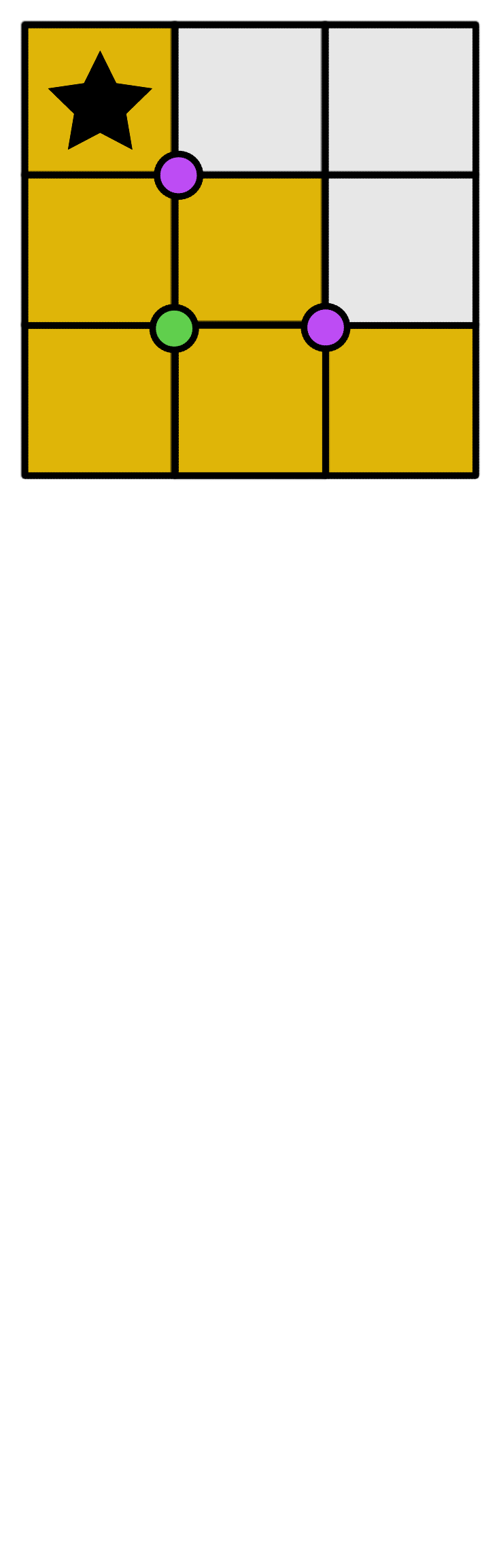} }
  \hspace{0.2cm}%
  \subfloat[\label{fig:fractal_move1}]{
  \includegraphics[height=7cm]{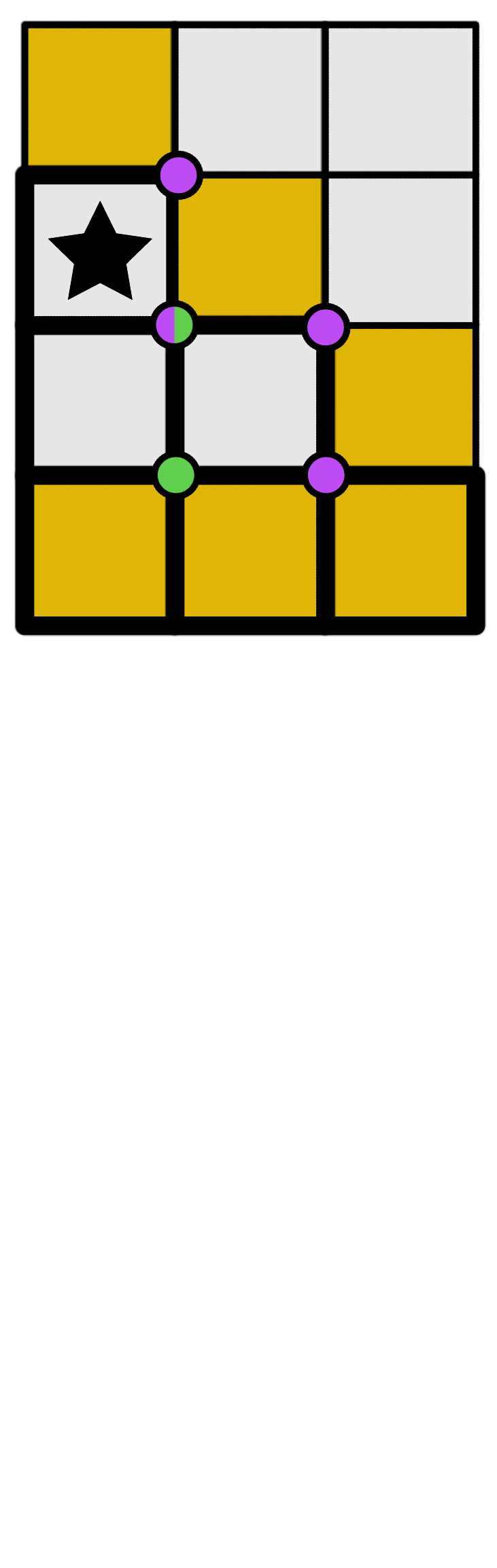} }
  \hspace{0.2cm}%
  \subfloat[\label{fig:fractal_move2}$\Delta = 1$]{
  \includegraphics[height=7cm]{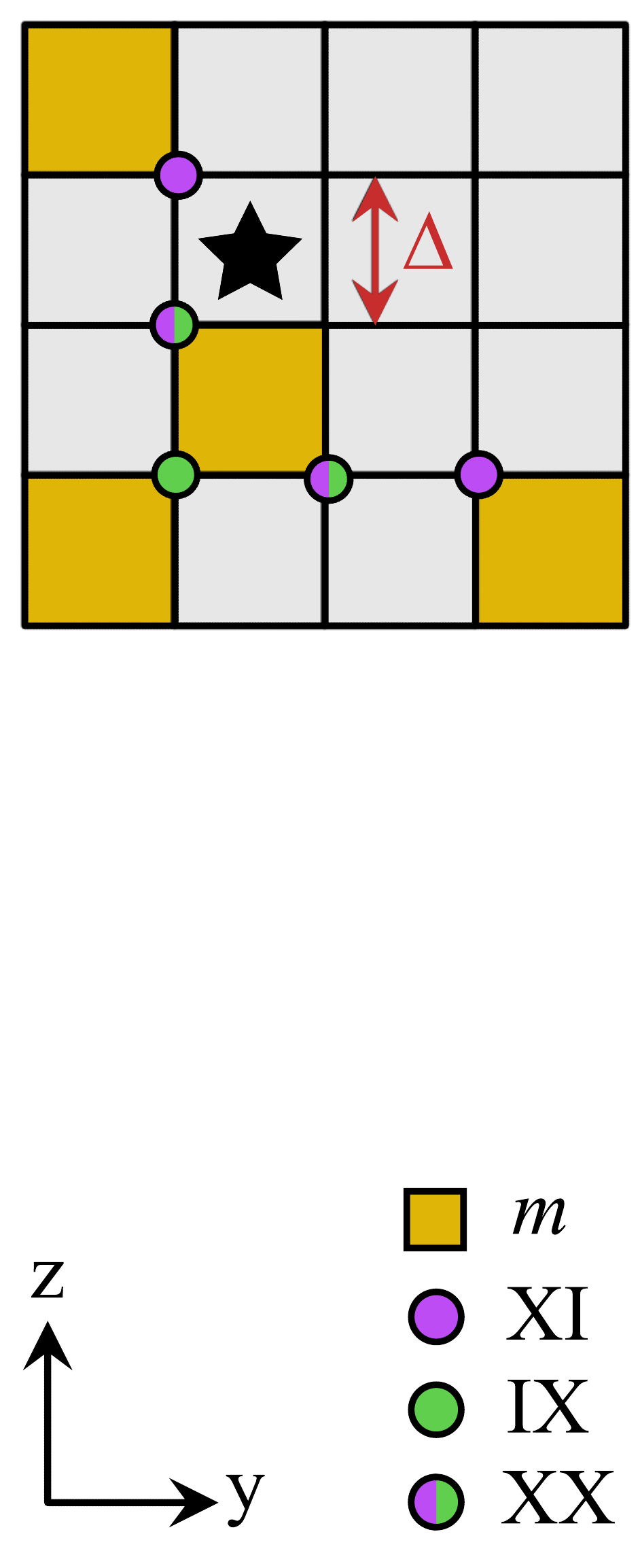} }
  \hspace{0.2cm}%
  \subfloat[\label{fig:fractal_move3}$\Delta = 7$]{
  \includegraphics[height=7cm]{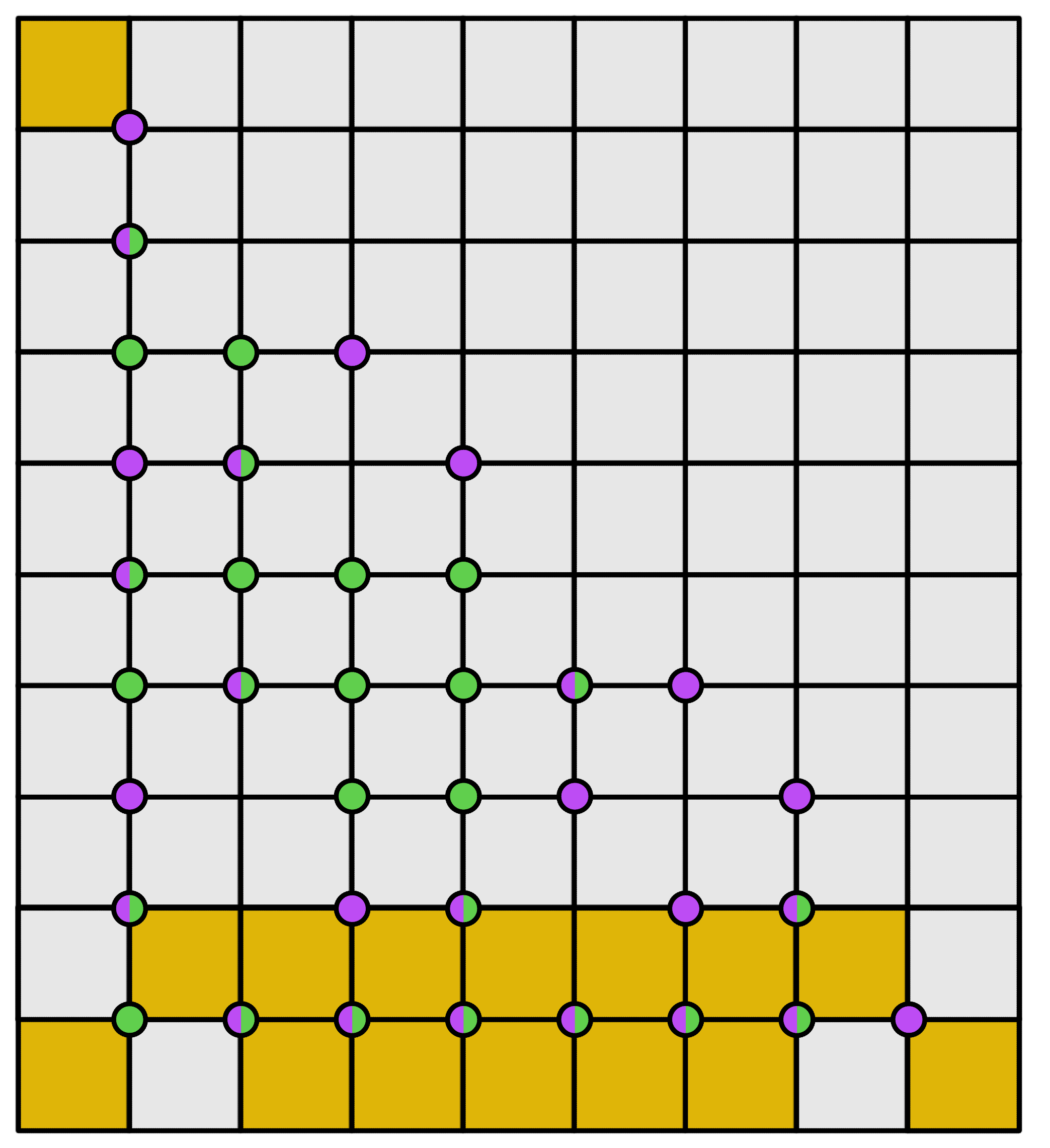} }
  \caption{The ``cascading procedure'': The $F_m^{\bar y \bar z}$ operator in (a) can move excitations in the $-\hat{z}$ direction. Starting with the excitation pattern in (a), we translate and re-apply $F_m^{\bar y \bar z}$ so that the $\star$ is aligned with the $\star$ in (b). In doing so, the stabilizers marked by the black squares are flipped. Repeating this at the $\star$ in (c) removes all excitations in the second row. We can repeat this process in each row to separate $m$ charges in the $-\hat{ z}$ direction. Larger separations $\Delta$ create additional excitations and the weight of the operator (number of single-qubit Pauli operators) scales superlinearly with $\Delta$.}
  \label{fig:fractal_move}
\end{figure*}

\begin{figure*}
  \centering 
  \subfloat[\label{fig:fractal_weight}Pauli weight (number of single-qubit Pauli
  operators) of repeated $F$ operators as the separation between charges
  ($\Delta$) is increased. A slope of $>1$ indicates superlinearity, due to the
  $\log$-$\log$ axes.]{
    \includegraphics[width=0.4\linewidth]{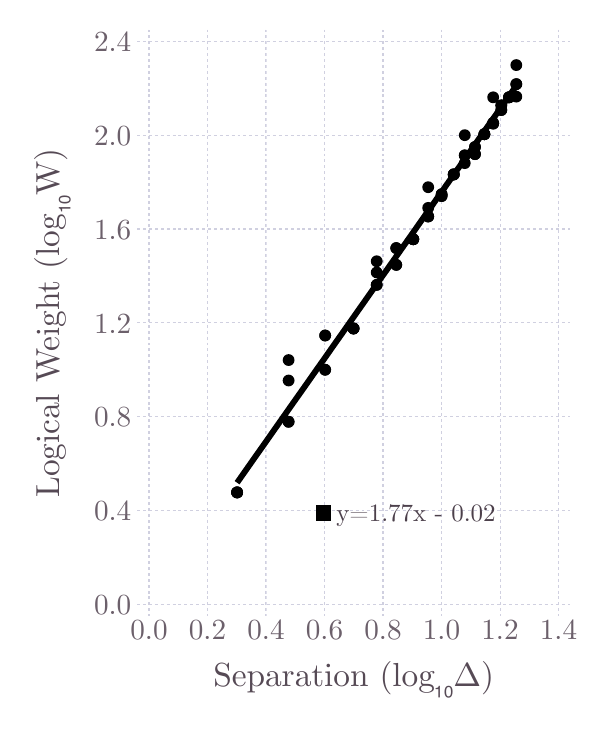}
    }
  \hspace{0.03\textwidth}
  \subfloat[\label{fig:fractal_energy}Upper bound on the energy cost associated
  with tunneling a charge between two boundaries. The data point at the bottom
  right indicates that the excess charges have condensed at the other boundary.]{
    \includegraphics[width=0.4\linewidth]{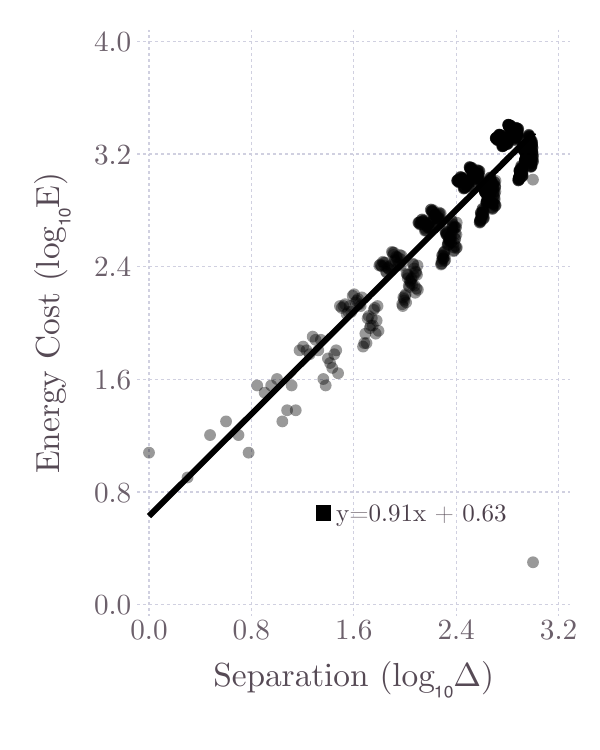}
  }
  \caption{Scaling properties of the $F$ operators as the separation between
  charges ($\Delta$) is increased, computed by numerically continuing the
process in Fig.~\ref{fig:fractal_move}. This demonstrates a superlinear upper
bound on the weight and a near-linear polynomial upper bound on the energy barrier of a
logical operator.}
  \label{fig:fractal_scaling}
\end{figure*}

\paragraph{\label{sec:cascade}Cascade Operators}
Secondly, excitations can be moved in a \emph{cascading} operation (using the
terminology from Ref.~\cite{bulmashBraidingGappedBoundaries2019}) that creates
additional excitations at each step of the motion. To highlight this, we
introduce three operators for $e$ and $m$, as in Fig.~\ref{fig:f_operators}.
Using these operators repeatedly creates the cascading procedure shown in
Fig.~\ref{fig:fractal_move}. Importantly, this process moves excitations through
the bulk while requiring a Pauli weight that scales superlinearly with the
distance $\Delta$, and creates additional excitations whose number scales
linearly with $\Delta$. Numerically continuing this process for larger
separations produces the results in Fig.~\ref{fig:fractal_scaling}.

\begin{figure}
  \centering 
  \includegraphics[width=\linewidth]{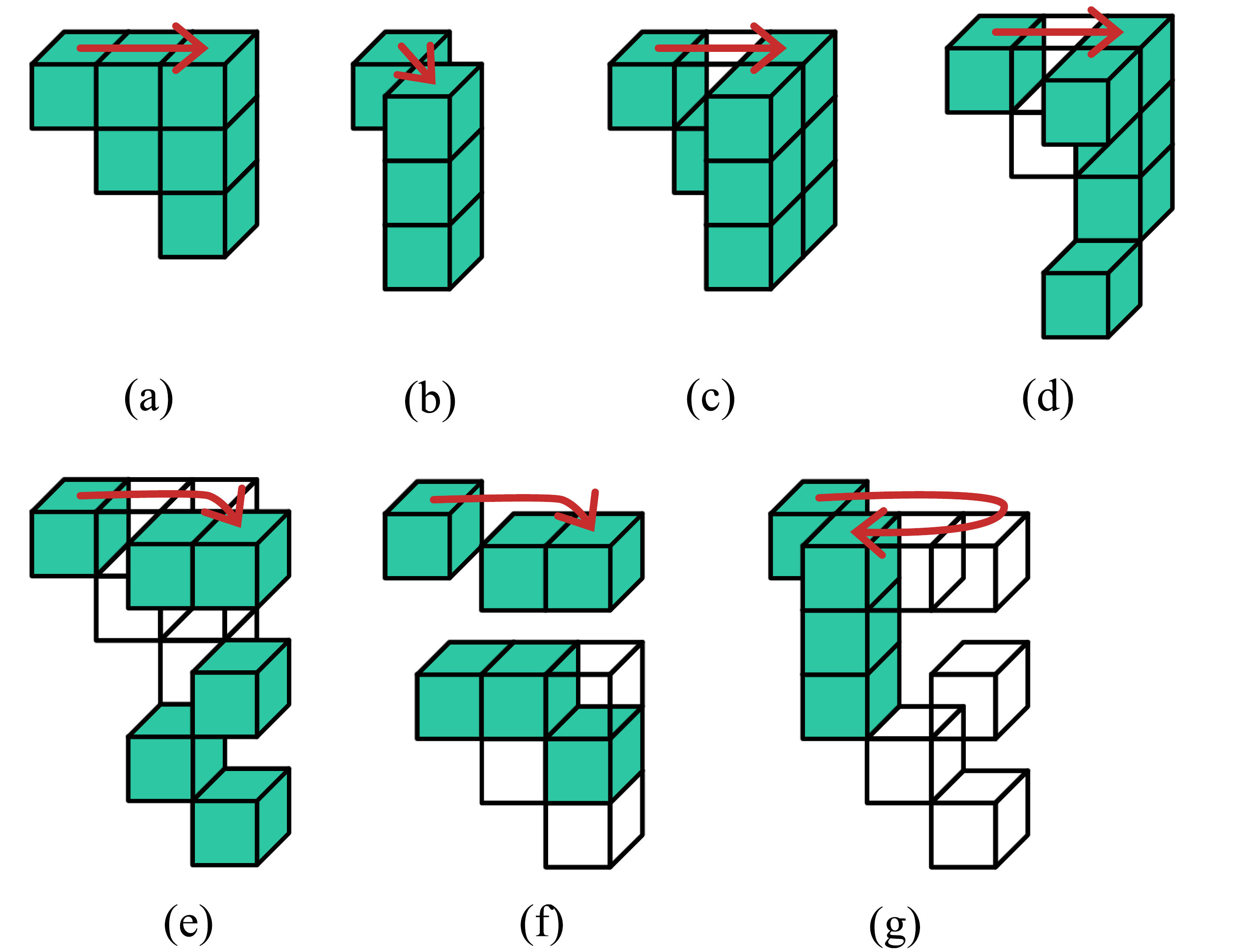}
  \caption{Using $F_e^{yz}$ and a new operator, $G$, to move excitations around a
    closed path (``cage'') in the $xy$ plane. {(a)} The $F_e^{yz}$ operator that we
    use to move excitations in the ${\hat y}$ direction. {(b)} A new
    operator, which we call $G$ (see Fig.~\ref{fig:zmax_a} in the appendix),
    that can be used to move excitations in the ${\hat x}+{\hat y}$ direction. {(c)}
    Starting with $F_e^{yz}$, we move an excitation in the second column
    into the ${\hat x}+{\hat y}$ direction using the tip of the new
    operator, $G$. Wireframe cubes indicate the annihilated charges from the previous
    step. {(d)} We then repeat this, using $G$ to move the second
    excitation in the second column. {(e)} We now move all three
    excitations in the third column into the ${\hat x} +
    {\hat y}$ direction, using $G$. {(f)} We now use $F_e^{yz}$ to begin
    to move these charges back in the $-{\hat y}$ direction. {(g)} We
    continue this process to move the remaining charges back in the
    $-{\hat y}$ direction. To complete the loop, we would use $G$ once
    again. This process can be generalized to larger loops, in other planes, and
  also with $m$ charges.} 
  \label{fig:loop_path}
\end{figure}

\paragraph{\label{sec:cage}Cage Operators}
Thirdly, following the terminology of Ref.~\cite{bulmashBraidingGappedBoundaries2019}, a
\emph{cage} operator is a generalization of a Wilson loop that moves excitations
around a closed path in the bulk, starting and ending from the vacuum state (ground state). This process is shown in Fig.~\ref{fig:loop_path}. Notably, the locations of Pauli operators, as well as the intermediary excitations, occupy a cylindrical shell with the required height of the cylinder scaling linearly with the radius. When viewed from a particular direction ($\hat z$ in Fig.~\ref{fig:loop_path}), this height can be compacted, leaving a $2$D loop. Cage operators are particularly relevant for the discussion of defects in Section \ref{sec:defects}.

\subsection{\label{sec:methods}Numerical Methods}
In this work, we aim to derive formulae for the number of encoded qubits, $k$,
and the distance, $d$, of variants of the cubic code with defects and modified boundary conditions. In general, however, it
is difficult to rigorously derive exact equations for these properties. Therefore, we present motivating discussions, backed up by numerical computations 
of small system sizes where we can determine empirical formulae for these
results. We then assume a consistent extrapolation for larger models. 

To perform these calculations, the stabilizers in a system of $n$ qubits are
represented by a binary vector in $\mathbb Z_2^{2n}$, with a $1$ corresponding
to a Pauli $X$ or $Z$ acting on a particular qubit~\cite{gottesmanStabilizerCodesQuantum1997a}. Commutation is a bilinear
map between two such vectors, and finding nontrivial logical operators reduces
to determining the kernel of this map when applied to the stabilizers. In this
way, the number of logical operators (or equivalently, the ground-state
degeneracy), and also examples of particular logical algebras, can be computed
exactly for system sizes up to the order of $20^3$ qubits. By restricting to a
subset of the physical qubits, we can also determine the properties of the
support of these logical operators. These results form the basis for the
conclusions drawn in the following sections. Code for these calculations is provided in Appendix \ref{sec:appcode}.

%% file: boundaries/boundaries.tex
In this section, we present and analyze formulations of the cubic code that
incorporate combinations of open and periodic boundaries. By doing so, we
identify the behavior of excitations on and near these boundaries, to inform the
discussion of encoding properties in Section \ref{sec:superbdries}. We first
consider isolated open boundaries in a semi-infinite system, before combining
multiple boundaries via edges and vertices. 

\label{sec:notation}
To characterize these boundaries, we introduce the following notation: 
Consider a rectangular prism centered at the origin of a Cartesian $(3+1)$D
coordinate system, with the terminating boundaries normal to the axes. In
similar notation to Miller indices, we use $(100$) to denote the boundary that
forms across the positive $x$ side of the prism and $(\bar100$) to denote the
boundary on the negative $x$ side. Additionally, define the \emph{sign} of a
boundary to be \emph{positive} for $(100)$, $(010)$, and $(001)$ orientations
and \emph{negative} for $(\bar100)$, $(0\bar10)$, and $(00\bar1)$. An additional
notation for specifying the types of stabilizers on these boundary faces is
introduced in Section \ref{sec:edges}. It is possible to consider
other more general boundaries, such as those with Miller index $(110)$. We present one such example in Appendix \ref{sec:colorcode} but will
defer the systematic study of these boundaries to future work.

\subsection{\label{sec:2dinf}Semi-Infinite Boundaries}

\begin{figure}
  \centering 
  \includegraphics[width=.7\linewidth]{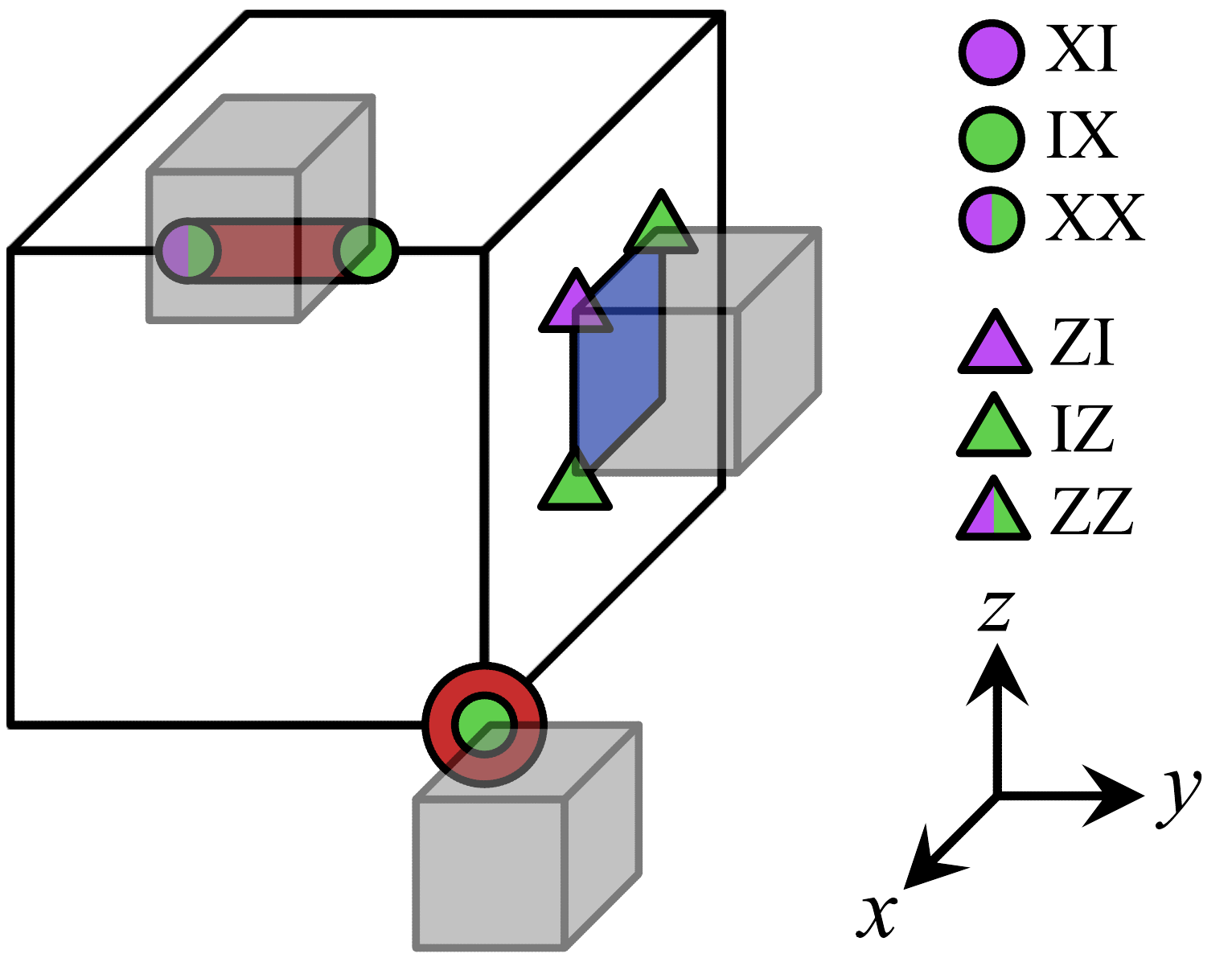}
  \caption{From top to bottom: Truncated stabilizers on the edges, plaquettes, and vertices
  of a finite lattice, formed by taking a section of the bulk $C_X, C_Z$ stabilizers.} 
  \label{fig:truncated_stabilisers_seams}
\end{figure}

We first consider the family of semi-infinite systems, such as $\mathbb R^2
\times (-\infty, 0]$ with a single terminating boundary corresponding to a
crystallographic plane, and propose a set of maximal stabilizer generators to populate this boundary. There are two possible constructions of translation-invariant Hamiltonian terms that maintain the required commutation relations with both the neighboring bulk and boundary stabilizers.
These correspond to truncated $C_X$ and $C_Z$ operators, as shown in
Fig.~\ref{fig:truncated_stabilisers_seams}, where the bulk operator is continued
outwards and all terms that lie beyond the system boundaries are ignored. We
denote these single-face (or plaquette) operators as $P_X$ and $P_Z$.

Let $X$-type denote a boundary Hamiltonian consisting of only $P_X$,
and $Z$-type for $P_Z$. More general configurations are possible, such as the
\emph{triangular} code in Table \ref{table:boundary_codes}, but their properties
can be explained solely by a discussion of the purely $X$ or $Z$-type
boundaries. 

Importantly, the orientation of a boundary affects its
behavior. That is, the boundary Hamiltonian of an $X$-type $(001)$ is not
equivalent to that of an $X$-type $(00\bar1)$. We see this by observing the two
inequivalent forms of truncated $C_X, C_Z$ operators, as well as the symmetries
noted in Section \ref{sec:symmetries}.

\begin{figure}[t]
  \centering 
  \includegraphics[width=\linewidth]{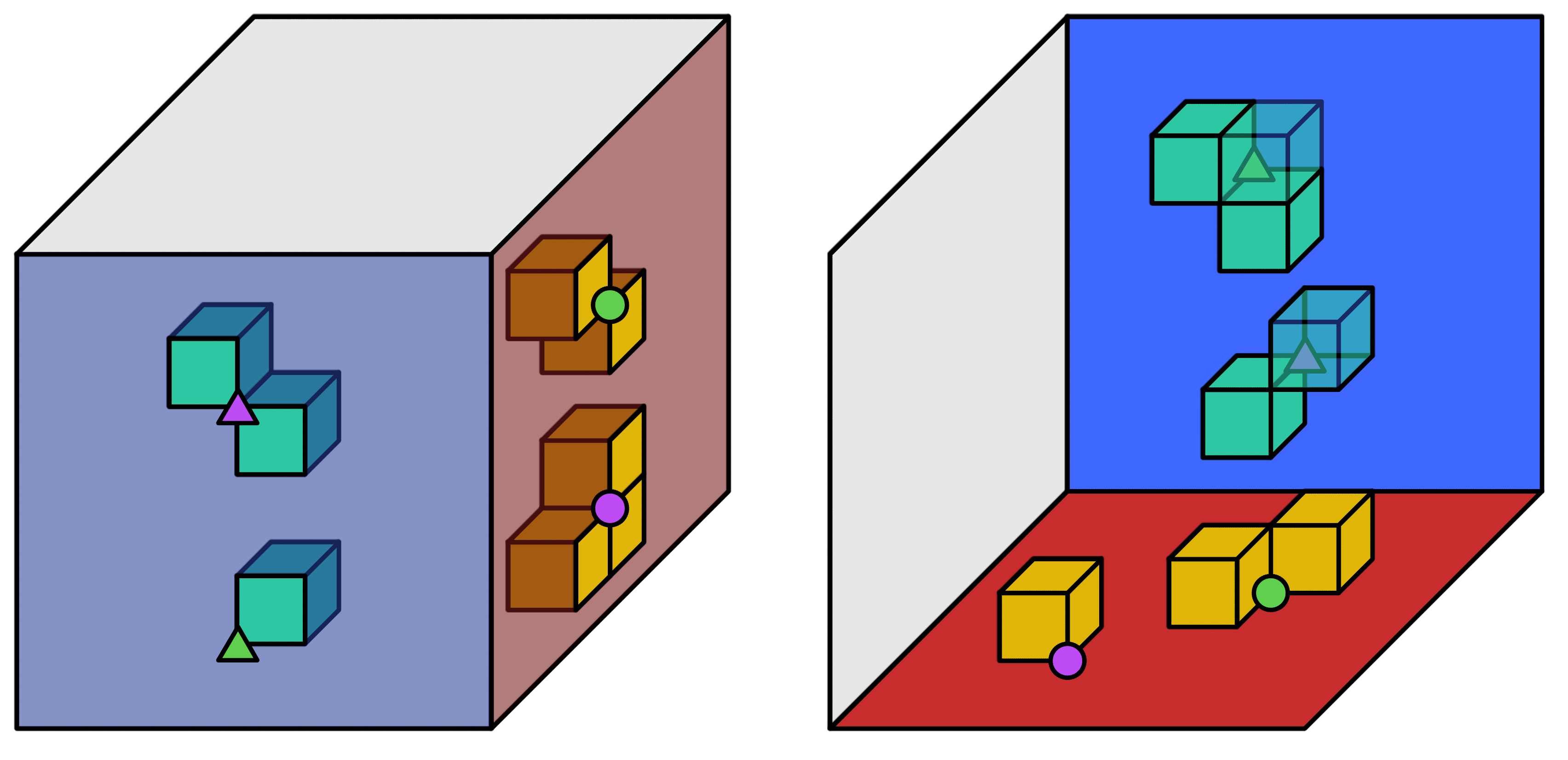}
  \caption{Positive (left) and negative (right), $X$-type (red) and $Z$-type
    (blue) boundaries, showing the excitation patterns for each combination of
  single-qubit Pauli operators.}
  \label{fig:boundary_excitations}
\end{figure}

\subsubsection{\label{sec:boundexc}Boundary Excitations}
Various applications of Pauli operators on a boundary are shown in
Fig.~\ref{fig:boundary_excitations}. When a single particle is able to be
created or destroyed in isolation by a local operator in the vicinity of the boundary layer, we refer to this
process as \emph{condensation}. Of note, positive $Z$-type boundaries
condense $e$ excitations, while negative $X$-type boundaries condense $m$
excitations; we refer to these boundaries as $(e)$ and $(m)$ type respectively.
However, on negative $Z$-type and positive $X$-type boundaries, single
excitations cannot be condensed. 
Instead, to describe their behavior, we introduce three new charges as subtypes
of both the $e$ and $m$: denoted $e_A, e_B, e_C$ and $m_A, m_B, m_C$.

\begin{figure}[t]
  \centering 
  \includegraphics[width=\linewidth]{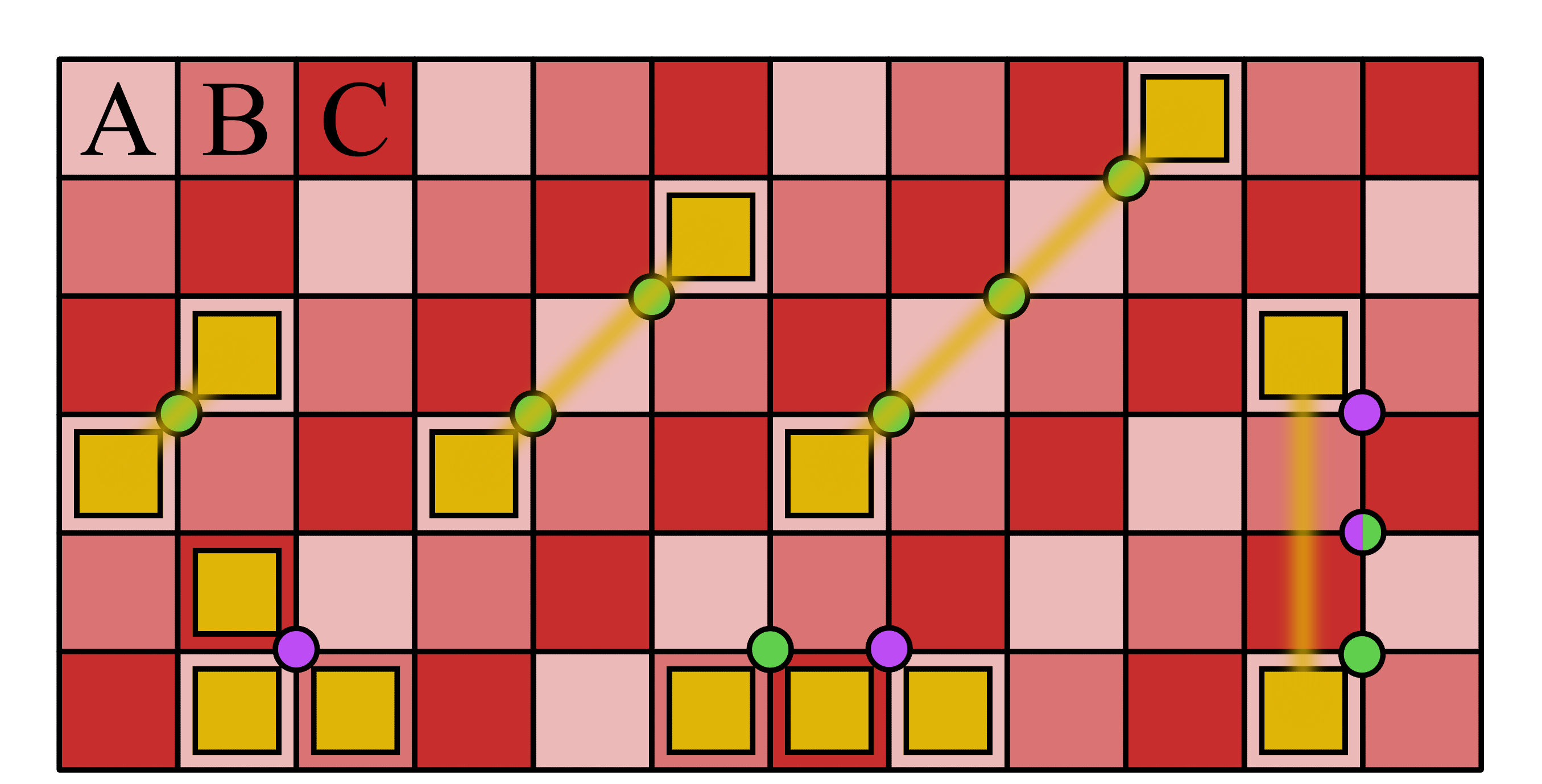}
  \caption{An $X$-type positive boundary, colored to indicate three types of
  $m$ charges: $m_A, m_B$ and $m_C$. Each charge is mobile within its particular set of
diagonals.}
  \label{fig:excitations_abc}
\end{figure}

Consider a positive $X$-type boundary in Fig.~\ref{fig:excitations_abc}, where
we striate the lattice to color squares along the diagonals as $A, B$, or $C$
(such that an $m$ excitation on an $A$-type square will have an $m_A$ charge,
etc.).
Importantly, a single $m_A$ (or an $m_A$ and an $m_B$, for example) cannot be
condensed in isolation. Instead, it gains $(2+1)$D mobility along the boundary
plane: $m_A$ charges are mobile along the $A$ diagonals via applications of
$IX$, and can hop to other $A$ diagonals using a combination of $IX, XI$. 
Equivalent results hold for $m_B$ and $m_C$. This increased mobility resembles phenomena observed on the boundaries of type-I fracton models, like the X-cube model~\cite{vijayFractonTopologicalOrder2016, bulmashBraidingGappedBoundaries2019}. 

Moreover, $XI$ creates a topologically nontrivial composite of $m_A$,
$m_B$, and $m_C$ excitations on the boundary. Motivated by these behaviors, we define the fusion rules
\begin{align}
  m_A \times m_A \sim 1,\quad m_B \times &m_B \sim 1,
  \quad m_C \times m_C \sim 1, \notag \\ 
  m_A \times m_B &\times m_C \sim 1 \label{eq:abc_fusion}
\end{align}
that describe how combinations of excitations can be created via
local operators. We note that these are $\mathbb{Z}_2\times\mathbb{Z}_2$ fusion rules.  An analogous result holds for $e_A, e_B$, and $e_C$ by substituting $m \mapsto e$. Given this
fundamentally different behavior to $(m)$ boundaries, we denote these positive $X$-type boundaries as
$(m_{ABC})$, and similarly $(e_{ABC})$ for negative $Z$-type boundaries. A summary of the new boundary notation is given in Table
\ref{table:boundary_condensations}. 

For these charges to be considered topologically distinct, there must not be a local operator that can fuse $m_A \times m_B\sim1$, for example. By considering the action of $IX$ and $XI$, this is trivially true using only operators with support on the boundary. Since such a fusion operator will create
excitations only along the boundary, we complete the
argument by using the cleaning process from
Refs.~\cite{haahLocalStabilizerCodes2011, duaSortingTopologicalStabilizer2019} to reduce the support of any bulk
operator to just terms on the boundary; we refer the reader there for a more
detailed description.
This argument holds when we consider operators stretching from the $(m_{ABC})$
boundary into the bulk, as long as the operator does not have support on other
boundaries (such as an opposing $(m)$). Therefore, it is valid to consider
$m_A$, $m_B$, and $m_C$ as distinct charges with the boundary fusion properties
above, if they are separated from additional boundaries by distances larger than the correlation length of the ground state.

\begin{table}
  \centering
  \begin{tabular}{ccccc}
    \toprule
    Boundary & Stabiliser & Sign & Color & Behavior 
    \\
    \cmidrule(r){1-1} \cmidrule(lr){2-2} \cmidrule(lr){3-3} \cmidrule(lr){4-4} \cmidrule(l){5-5}
    $(m)$ & $X$ & $-$ & \raisebox{-0.3\totalheight}{\includegraphics[height=12pt]{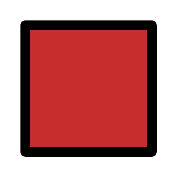}} & $m$ condenses \\ 
    $(m_{ABC})$ & $X$ & $+$ & \raisebox{-0.3\totalheight}{\includegraphics[height=12pt]{boundaries/Red.pdf}} &  Eq.~(\ref{eq:abc_fusion}) \\ 
    $(e)$ & $Z$ & $+$ & \raisebox{-0.3\totalheight}{\includegraphics[height=12pt]{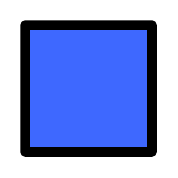}} &  $e$ condenses \\ 
    $(e_{ABC})$ & $Z$ & $-$ & \raisebox{-0.3\totalheight}{\includegraphics[height=12pt]{boundaries/Blue.pdf}} & 
 Eq.~(\ref{eq:abc_fusion}) (for $e$) \\
  \bottomrule
\end{tabular}
\caption{Boundary labels based on the type of topological charge that condenses.
Boundary sign is defined as per Section \ref{sec:notation}. Color refers to the convention used in the figures.}
\label{table:boundary_condensations}
\end{table}

\begin{figure}
  \centering 
  \includegraphics[width=.9\linewidth]{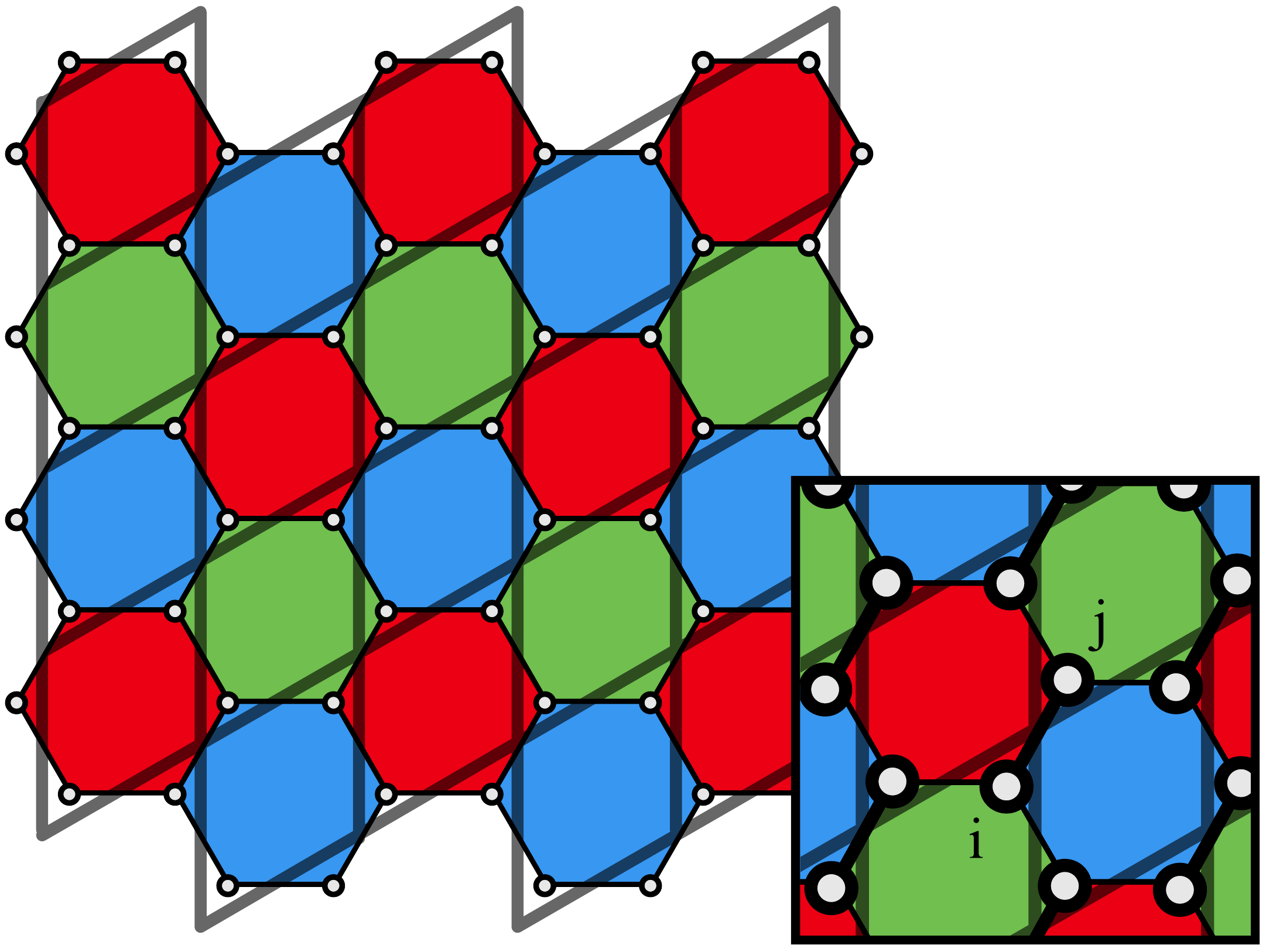}
  \caption{A hexagonal color code, with a qubit at each white circle.
  Overlayed is a rhomboidal lattice, with vertices corresponding to the lattice
  points in the boundary layer of the cubic code. As per the inset, at each
  vertex the adjacent two color code qubits $i,j$ are identified with the two
  qubits at each cubic code lattice site. To map onto the cubic code
  stabilizers a \texttt{CNOT} is used on each pair, controlled on qubit $j$.}
  \label{fig:colorcode}
\end{figure}

Due to the triplet $\mathbb{Z}_2\times\mathbb{Z}_2$ fusion nature of these charges, these fusion rules bear a
resemblance to that of the \emph{color code}~\cite{bombinTopologicalQuantumDistillation2006, bombinGaugeColorCodes2015a,
kesselringBoundariesTwistDefects2018}. In
fact, the color code can be directly transformed into this $(2+1)$D layer:
Consider the hexagonal lattice of the 6-6-6 color code in
Fig.~\ref{fig:colorcode}, with a qubit at each lattice point (white circle). The
stabilizers of this code are the product of $X$ or $Z$ on all $6$ qubits of each
hexagon. We begin the transformation by first overlaying a rhomboidal lattice such that its vertices lie between
exactly two qubits $i,j$, as shown in the inset. Note that acting on both qubits
with $X_iX_j$, for example, excites the two adjacent stabilizers on the green
hexagons. This is equivalent to exciting two $A$ squares along the diagonal of the
cubic code's $(m_{ABC})$ using $IX$ in Fig.~\ref{fig:excitations_abc}. Moreover,
acting with just $X_iI_j$ excites the three adjacent red, green, and blue
hexagons - equivalent to exciting $m_A, m_B$, and $m_C$ using $XI$ on the cubic
code. Corresponding similarities also apply for $Z$ operators acting on an $(e_{ABC})$ boundary. We, therefore, have the following map relating excitations of the color code to excitations of the cubic code boundaries: 
\begin{equation}
  X_iX_j \mapsto IX,\, X_iI_j \mapsto XI,\, Z_iZ_j \mapsto ZI, \, I_iZ_j
  \mapsto IZ
\end{equation}
Notably, in the Heisenberg representation this transformation is equivalent to
acting on each $i,j$ pair with a \verb`CNOT` gate, controlled on qubit $j$~\cite{gottesmanHeisenbergRepresentationQuantum1998}. If we consider
the $(m_{ABC})$ boundary on $(100$) and $(e_{ABC})$ on $(\bar100)$, then this
transformation directly maps the truncated boundary $X$ and $Z$ stabilizers of the cubic code onto the $X$ and $Z$ stabilizers of the color code. This property is discussed further
in Appendix \ref{sec:colorcode}. It remains an open question as to how or whether this correspondence generalizes to the bulk stabilizers of the cubic code, and moreover for the other cubic codes proposed by Haah~\cite{haahLocalStabilizerCodes2011}.

\subsubsection{\label{sec:periodic}Periodic Behavior}
Consider a lattice that is periodic in $x, y \in [0,L]$ and occupies $z \in
[0,\infty)$. Along the $(00\bar 1)$ face we construct an $(e_{ABC})$ boundary using $P_Z$ stabilizers. If $3|L$, then the diagonal striation (as in Fig.~\ref{fig:excitations_abc}) of this boundary
layer is self-consistent and we have distinct $e_A, e_B$, and $e_C$ charges that
are mobile within their $(2+1)$D diagonal subspaces. That is, if an $e_A$ were
to move around the periodic boundary, it would retain its $e_A$ charge. We can
thus consider an operator that creates two $e_A$ out of the vacuum, hops one
around the periodic boundary, and annihilates it with the other to create a
logical operator - comparable to those in the $(2+1)$D toric code, for example,~\cite{kitaevFaulttolerantQuantumComputation2003}. Due to the two unique charges (as $A\times B \sim C$ by the
fusion rules), these surface string operators define two independent logical
operators along an $(e_{ABC})$ or $(m_{ABC})$ boundary. When $3 \! \nmid \! L$, it can be checked that hopping an excitation three times around a boundary
(to return it to its original labeling) produces the identity operator. 

However, these string operators can be extended to additional layers beyond the boundary,
creating operators that extend into the bulk while requiring longer periods. 
To describe this procedure, we consider each $xy$ plane to be a generalization of the
boundary $(e_{ABC})$ layer. That is, acting with $IZ$ or $ZI$ at $z=1$ creates $e_{ABC}$ charges in both the $z=0$ and $z=1$ layers. Repeating the
hopping process at $z=1$ to move an $e_A$ around the periodic boundary will
remove all charges in that layer while introducing \emph{residual} charge at $z=0$.
 If this residual charge is
\emph{trivial} - that is, it is equivalent to the vacuum state up to the fusion
operators in Eq.~(\ref{eq:abc_fusion}) - then the charge in that layer can also
be cleaned away. Importantly, these processes only
deposit residual charge in layers at smaller $z$, not affecting layers that have already been cleaned of excitations. Since $(e_{ABC})$ restricts any remaining charge from forming beyond the $(00\bar1)$ face itself, completing the process by annihilating charges at $z=0$ results in a (nontrivial) logical operator. 

If this process were started instead at $z>1$, then this cleaning can be continued
downwards until either the topmost remaining layer has a nontrivial residual
charge or all charges are annihilated or condensed into the $(e_{ABC})$ boundary. In this way, nontrivial logical operators can be constructed at
varying depths within the bulk, which wrap around the periodic boundary and iteratively clean layers of charge down to $z=0$. 

\begin{figure}
  \centering 
  \includegraphics[width=\linewidth]{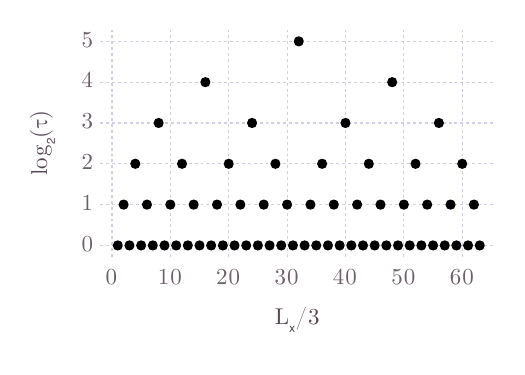}
  \vspace{-12mm}
  \caption{Numerical calculation of $\tau(L_x, L_\infty) \equiv \zeta(L_x/3)$
    from Eq.~(\ref{eq:zeta}) for increasing linear system sizes $L_x$ with
    $L_\infty \gg L_x$. There is a clear fractal nature to this behavior.}
  \label{fig:tau}
\end{figure}

It can be shown (see Appendix \ref{sec:periodicproof}) that for a given lattice
that is periodic in linear system size $L$, the maximum number of $z$ layers that can be cleaned until
a nontrivial residual charge is created is 
\begin{equation}
  z_\text{max}(L) = \zeta(L/3)
  \label{eq:zmax}
\end{equation}
using the notation in Eq.~(\ref{eq:zeta}). This function is plotted in
Fig.~\ref{fig:tau}. Each layer introduces two additional nontrivial mutually commuting logical
operators, giving
\begin{equation} 
  k = 2\zeta(L/3)
\end{equation}
Note that if $L_z < z_\text{max}(L)$ and the $(001)$ boundary is of type $(m)$
or $(m_{ABC})$, then $k=2L_z$ since the maximum number of layers cannot fit in
the given lattice. Incorporating this, we define the function 
\begin{equation}
  \tau(L_1;\; L_2) = \min \{\zeta(L_1/3),\, L_2\}
  \label{eq:tau}
\end{equation}
For rectangular lattices with $L_x \neq L_y$, then we
instead have 
\begin{equation}
  k=2\min\{\tau(L_x;\; L_z),\,\tau(L_y;\; L_z)\}
\end{equation}
defined by the shortest path around the periodic boundary. As a result, $k=0$ if
one of $L_x$ or $L_y$ is not a multiple of $3$.

\subsubsection{\label{sec:translation}Translational-Symmetry Violations}
\begin{figure}
  \centering 
  \includegraphics[width=\linewidth]{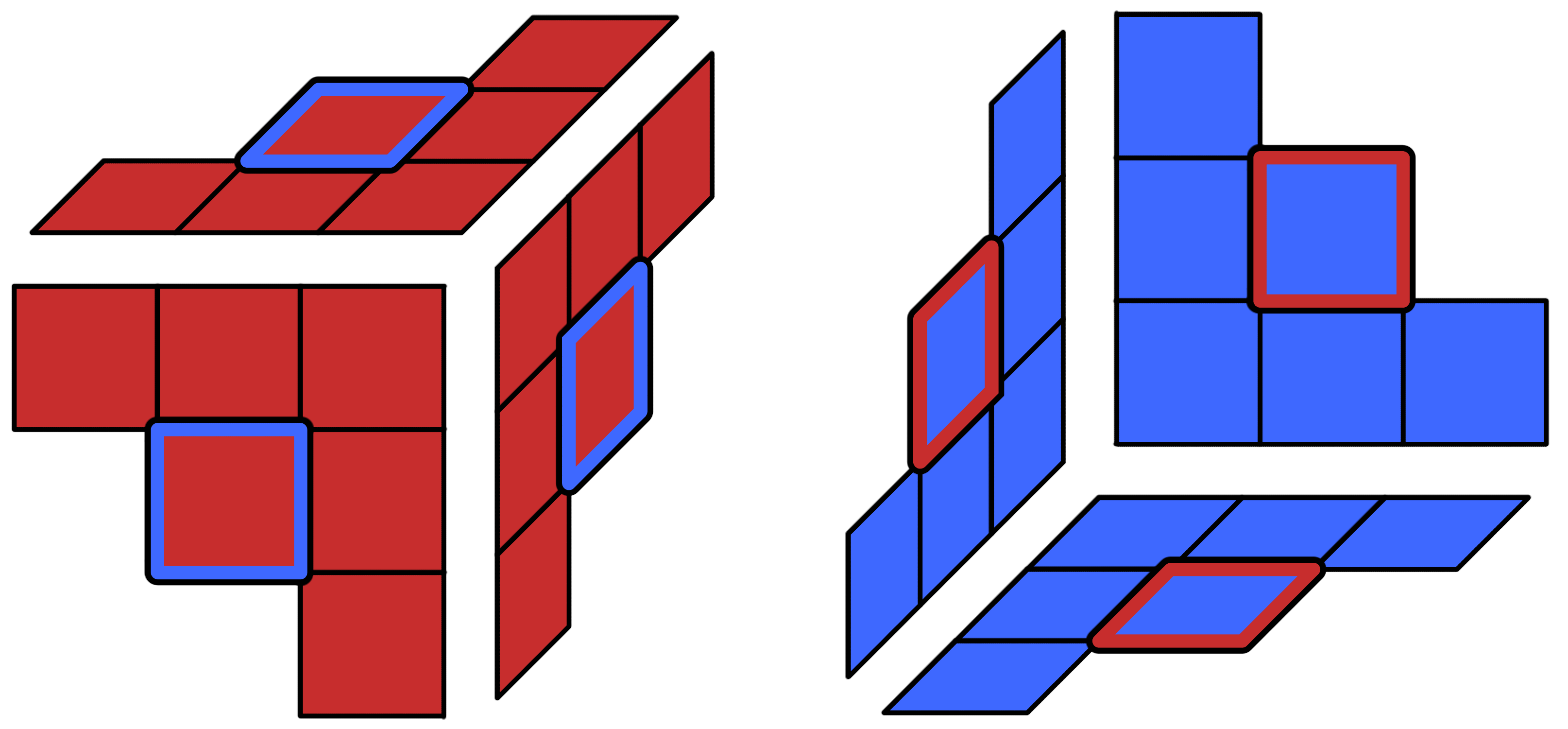}
  \caption{Anti-commutation relations for
    plaquette operators on the boundary. {(Left)} When $P_Z$ is applied to
    the blue square on the positive faces of the lattice, the red
    squares on the same face correspond to $P_X$ that anti-commute. {(Right)} When $P_X$ is
    applied to the red square on the negative faces of the lattice, the blue squares correspond to $P_Z$ that anti-commute.}
  \label{fig:anticom}
\end{figure}

If translational invariance is relaxed, there are additional possible commuting
configurations of the boundary Hamiltonian. The commutation relation between
neighboring plaquette terms is given in Fig~\ref{fig:anticom}. Notably, these
relations allow for the construction of a natural \emph{diagonal} commuting
interface between $P_X, P_Z$ operators, such as the
configuration labeled \emph{triangular} in Table \ref{table:boundary_codes}. 

As with the other boundaries discussed so far, each sector of the mixed boundary
is associated with an $(e), (m), (e_{ABC}),$ or $(m_{ABC})$ behavior. Due to this equivalence, this paper focuses on pure $X$ and $Z$ boundary configurations. 

\subsection{\label{sec:edges}Boundary Seams}
As in Fig.~\ref{fig:truncated_stabilisers_seams}, two adjacent boundaries can
be joined along an edge, and three boundaries at a vertex. These features are required to construct the full codes discussed in Section \ref{sec:superbdries}. 

\begin{figure}
  \centering 
  \includegraphics[width=0.8\linewidth]{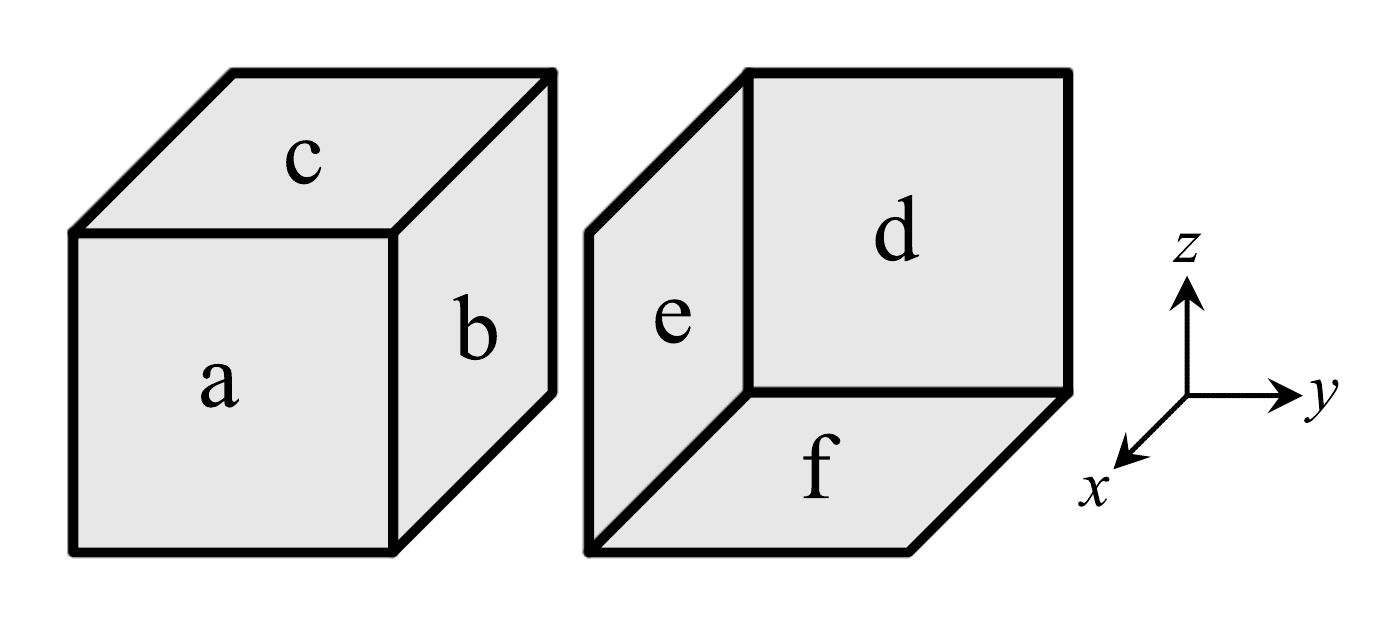}
  \caption{Notation used to specify the choice of boundaries, $(abc;def)$, where
    each letter is $e$ or $m$ (as defined in Table
\ref{table:boundary_condensations} and the $ABC$ subscript is implied), or $p$
for a periodic boundary.}
  \label{fig:boundary_notation}
\end{figure}

Notably, the
plaquette-plaquette commutation relations in Fig.~\ref{fig:anticom} are
equivalent for edge-plaquette commutation since geometrically, neighboring
plaquettes only share support on at most two sites - identical to plaquettes.
Given these constraints, we can thus specify a fully-gapped configuration for
a finite prism such as in Table \ref{table:boundary_codes}. To describe these configurations, the notation $(x y
z;\bar x \bar y \bar z)$ is used to specify the boundary type on each of the six
lattice boundaries (see Fig.~\ref{fig:boundary_notation}). For open boundaries,
we use $e$ and $m$ as defined in Table \ref{table:boundary_condensations}, where
the $ABC$ subscript is dropped for readability. Periodic boundaries
are denoted by $p$. It is assumed in each case that the edge and vertex
stabilizers are chosen to create a maximally commuting stabilizer group of local
operators (that is, where possible there are no local Pauli operators with
support on a single plaquette, edge, or vertex that commute with all the
stabilizers).

Using this notation, we can revisit the symmetries described in Section \ref{sec:symmetries}:
\begin{enumerate}
  \item $(abc;def)$ is equivalent to $(cab;fde)$ and $(bca;efd)$. 
  \item $(abc;def)$ is equivalent to $(bac;edf)$.
  \item $(abc;def)$ is equivalent
    to $(def;abc)$ with all $e \leftrightarrow m$ swapped. 
\end{enumerate}
Combined, the first two symmetries imply that for $(abc;def)$, all permutations of
$abc$ produce equivalent codes, given that the same permutation is also applied to $def$.

%% file: boundaries/bdries_super.tex
A core motivation for this work is to maintain the partial self-correction of the cubic code, without the requirement of nonlocality to implement a $3$-torus. Given the
discussion in Section \ref{sec:bg}, operators with superlinear weight (necessary for self-correction) arise when
excitations \emph{cascade} through the bulk of the lattice, while unwanted string-like
operators appear near certain boundary configurations. Motivated by this,
we thus consider codes that contain opposing $(e), (e_{ABC})$ boundaries and
opposing $(m), (m_{ABC})$ boundaries. This specifies four of the six faces of the lattice,
leaving (up to symmetries) a potential three unique configurations. In the
following section, we highlight one such case, dubbed \emph{tennis ball 1}, while the others are discussed in Appendix \ref{sec:otherbdries}.

\subsection{\label{sec:tennis1}Tennis Ball 1}
This configuration is constructed with
$(e_{ABC})$ and $(m_{ABC})$ on the remaining two unspecified faces. For example,
$(mem;mee)$ as shown in Table \ref{table:boundary_codes}. 

\begin{figure}
  \centering 
  \subfloat[The $\bar X$ logical operators that condense $m$ charges on the
    $(\bar100)$ and $(100)$ faces become linear in weight when near the $(001)$ boundary.]
  {\includegraphics[width=0.8\linewidth]{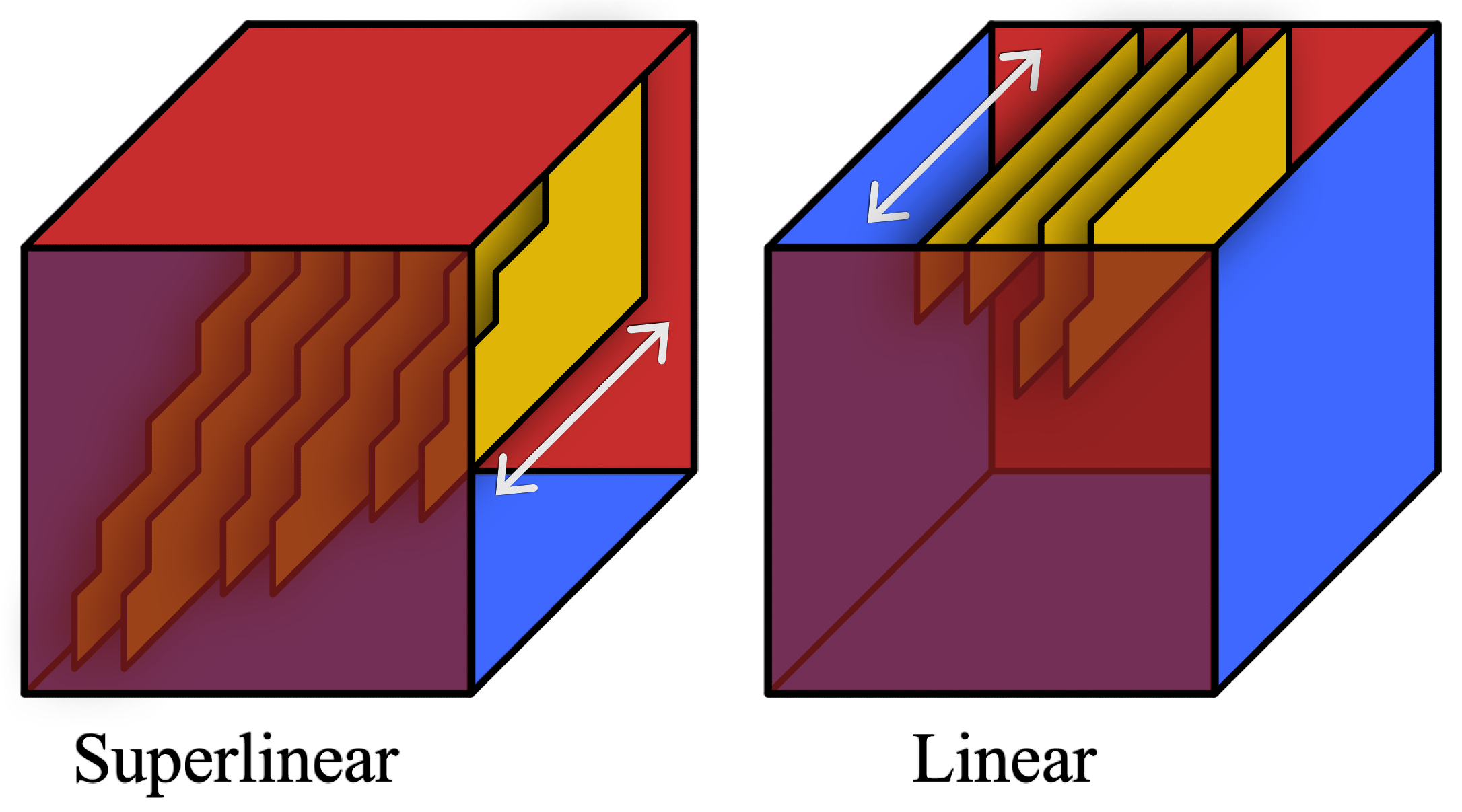}}
  \hspace{12pt}
  \subfloat[The $\bar Z$ logical operators that condense $e$ charges on the
    $(0\bar10)$ and $(010)$ faces become linear in weight when near the $(00\bar1)$ boundary.]
  {\includegraphics[width=0.8\linewidth]{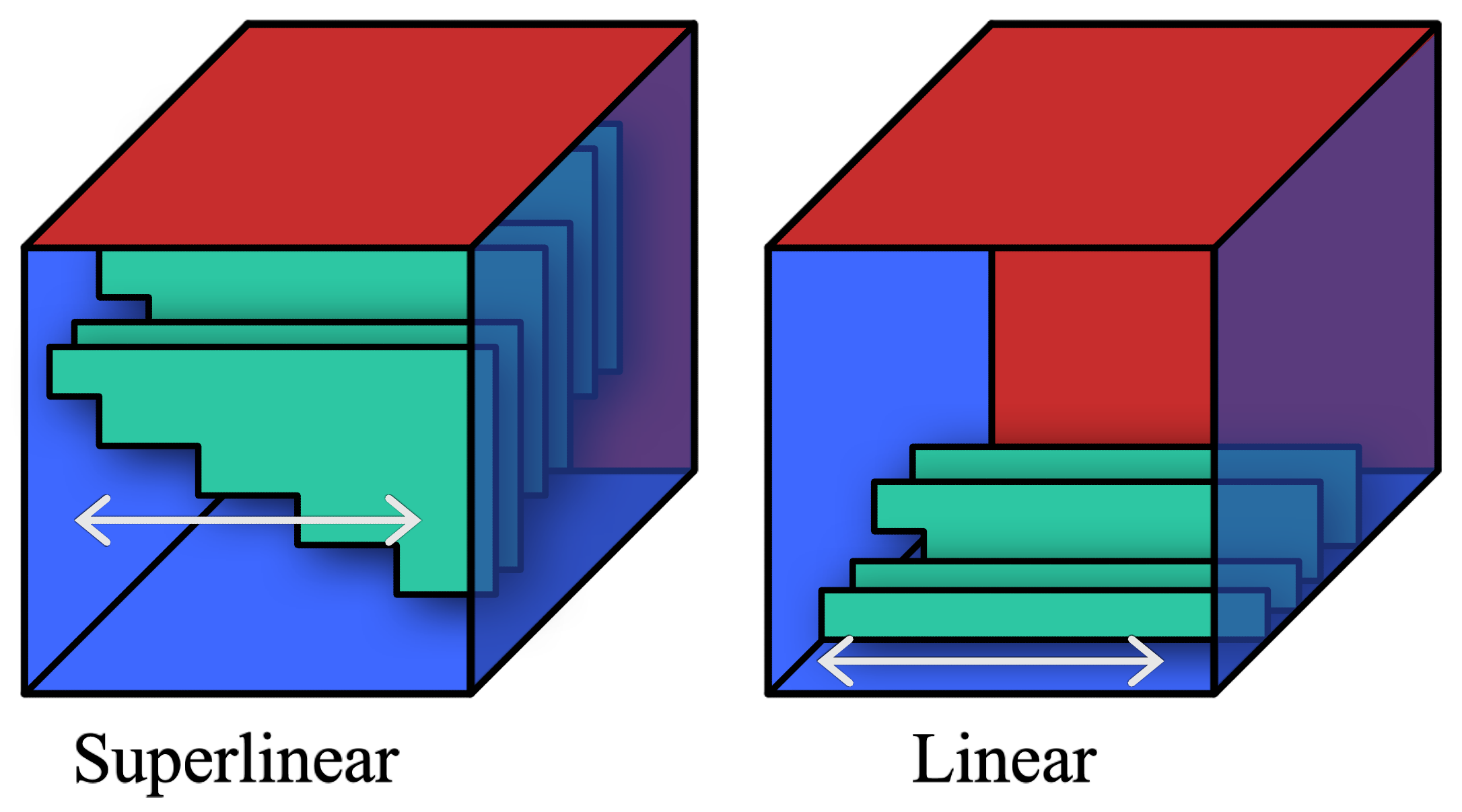}}
  \caption{Logical operators on the first \emph{tennis ball} configuration,
    $(mem;mee)$. Solid shapes indicate repeated applications of the $F$
    operators to cascade $m$ (yellow) and $e$ (cyan) charges. Red faces
  represent $X$, and blue for $Z$, stabilizer choices on the boundaries.}
  \label{fig:tennis_logicals}
\end{figure}

To construct $\bar X$ logical operators, $m$ charges must cascade from $(100)$
and condense at $(\bar100)$ using $F_m^{\bar x\bar y}$ and $F_m^{\bar x \bar
z}$, as shown in Fig.~\ref{fig:tennis_logicals}. If the $m$ excitations
produced in the cascade cannot appear on the $(001)$ boundary, this procedure produces an operator with a weight superlinear in $L_x$ and an increasing energy
barrier, since excitations must cascade through the bulk by a distance $L_x$.
However, if $m$ excitations \textit{can} be produced on the $(001)$ boundary
during the cascade, these excitations become mobile on the surface
and can hop to the $(\bar100)$ face using a string operator. Hence, any cascade
that begins less than a distance of order $\mathcal O(L_x)$ from the $(001)$ boundary
creates a string-like operator: the height is constant in $L_x$, the
weight is linear, and the energy cost is constant. An analogous argument holds
for producing $\bar Z$ logical operators by cascading $e$ excitations; only
cascades which begin a distance at least $\simeq L_y$ from the $(00\bar 1)$
boundary produce logical operators with superlinear weight.

\begin{figure*}
  \centering 
  \subfloat[$\bar X_1$]
  {\includegraphics[width=0.45\textwidth]{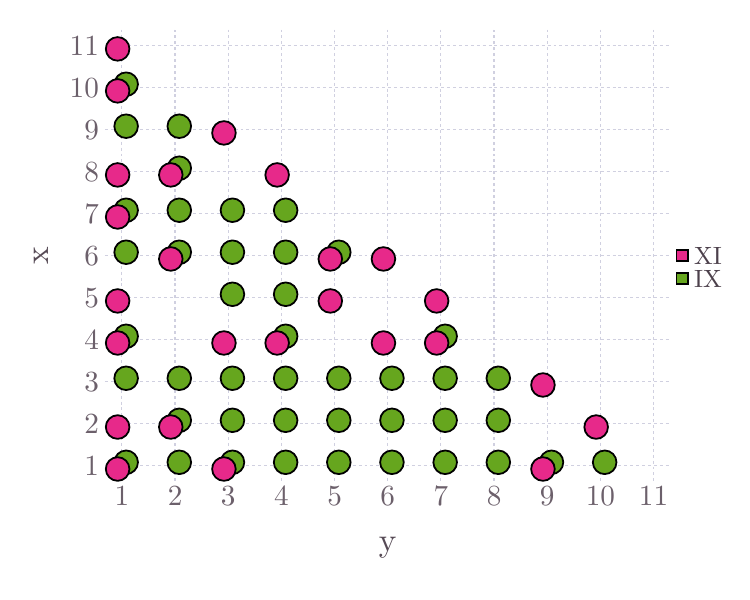}}
  \subfloat[$\bar Z_1$]
  {\includegraphics[width=0.49\textwidth]{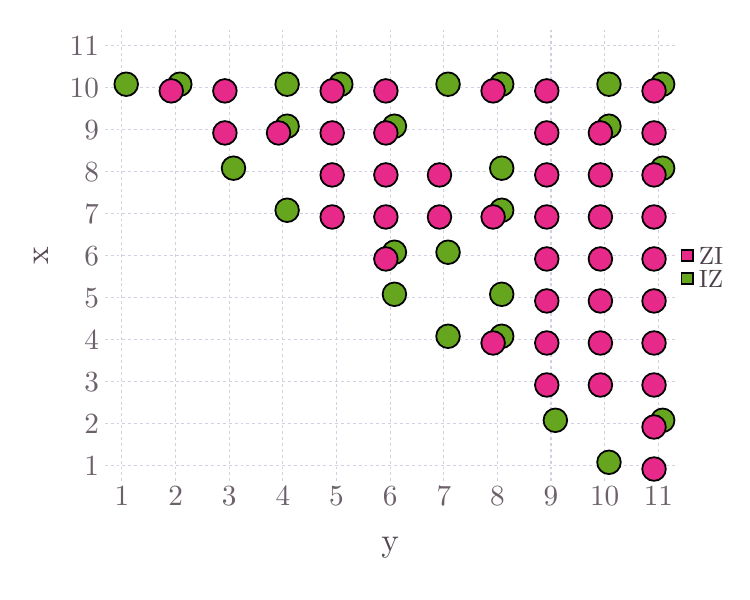}}
  
  \subfloat[$\bar X_2$]
  {\includegraphics[width=0.49\textwidth]{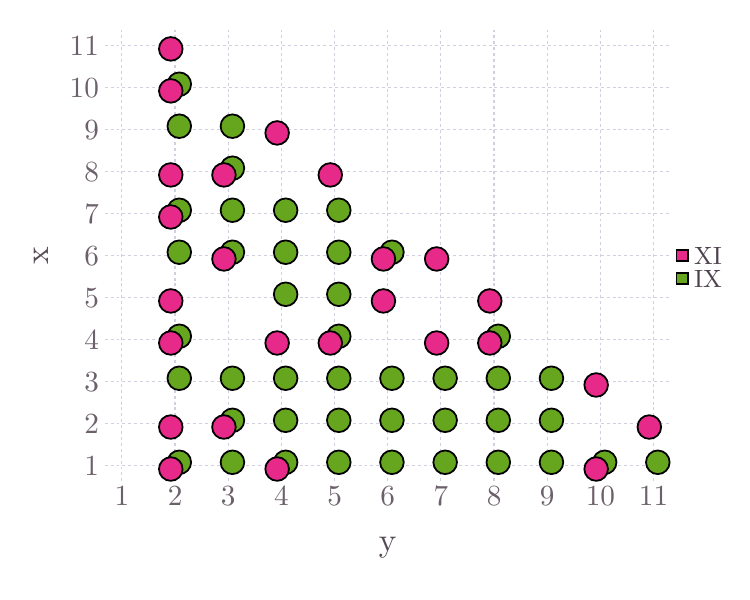}}
  \subfloat[$\bar Z_2$]
  {\includegraphics[width=0.49\textwidth]{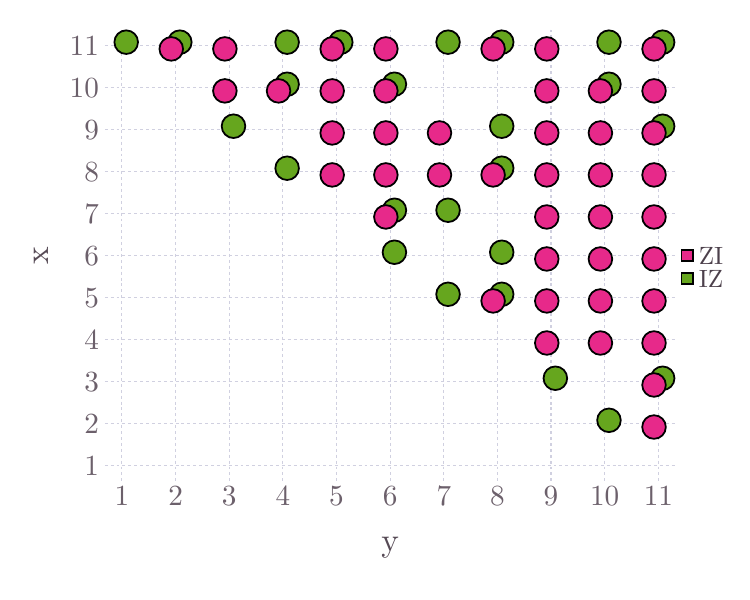}}

  \caption{Two pairs of logical operators on the first \emph{tennis ball} code,
    $(mem;mee)$, for a linear system size $(11,11,L_z)$, considering a slice of the
    $xy$ plane at some $z$ far from the $(001)$ and $(00\bar1)$ boundaries.
    These were calculated and plotted computationally; A slight jitter is
    applied to each point to distinguish when $XI$ and $IX$ occur on the same
    lattice point. Note that there are only two such pairs of logical operators
    that can be cleaned onto just this particular $xy$ plane. If we consider the
    plane at $z+1$ we would find an analogous result of two new independent
    pairs of logical operators; in this way, the \emph{tennis ball 1} code has $k=2L_z$ logical pairs.} 
\label{fig:tennis_logicals_additional} 
\end{figure*}

Notably, the $\bar X$ logical operators extend in the $\hat z$ direction and $\bar Z$ in the $-\hat z$ direction when cascading. Pairs of $\bar X$ and $\bar Z$ must then anti-commute along their shared $xy$-plane. 
To determine the number of independent logical pairs, first note that using the \emph{long} edge of the $F$ operators to move charges through the bulk is equivalent to the hopping operator in Fig.~\ref{fig:excitations_abc}. We can then identify this cascading procedure with hopping an $A,B$ or $C$ charge from one boundary to the other, creating neighboring intermediary excitations that must also be hopped into a condensing boundary. In a similar argument to Section \ref{sec:periodic}, we, therefore, expect each layer to contribute $2$ encoded qubits. Since only one orientation, namely the $xy$ planes, have opposing $e$ and $m$-condensing boundaries - and therefore can support mutually anti-commuting pairs of both $e$ and $m$ operators - the scaling
is thus expected to be $k = 2L_z$. We confirm this by numerically
computing the ground state degeneracy for small values of $L_x,L_y,L_z$; an
example of the computed logical algebra cleaned into a single $xy$ plane is shown in
Fig.~\ref{fig:tennis_logicals_additional}. This same analysis cannot
be conducted with $xz$ planes, for example, since these will have three
adjoining $m$-type boundaries and only one $e$-type. As with the surface code,
such a configuration does not support any nontrivial logical operators.

Notably, this scaling confirms the results in
Ref.~\cite{duaCompactifyingFractonStabilizer2019}, where Dua et al. showed that
the periodic cubic code can be interpreted as interwoven layers of toric code.
Specifically, a model of linear system size $(L_x, L_y, L_z)$ is equivalent to $2L_z$ copies of
toric code, up to a unitary that is non-local in $z$. When placed on open boundaries as done here, the corresponding
surface codes should each contain one encoded qubit, thus giving $k=2L_z$ total.

\subsection{\label{sec:subsystem_codes}Subsystem Codes}

The presence of the string-like operators initially appears detrimental to the desired self-correcting behavior of the \emph{tennis ball} code. However, we can still ensure a superlinear-distance encoding by using a subsystem code \cite{bombinTopologicalSubsystemCodes2010}. 
By treating all logical qubits with linear distance as gauge qubits, the subsequent dressed logical operators
should remain superlinear in weight. We argue this claim as follows: 

First, note that the support of the bare superlinear-weight logical operators extends further in the $z$ direction than the bare linear-weight logical operators, by definition. This extra support includes a region in the bulk where the $\bar X$ and $\bar Z$ pairs intersect and anti-commute. Since $\bar X$ commutes with all stabilizers and all other $\bar Z$ operators apart from its pair, this anti-commutation relation must be maintained when multiplied with these other operators. 
Therefore, the product of bare superlinear and bare linear weight operators will always contain support on a region in the bulk of the lattice.
This region is either entirely isolated from the boundary or extends to the boundary. In the former case, by the properties of the original cubic code, there are no string-like operators in the bulk of the lattice. For the latter,
the operator now occupies a support that is larger than the original superlinear-weight support, which didn't extend to the boundary. In both situations, the dressed operator must still have superlinear weight. 

With \emph{tennis ball 1}, we can, for example, label all logical operators that can be supported solely in $z \geq 2L_z/3$ or $z \leq L_z/3$ as ancillary.
Computing the remaining ground state degeneracy numerically, the number of logical qubits is
\begin{equation}
  k = 2\left(\left\lfloor \frac{L_z}{3} \right\rfloor + (L_z\text{ mod }3)\right) 
\end{equation}
The periodic scaling is a result of specifying discrete lattice points for the cut-off regions when $3$ may not divide $L_z$. 

It is therefore possible to modify the family of cubic codes such that we maintain the
superlinear distance and partial self-correction while improving $k$ to be a simple linear
function of $L_i$, with only a constant-order periodic correction. Moreover, these codes do not
require any periodic boundaries and are thus comparably more realistic for implementations in physical systems.

%% file: defects/defects.tex
Having presented the key properties of boundaries in the cubic code in the previous section, in the following section we similarly introduce and characterize three defect types: vacancies, edge dislocations, and screw dislocations. A discussion of their use in QEC codes is then provided in Section \ref{sec:superdefects}. Defects have been used in previous work to encode additional quantum information and to perform logical Clifford operations~\cite{bombinTopologicalOrderTwist2010, brownTopologicalEntanglementEntropy2013,barkeshli2014symmetry,
  youNonAbelianDefectsFracton2019, barkeshliCodimensionDefectsHigher2022,
  krishnaTopologicalWormholesNonlocal2020,
  kesselringBoundariesTwistDefects2018, brownPokingHolesCutting2017}. 

From a condensed matter perspective, defects are an important consideration when forming a complete understanding of a given phase of matter. For example, encircling a defect can induce a lattice translation. Since fractons are fundamentally immobile, this translation has the potential to cause nontrivial changes to the fracton topological order~\cite{manojScrewDislocationsXcube2021}, such as in the form of increased mobility.

\subsection{\label{sec:vacancies}Vacancies}

Vacancies, the removal of lattice sites (and qubits) within the bulk of the
model, behave similarly to boundaries. 
Stabilizers obtained by truncating the operators acting on the vacant site still
commute with all stabilizers away from the vacancy and with other truncated
stabilizers of the same type. For simplicity,
we consider only vacancies with single choices of truncated stabilizers, which
we denote as $\braket m$ and $\braket e$ for $X$ and $Z$ respectively. As with
the boundaries, these results could be generalized to more complex
constructions, such as $(110)$ orientations or combinations of $e$ and  $m$
faces, in future works. 

\begin{table*}
  \centering
  \begin{tabular}{ccccc}
    \toprule
    Vacancy Type & Vacancy Boundary & Stabilizer & Boundary Sign & Behavior
    \\ 
    \cmidrule(r){1-1} \cmidrule(lr){2-2} \cmidrule(lr){3-3} \cmidrule(lr){4-4}
    \cmidrule(l){5-5} 
    $\braket m$ & $(m)$ & $X$ & $+$ & $m$ condenses \\ 
    $\braket m$ & $(m_{ABC})$ & $X$ & $-$ & Eq.~(\ref{eq:abc_fusion}) \\ 
    $\braket e$ & $(e)$ & $Z$ & $-$ & $e$ condenses \\ 
    $\braket e$ & $(e_{ABC})$ & $Z$ & $+$ & Eq.~(\ref{eq:abc_fusion}) (for $e$)
    \\
  \bottomrule
\end{tabular}
\caption{Terminology for vacancies, analogous to Table
\ref{table:boundary_condensations} for open boundaries.}
\label{table:vacancy_condensations}
\end{table*}

We identify here three key ways in which bulk excitations interact with nearby
vacancies, the first two of which are equivalent to the boundary interactions.

Firstly, fractons can condense into certain faces of $\braket m$ and $\braket e$ vacancies, just as with the exterior lattice
boundaries. The analogous results to Table \ref{table:boundary_condensations}
are summarized in Table \ref{table:vacancy_condensations}. 

Secondly, if the vacancy extends in a particular direction, such as $\hat{z}$, mobilized charges on $(m_{ABC})$ and $(e_{ABC})$ faces of the vacancy are able to
repeatedly \emph{hop} along $\hat{z}$. If the vacancy extends around a lattice periodic in $z$, and with $3|L_z$, then an operator can create two $m_A$ out of
the vacuum, hop one charge around the periodic boundary and annihilate it with the other. This is equivalent to the boundary behavior in Section \ref{sec:periodic}.

Finally, by employing a \emph{cage operator}, fractons can encircle a vacancy.
As described in Ref.~\cite{bulmashBraidingGappedBoundaries2019} and Section
\ref{sec:cage}, these operators create a set of fractons out of the vacuum,
separate them via a cascading procedure in two directions, before inverting the
cascading operation to bring the excitations back together to annihilate - returning to the vacuum state. Such an operation in the bulk will typically be
trivial, resulting in a product of stabilizers. However, the vacancy removes
terms from the stabilizer group, causing a cage that encircles a vacancy to be potentially nontrivial.

\begin{figure}
  \centering 
  \subfloat[\label{fig:twist1}At the ends of the dislocation are two
  \emph{twist} stabilizer lines (blue), shaped like trapezoidal
  prisms.]{\includegraphics[height=3.5cm]{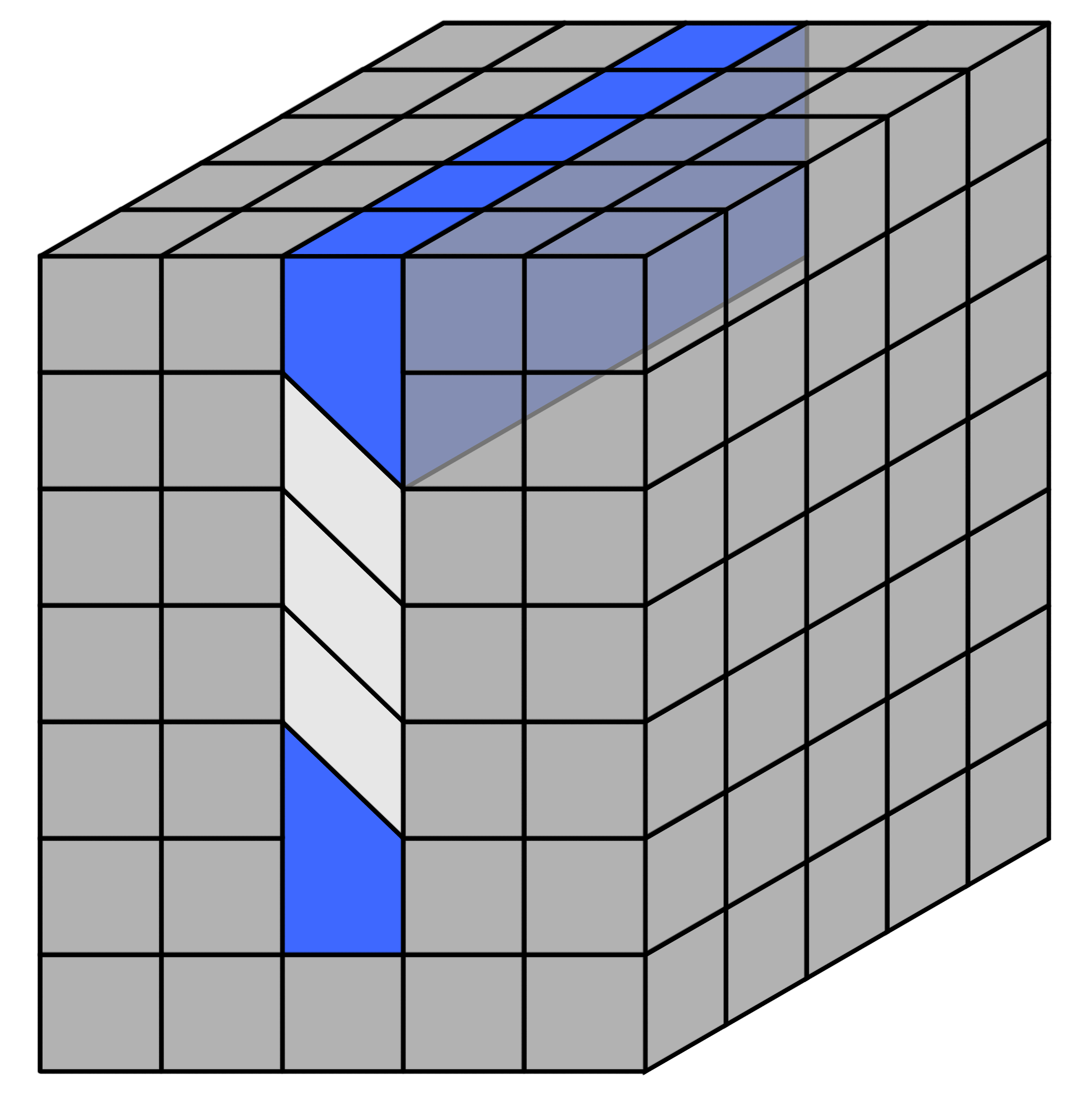}}
  \hspace{0.03\textwidth}
  \subfloat[\label{fig:twist2}Closed loops in the bulk and around a twist.]
      {\includegraphics[height=3.5cm]{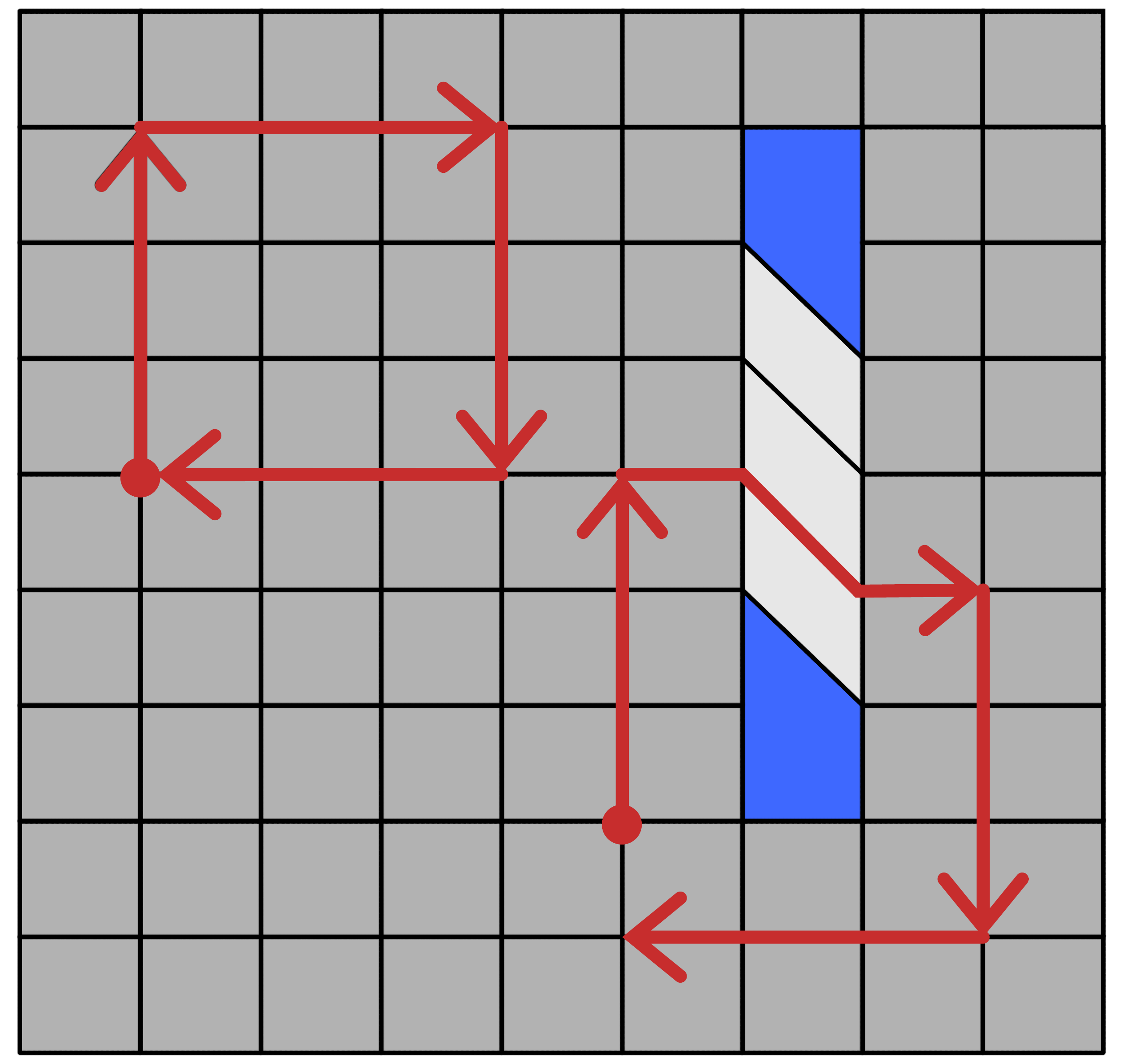}}
  \caption{Edge dislocations on the cubic code contain twists (blue). An
    operator that moves a fracton around a closed cage (red loop) in the bulk
    will no longer be closed when encircling a twist. This translation
    potentially increases the mobility of fractons when near a defect.}
  \label{fig:twist1_2}
\end{figure}

\begin{figure*}
  \centering
  \includegraphics[width=0.7\linewidth]{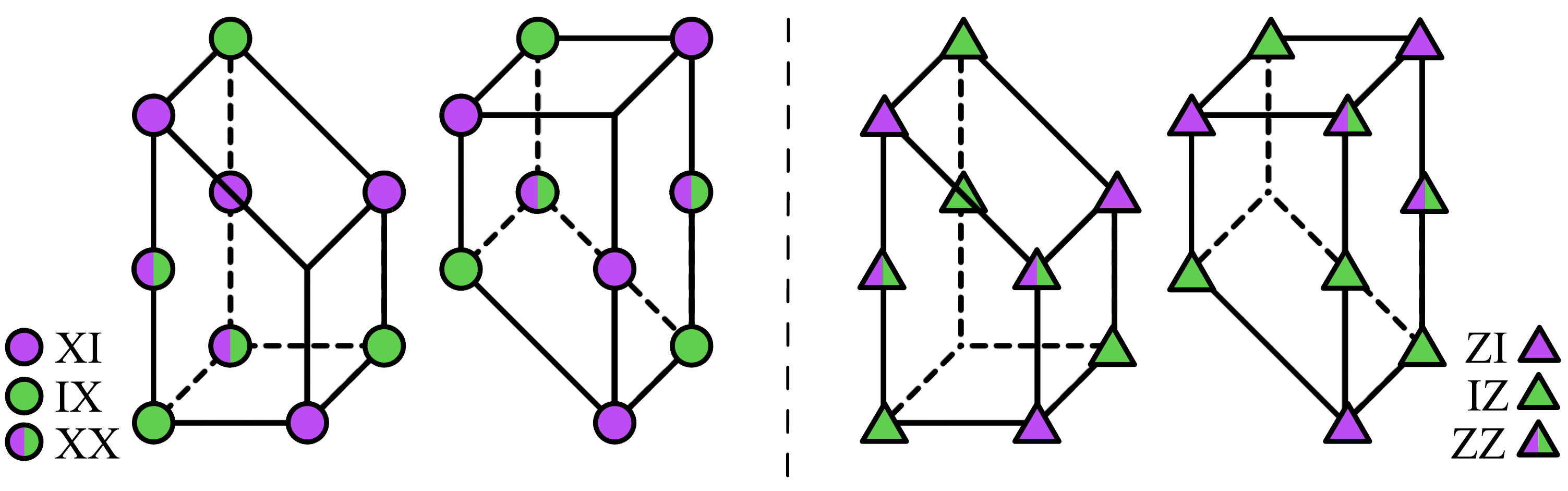}
  \caption{Stabilizers for the twists occurring at the ends of edge dislocations.
  There are four types, corresponding to $T_X$ (left pane) and $T_Z$ (right
pane), as well as at the bottom and top of the edge dislocation. The stabilizers
were found by considering the commutation with the neighboring bulk $C_X$ and
$C_Z$ stabilizers.}
\label{fig:twist_stabs}
\end{figure*}

\subsection{\label{sec:twists}Edge Dislocations}

Edge dislocations in (3+1)D are analogous to those in the (2+1)D surface code~\cite{bombinTopologicalOrderTwist2010}, with the additional feature that the dislocation becomes a line extending into the third spatial dimension, as in Fig.~\ref{fig:twist1_2}. Along the
dislocation line, we include trapezoidal prism stabilizer terms - known as
\emph{twists} in Refs.~\cite{bombinTopologicalOrderTwist2010,
brownTopologicalEntanglementEntropy2013}. Joining the twists are a region of
slanted $C_X$ and $C_Z$ operators (shown as white in Fig.~\ref{fig:twist1_2}) that naturally
commute with the adjoining bulk stabilizers. For the twists themselves, however,
there are only two choices of stabilizers: one from $X$ terms (denoted $T_X$)
and another from $Z$ ($T_Z$). These are provided in Fig.~\ref{fig:twist_stabs}.
Importantly, although $T_X$ and $T_Z$ commute at the same location, adjacent
$T_X$ and $T_Z$ anti-commute. Constructing a gapped edge dislocation, therefore,
requires taking either pure $T_X$, pure $T_Z$, or $T_X$ and $T_Z$ jointly on
every second stabilizer (or some combination of the above).
We could further consider a non-commuting, potentially gapless, Hamiltonian along the defect including all $T_X$ and $T_Z$ terms with arbitrary weights. Due to the commutation relations, this Hamiltonian is seen to be equivalent to two copies of the $(1+1)D$ quantum Ising model. 
We proceed to consider only the pure $T_X$ or $T_Z$ defects, leaving more complicated configurations to future work.

As sketched in Fig.~\ref{fig:twist2}, using a cage operator to propagate a fracton around a twist does not return the fracton to its original position; instead, the fracton moves by the Burgers vector of the dislocation.  The once-immobile fractons thus gain limited mobility in the vicinity of a twist. This behavior has the potential to introduce nontrivial logical operators. 

As with boundaries and vacancies, bulk excitations can condense onto particular
twists. A $T_X$ twist condenses $m$ charges, while $T_Z$ twist condenses $e$.
Since an edge dislocation is characterized by two twists, we denote, for example,
$\braket{em}$ to be a $T_X$ twist on the positive side of the pair, and $T_Z$ on
the negative side.

\subsection{\label{sec:screws}Screw Dislocations}

\begin{figure}[h]
  \centering 
  \includegraphics[width=0.4\textwidth]{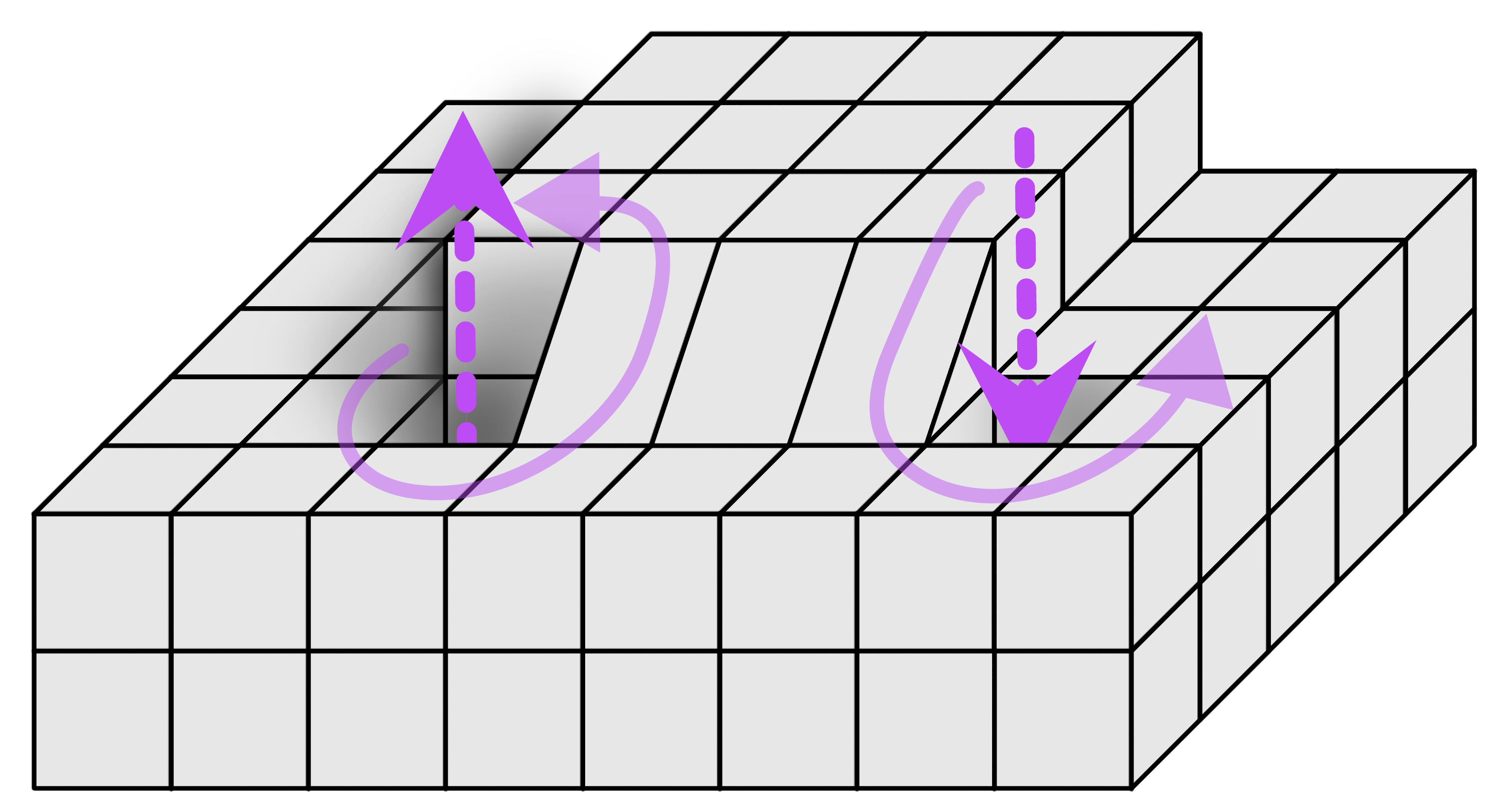}
  \caption{The two independent forms of screw dislocations in a (3+1)D lattice,
  such as the cubic code. The screws are labeled by the handedness of the
  Burgers vector when traversing around the dislocation. On the left is a
  right-handed screw, $\braket R$, and on the right is a left-handed screw,
$\braket L$.}
  \label{fig:screw}
\end{figure}

The final defect type we consider is the \emph{screw dislocation}, shown in
Fig.~\ref{fig:screw}. We label the screw as $\braket L$ or $\braket R$ by the
handedness of the Burgers vector when traversing around the dislocation. 
Unlike the edge dislocation, the screw invokes a
translation vector along the defect line itself. Importantly, this means that a
fracton can continuously wind around the defect via cage operators while
translating along the screw. Fractons thus gain 1D mobility in the
vicinity of screw dislocations.

Unlike vacancies and twists, however, for the screw dislocations with a Burgers vector of $1$ considered here, there is no local operator supported on the dislocation line that commutes with the bulk stabilizers. Because of this, $\braket L$ and $\braket R$ screws have no inherent stabilizer and can condense both $m$ and $e$ excitations. 
Notably, this condensation behavior largely negates any
change to the mobility of fractons near this screw, since individual $e$ and $m$ can be created or destroyed at any location along the defect, and thus become trivial.

%% file: defects/defects_super.tex
Similar to the discussion of boundary codes in Section \ref{sec:superbdries}, in
this section we examine how defects can be used with the cubic code to create QEC codes with superlinear distances. Further configurations without this property are
discussed in Appendix \ref{sec:otherdefects}. Notably, although there are codes using screw dislocations with a superlinear distance, the number of encoded qubits there is constant. We therefore defer the discussion of screw-dislocation codes to Appendix \ref{sec:enc_screw}.

\begin{figure*}[t]
  \centering 
  \subfloat[\label{fig:twist_logicals}Logical operators between pairs of
  $\braket{mm}$ twists. One traverses the gap between two twists on the same
  dislocation, while the other joins one twist from each pair. The coloring in
  this picture was arbitrary: $e$ and $m$ operators can be found in both
  configurations.]
  {\includegraphics[height=6cm]{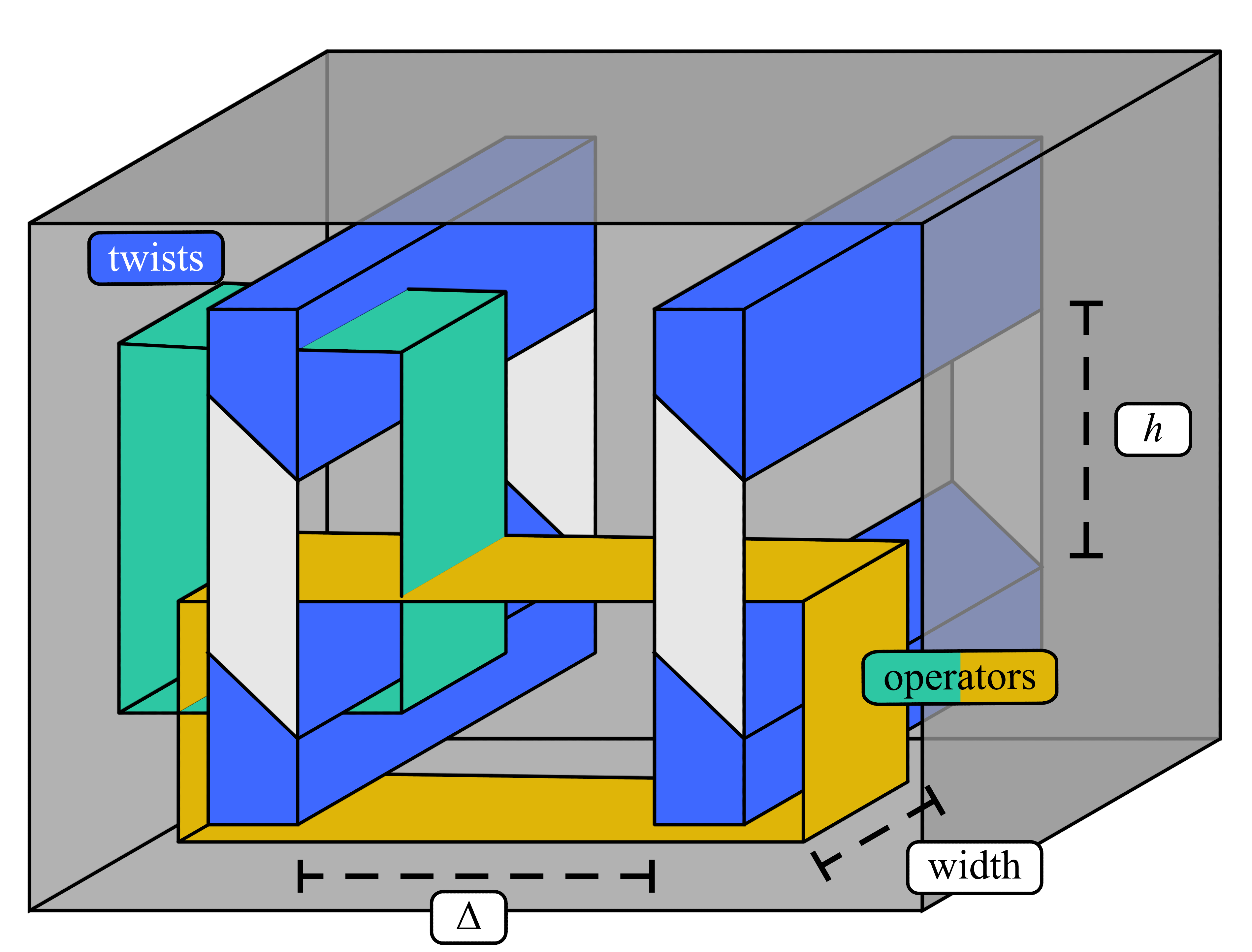}}
  \hspace{0.02\textwidth}
  \subfloat[\label{fig:twist_sep}Minimum $x$ width required to support at least
  one logical operator, as the height $h$ and separation $\Delta$ increases.
  Colors show results for different twist types. A linearly-increasing width
  indicates that the weight of the operator should be superlinear in $\Delta =
  h$.]
  {\includegraphics[height=7cm]{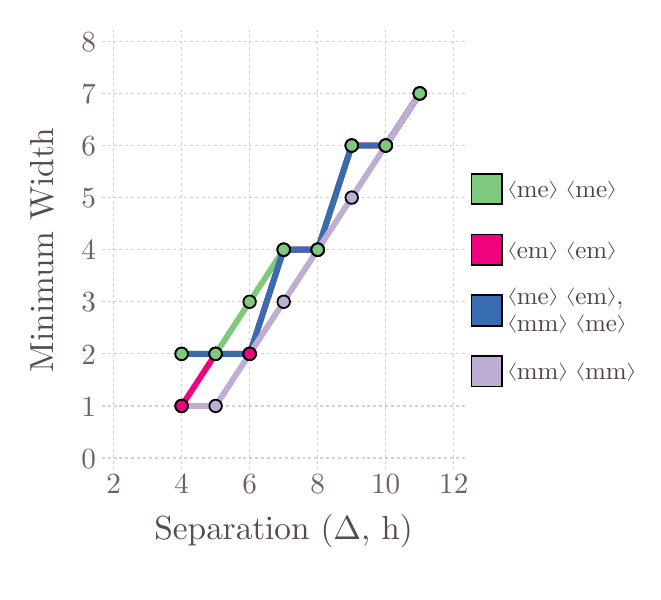}}
  \caption{Pairs of edge dislocations support
  superlinear-weight logical operators.}
  \label{fig:twist_pairs}
\end{figure*}

\subsection{\label{sec:supervac}Vacancy Encodings}
Consider a lattice of size $(L_x, L_y, L_z) = (L_\infty, L_\infty, L_z)$, where
$L_\infty \gg L_z, \kappa$ with $\kappa$ the correlation length of the ground state,
such that interactions with the $x, y$ boundaries are negligible. We also place
periodic boundary conditions in the $z$ direction. Within this, there is an
$\braket m$ vacancy of width $(w_x, w_y, L_z)$ such that it extends around the
periodic boundary. We expect that a single vacancy in this configuration does
not modify the ground state degeneracy, since creating and annihilating $m$ on
the same vacancy is trivial. Indeed, this result is confirmed by numerical
computation, in all cases except when $3 | L_z$. As discussed in Section
\ref{sec:periodic}, $m$ charges gain 1D mobility along a periodic boundary,
producing additional logical operators. This gives a number of encoded qubits 
\begin{equation}
  k = 4\tau(L_z;\;L_\infty)
\end{equation}
where $\tau$ is defined in Eq.~(\ref{eq:tau}), with the factor of $4$ arising
from the two unique string-like operators on the two $(m_{ABC})$ boundaries of the vacancy.
These logical operators all have a linear weight. 

Once additional $\braket m$ vacancies are introduced, however, additional logical operators arise. Numerically computing the ground state degeneracies, the number of encoded qubits scales with
\begin{equation}
  k = 2(v-1)L_z + 4\tau(L_z;\; L_\infty)
  \label{eq:kvacancies_kn}
\end{equation}
where $v$ is the number of vacancies.

$m$ excitations can now cascade through the bulk from one vacancy to another, creating an $\bar X$
logical operator with a weight that is superlinear in the vacancy separation $\Delta$ (using the Manhattan distance). Each additional vacancy introduces an additional
independent site for condensation, thus increasing $k$. Moreover, since
translation is a nontrivial action in the bulk of the model, performing the
cascading procedure at different values of $z$ produces independent operators.
This gives the dependence on $L_z$.

On the other hand, the $\bar Z$ logical operators are formed from cages of $e$
excitations encircling a vacancy. Importantly, because these cage operators
are moving charges through the bulk around the vacancy, these must also have a weight that scales superlinearly with the widths $w_x,w_y$. If these widths were scaled with the
separation between the vacancies, this encoding has the potential to be partially
self-correcting (when $3\! \nmid\! L_z$), while also supporting a number of encoded qubits that scale
linearly in $L_z$ and the number of vacancies. 

If the assumption of $L_x, L_y \gg L_z$ is relaxed, such as for an $(mmp;mmp)$ code with $v$ number of $\braket m$ vacancies wrapping around the periodic boundary, we
get two additional logical operators per $L_z$, giving 
\begin{equation}
  k = 2vL_z + 4\tau(L_z;\;L_\infty)
\end{equation} 
in comparison to Eq.~(\ref{eq:kvacancies_kn}). This is because $m$ charges can now also cascade from the vacancies to a boundary. 

Overall, this code is a notable improvement over the original cubic code model and does not require a subsystem code as in Section \ref{sec:superbdries}. Unlike vacancies in the surface code, there is also no significant trade-off between the code distance and the number of encoded qubits, since $v$ can be kept constant while using $L_z$ to increase $k$, and using the $x,y$ dimensions (vacancy width and linear system size) to increase the code distance. However, this construction does require a topology with one periodic direction.

\subsection{\label{sec:superdefect}Edge Dislocation Encodings}
A single edge dislocation in the bulk, far from any boundary, encodes no
additional logical qubits. 

However, more appealing behavior arises using multiple edge dislocations.
Consider two dislocations, in Fig.~\ref{fig:twist_logicals}, each of height $h$
in the $z$ direction and separated by perpendicular distance $\Delta$ in the $y$
direction. The dislocation line extends along $x$. Numerically computing the
ground space degeneracy yields
\begin{equation}
  k = 4L_x + \mathcal O(1)
  \label{eq:kdefects}
\end{equation}
for some additional constant arising depending on the exact choice of twist
stabilizers, $\Delta$, and $h$. 

Effectively, these logical operators consist of movement of $e$ and $m$
excitations around and between two of the four twists, as in
Fig.~\ref{fig:twist_logicals}. Importantly, because these cage operators are
moving charges through the bulk, the prohibition of string-like operators means
that the width of the support in the $x$ direction
must increase with $h$ and $\Delta$. This result is confirmed by numerically
computing the minimum width that supports logical operators as we scale $\Delta
= h$ in Fig.~\ref{fig:twist_sep}. Slight variations in the widths are due to the
particulars for constructing the operators around the twists. However, in all
cases, we observe an overall linear trend. 

If the analysis above is repeated except with open or periodic boundary
conditions in $x$, we observe the same scaling as in Eq.~(\ref{eq:kdefects}).
There is an additional, constant number of logical operators that have support
solely on a single twist and are string-like in the $x$ direction. Unlike the case with vacancies, these operators remain even if $3 \! \nmid \! L_x$. As with the
case for \emph{tennis ball 1}, considering a subsystem code to ignore these extra
logical operators may be sufficient to ensure the dressed code has a superlinear distance as $\Delta$ and $h$ are increased. A rigorous proof of this solution is deferred to future works. 

As with periodic vacancies, pairs of twists provide a promising approach to improve the cubic code: maintaining 
the desirable code distance while encoding a number of qubits that scales linearly with $L_x$.

%% file: conclusion.tex
In this work, we have presented a systematic study of open boundaries and defects in Haah's cubic code. We focused on planar $(100)$-like boundaries normal to the crystallographic axes and constructed $X$- and $Z$-type open boundary conditions using truncated plaquette, edge and vertex stabilizer terms. The interaction of these boundaries with fractonic topological excitations depends intrinsically on their orientation: $X$-type negative faces and $Z$-type positive faces condense single fractons of the respective type, while the opposing faces lead to increased fracton mobility within their vicinity. These otherwise fractonic excitations become mobile within a $(2+1)$D diagonal subsystem along the surface. This implies that the fundamental no string-like operator property of the original cubic code is violated in the vicinity of these boundaries. 

Similar behavior was observed in the vicinity of vacancies, edge dislocations, and screw dislocations: patterns of fracton condensation that lead to increased mobility were seen to depend on the orientation and stabilizer type of each defect. The nontrivial action of translation symmetry on a type-II fracton topological phase means that encircling a dislocation defect with a cage operator can also lead to increased topological excitation mobility. We found that dislocation defect encodings were therefore able to support additional forms of logical operators that do not appear for encodings constructed from boundaries and vacancies. 

The cubic code is known to form a partially self-correcting quantum memory on periodic boundary conditions~\cite{bravyiAnalyticNumericalDemonstration2013}. 
The absence of any string-like logical operator is essential to enable such a quantum memory~\cite{haahLocalStabilizerCodes2011}. 
With this in mind, we aimed to determine if it is possible to retain the superlinear distance of the original cubic code model while making the number of encoded qubits scale as a simple linear function of the linear system size, without sporadic fluctuation. 
We have shown that it is possible to achieve this using a combination of open and periodic boundaries, vacancies, and defects, despite the no string-like operator property potentially being lost when translation-invariance is violated by the introduction of defects. 
For cubic codes with open boundary conditions - which are typically easier to realize in a physical implementation - it is possible to achieve a superlinear code distance by restricting the encoded states to a subsystem, with all dressed logical operators maintaining a weight that scales superlinearly with the linear system size. We have shown this to be possible with the \emph{tennis ball 1} code construction. It is an open question whether this can be generalized to other open boundary condition codes, such as those in Appendix~\ref{sec:otherbdries}. Our results also focused on planar $(100)$-like boundaries; it remains an open question as to whether alternate constructions such as $(110)$ boundaries yield fundamentally different behavior. 

We showed that emergent $(2+1)$D topological order arises on certain open boundary conditions,
supporting particles with increased mobility. Interestingly, this phase can be unitarily
transformed into the $\text{6-6-6}$ color code. 
We leave a further investigation of this correspondence, such as how it extends into the bulk, to future research. Similarly, it would be interesting to relate the bulk cubic code defects to defects of the color code~\cite{kesselringBoundariesTwistDefects2018} on the boundary.  Presumably, the emergent color code on the boundary has an associated non-invertible anomaly~\cite{Ji2019} which is a property of the type-II fracton bulk phase. Understanding the nature of the surface topological order could uncover further insights into type-II fracton topological order via a bulk-boundary correspondence.

In this work we have only explored Pauli-$X$ or $Z$ type boundaries of the cubic code, rather than more complicated mixed or twisted boundary conditions. 
We suspect that such boundary conditions exist and are inequivalent to those we have studied. 
Our reasoning uses the construction of twisted boundaries via gauging symmetry-protected topological (SPT) domain walls~\cite{yoshida2015gapped}. 
For the cubic code, this allows a construction of boundaries via the gauging duality to a 3D fractal Ising model~\cite{vijayFractonTopologicalOrder2016,williamsonFractalSymmetriesUngauging2016}. 
For one type of boundary condition of the fractal Ising model, the symmetry action on the boundary is simply the action of the bulk fractal symmetry restricted to the boundary plane. 
This symmetry action on the boundary appears to also be a fractal of the form considered in Ref.~\cite{devakulClassifyingLocalFractal2019}. 
Following Ref.~\cite{devakulClassifyingLocalFractal2019}, one can consider stacking the fractal Ising model with a reflected copy of itself such that the resulting boundary fractal symmetry supports a nontrivial fractal SPT phase. 
We can then twist the boundary condition by the associated SPT entangler and gauge the resulting model to obtain a twisted boundary condition of a cubic code stacked with a reflected cubic code. 
We leave the detailed study and classification of such twisted boundary conditions to future work. We anticipate that existing constructions of type-II fracton models from layers of fractal SPTs may prove advantageous for this study~\cite{williamsonTypeIIFractonsCoupled2021}. 

We now highlight several further directions for future consideration. It remains to be seen how the results in this paper can be generalized to other type-II fracton topological phases, such as the additional codes introduced in Ref.~\cite{haahLocalStabilizerCodes2011} or the fractal spin liquid codes~\cite{yoshidaExoticTopologicalOrder2013}. 
It would be interesting to relate defects in the latter codes to known defects in $(2+1)$D topological codes via the fractalization procedure of Ref.~\cite{devakulFractalizingQuantumCodes2021}. 
Our work, together with previous work on type-I fracton phases~\cite{bulmashBraidingGappedBoundaries2019}, raises the question of developing a general theory of boundaries and defects for fracton topological orders. 
A promising approach to this is the inclusion of conventional topological boundaries and defects into the topological defect network framework for fracton topological order~\cite{Slagle2018FFT,aasenTopologicalDefectNetworks2020,songTopologicalDefectNetwork2023}. 
Another interesting open question is the development of a notion of Lagrangian algebra objects~\cite{levinProtectedEdgeModes2013,beigi2011quantum,Kitaev2012,Kong2014} for fracton topological orders that can be used to classify possible gapped boundaries in relation to fracton braiding statistics~\cite{Pai2019,Song2023}. 
A further challenge is the extension of these concepts to nonabelian fracton models~\cite{vijay2017generalization,Prem2019Cage,song2018twisted,Prem2019Gauging,Bulmash2019Gauging,Williamson2020Designer,Sullivan2021,Tantivasadakarn2021}.

Our work leaves open the question of superlinear code distances for subspace stabilizer encodings in type-II codes with open boundary conditions. It could be interesting to search for bounds on the best achievable parameters for a topological subspace stabilizer code with open boundary conditions in $(3+1)$D. 
Proving rigorous lower bounds on the code distance and energy barriers, as well as modifying the decoding algorithms from Ref.~\cite{haahLatticeQuantumCodes2013} and~\cite{brownParallelizedQuantumError2020} in the presence of boundaries and defects, would allow us to make a more definitive judgment on the feasibility of these codes as self-correcting quantum memories. 
It would also be useful to consider how lattice defects can be braided to produce Clifford gate sets, such as with the surface code~\cite{bombinTopologicalOrderTwist2010}. 
Another open direction for future work is calculating the fusion of multiple twist and screw dislocations to create defects with larger Burgers vectors, and the relationship between braiding and condensations on these defects. 
This raises the challenge of developing a theory of translation symmetry enrichment for fracton topological orders, extending the $(2+1)$D results of Ref.~\cite{barkeshli2014symmetry}. 
This is a promising lens through which to understand the general structure of fracton topological orders on crystal lattices. 
We remark that the possible phenomena exhibited by known fracton models reveal a far richer theory than the two-dimensional analog~\cite{Else2018}. 
In particular, a general theory should capture the interplay between subgroups of translation symmetry and bifurcating entanglement-renormalization group flows satisfied by fracton topological orders including the cubic code~\cite{Evenbly2014,haah2014bifurcation,shirley2017fracton,Dua2019Bifurcating}.

\acknowledgements
\noindent
The authors acknowledge fruitful collaboration with Tom Iadecola and Meng Cheng during the early stages of this work. 
AD and DB are supported by the Simons Collaboration on Ultra-Quantum Matter, which
is a grant from the Simons Foundation (651438, AD; 651440, DB). 
AD is also supported by the Institute for Quantum
Information and Matter, an NSF Physics Frontiers Center (PHY-1733907). 
ACD is supported by the Australian Research Council Centre of Excellence for Engineered Quantum Systems (EQUS, CE170100009). 
DW is supported by the Australian Research Council Discovery Early Career Research Award (DE220100625). 

%% file: appendix/app_bdries.tex
Configurations such as \emph{tennis ball 1}, vacancies around periodic
boundaries, and pairs of twists have the potential to be partially
self-correcting, while having a number of encoded qubits that scales linearly in
some macroscopic quantity. However, this behavior is not true for the majority
of other configurations. In this section, we present the remaining results from a
systematic study of all boundary configurations (up to symmetries, and with pure $(100)$-like planar boundaries). A discussion on defects is provided in Appendix \ref{sec:otherdefects}. 

We first consider models with open boundary conditions on all six faces, before
also addressing combinations with periodic boundaries. We
perform a systematic search over all configurations, up to symmetries, and their
key results are summarized in Tables \ref{table:boundary_codes} and
\ref{table:periodic_codes}. 

\subsection{\label{sec:simple_bcodes}Simple Cases}
Following the construction in Section \ref{sec:bdries}, we first verify that our
boundary conditions produce gapped surface Hamiltonians. To do so, we construct
an $(L_x, L_y, L_z)$ model with only $X$ or $Z$ stabilizers on all
faces: $(eee;eee)$ or $(mmm;mmm)$. Numerically, we verify that all
configurations (tested up to reasonable lattice sizes) contain a non-degenerate ground state - or equivalently, no encoded qubits. 
Intuitively, this agrees with our analysis so far. Consider $(mmm;mmm)$. Using either the cascading procedure or the fractal tetrahedra, one can consider
$m$ excitations moving between two or more opposing boundaries of the lattice,
creating an operator that commutes with the stabilizers. However, no product of
$Z$ operators can create and annihilate $e$ excitations to produce a nontrivial
$\bar Z$ logical operator. There are thus no encoded logical qubits and an
equivalent argument holds for $(eee;eee)$.

For similar reasons, if one face is replaced with the opposing boundary type,
such as $(mmm;emm)$, we again see no encoded logical qubits. As with other stabilizer codes, creating an excitation on a boundary and then condensing it back into the same boundary does not constitute a nontrivial operation~\cite{bravyiQuantumCodesLattice1998}.

\subsection{\label{sec:tennis2}Tennis Ball 2}
In addition to the first \emph{tennis ball} code discussed in Section
\ref{sec:tennis1}, there are two additional configurations that we argued may
have a superlinear distance. The second \emph{tennis ball} code is specified using
$(e)$ and $(m)$ on the remaining two faces. Consider the case shown in Table
\ref{table:boundary_codes}, with $(mee;mem)$.

\begin{figure}
  \centering 
  \subfloat[The $\bar X$ logical operators that condense $m$ charges on the
    $(\bar100)$ and $(00\bar1)$ faces can be constant-weight as the lattice size
  increases.]
  {\includegraphics[width=0.8\linewidth]{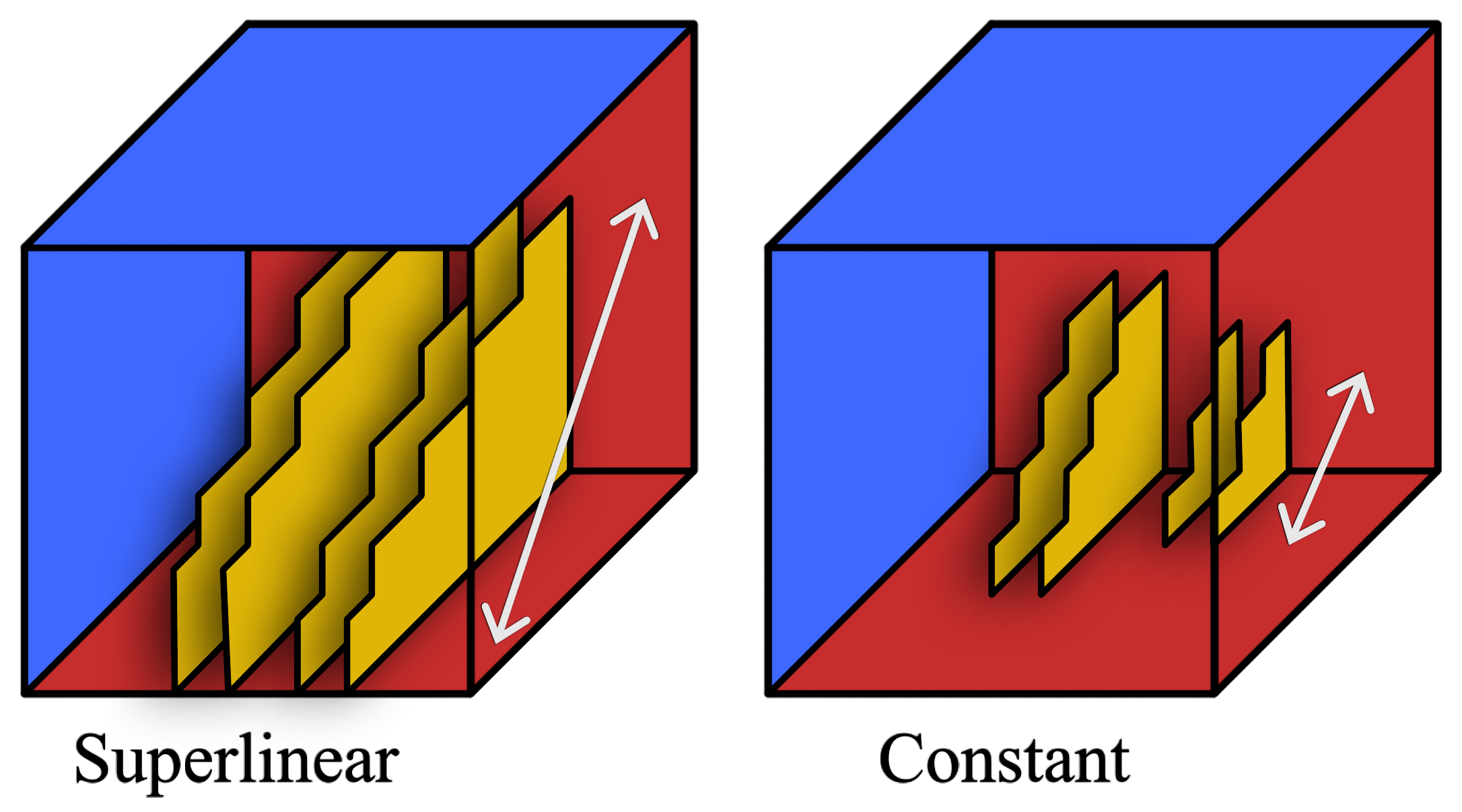}}
  \hspace{12pt}
  \subfloat[The $\bar Z$ logical operators that condense $e$ charges on the
    $(010)$ and $(001)$ faces can also be constant-weight as the lattice size
  increases.]
  {\includegraphics[width=0.8\linewidth]{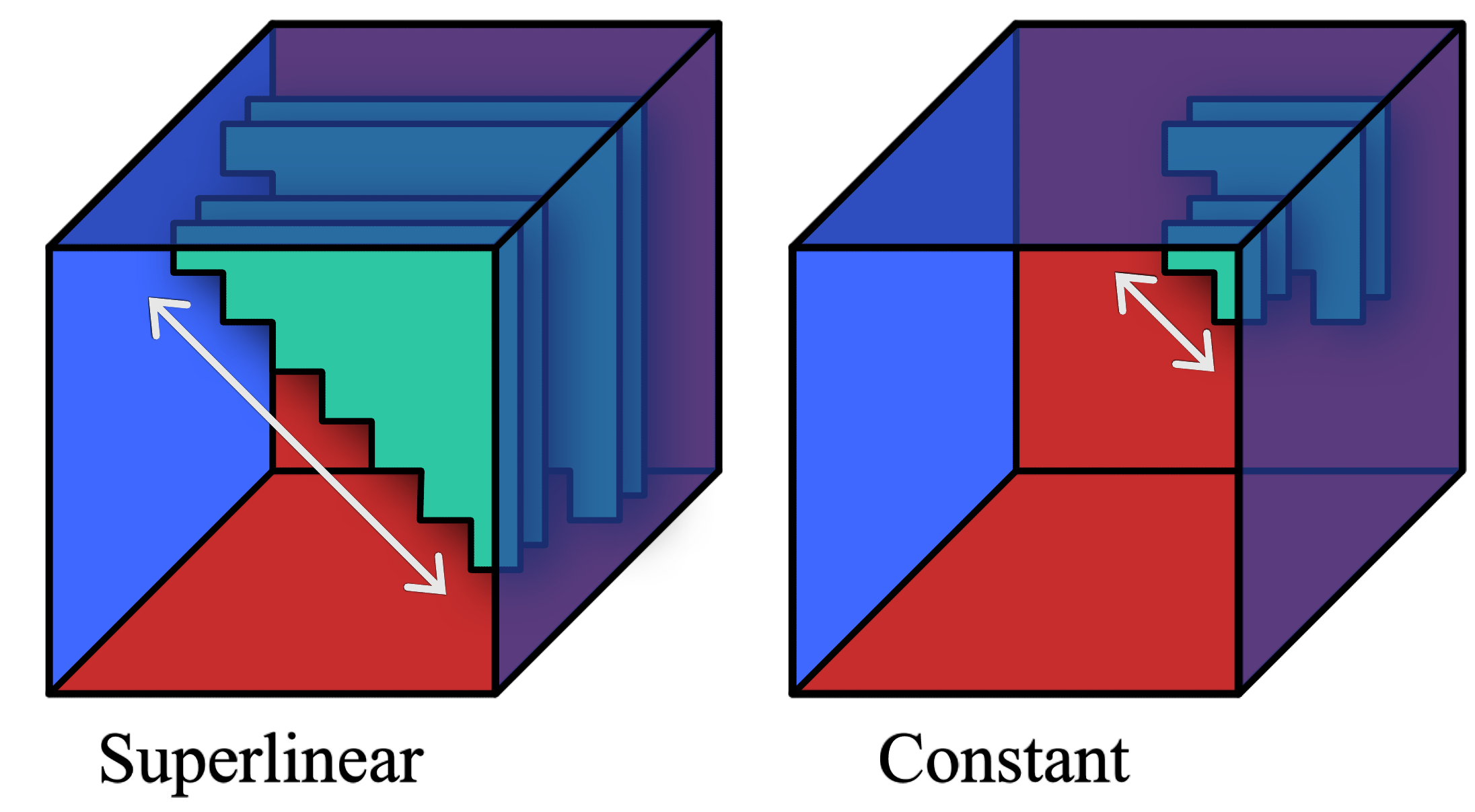}}
  \caption{Logical operators on the second \emph{tennis ball} configuration,
    $(mee;mem)$. Solid shapes indicate repeated applications of the $F$
  operators to cascade $m$ (yellow) and $e$ (cyan) charges. Red faces represent
$X$, and blue for $Z$, stabilizer choices on the boundaries.}
  \label{fig:tennis_logicals_2}
\end{figure}

The logical operators can be constructed by a similar cascading procedure to
\emph{tennis ball 1}, except now acting along a diagonal direction (see
Fig.~\ref{fig:tennis_logicals_2}). As with the first \emph{tennis ball}, not all such
logical operators are superlinear. Indeed, those localized near the
$(\bar1\bar10)$ and $(011)$ edges are constant-weight with respect to the linear system size. That is, as $L_x,L_y,L_z$ are increased, there will remain a
number of logical operators with constant support, localized near those edges.

Numerically computing the ground state degeneracy, we arrive at a number of
encoded qubits
\begin{equation}
  k =2L_z - 6
\end{equation}
with the constant offset in comparison to \emph{tennis ball 1} due to the inclusion of
vertex stabilizer terms, which would otherwise be incorporated as local logical
operators. Note that the analysis here can also be generalized to explain the
scaling of encoded qubits for $(eee;mem)$ and $(eee;mmm)$ in Table
\ref{table:boundary_codes} (the latter of which has a logical algebra
constructed from two independent sets of these diagonal operators).

As with \emph{tennis ball 1}, this code, therefore, does not generally have a
superlinear distance. However, it may be possible to generalize the arguments
for the subsystem code to also reduce these logical operators into superlinear-weight
dressed operators. The specifics of this approach are left as a future
consideration.

\subsection{\label{sec:thetube}Tube}
The third code, \emph{tube}, is specified with $(e)$ and $(e_{ABC})$ -
or equivalently, $(m)$ and $(m_{ABC})$ - on the remaining two faces. In Table
\ref{table:boundary_codes}, we have chosen $(mee;mee)$. 

The $\bar X$ logical operators must cascade $m$ charges between the $(m)$ and
$(m_{ABC})$ boundaries, as these are the only possible condensation sites. Such
a cascading procedure must incur a superlinear weight as it moves through the
bulk of the lattice. Unlike with the \emph{tennis ball} configurations, there is no
adjacent $(m_{ABC})$ face on which the cascade can be cut-off, and there are
thus no linear-weight operators for $\bar X$. 
On the other hand, $e$ charges can move as strings along the $(e_{ABC})$ boundary of $(00\bar1)$ as in \emph{tennis ball 1}, and can also form local operators
between the $(e)$ boundaries of $(010)$ and $(001)$ as in \emph{tennis ball 2}. This
configuration, therefore, improves the $\bar X$ distance while reducing the
distance for $\bar Z$ errors.

Numerically computing the ground state degeneracy for small system sizes yields
a number of encoded qubits 
\begin{equation}
  k = 2(L_y + L_z - L_x) - 3
\end{equation}
We can heuristically explain this by considering the cascading of $m$ charges
from $(100)$ to $(\bar100)$. Using $F_m^{\bar x \bar y}$, we can create $m$ on
$(100)$ and push them in the $-\hat x$ direction. Since the support of the operator
increases in width with distance, the maximum distance that the charge can
be pushed is $L_y$ before the operator encounters an $(e)$ boundary,
beyond which it is confined. Using $F_m^{\bar x\bar z}$, the accumulated charges
can be cascaded a further $L_z$ distance towards $(\bar100)$. If $L_x > L_y +
L_z$ we should thus get no logical operators. For each $L_x < L_y + L_z$, we
expect two additional independent logical operators, accounting for the two
unique condensation operators. The constant factor of $3$ again arises due to
the particular choice of edge and vertex stabilizers on the boundary seams.

\subsection{\label{sec:periodic_codes}Periodic Boundaries}
There are further configurations possible that include mixtures of periodic and
open boundary conditions. All possible combinations, up to symmetry, are
summarized in Table \ref{table:periodic_codes}.

As discussed in Section \ref{sec:bg} and
Ref.~\cite{haahLatticeQuantumCodes2013}, the ground state degeneracy of
$(ppp;ppp)$ is highly dependent on the specific value of $L$. This is
because the logical operators are formed from combinations of fractal tetrahedra
with excitations at the four vertices. When $L$ is a power of $2$, these
tetrahedra tessellate perfectly and the number of encoded qubits scales as
$k=4L-2$. For other values of $L$, the tetrahedra need to be combined in
particular ways, possibly wrapping around the periodic boundaries multiple
times, leading to a nontrivial function for $k$, as defined in
Eq.~(\ref{eq:cubic_k}).

For $(ppe;ppe)$, these tetrahedral operators in $Z$ remain as logical operators,
condensing $e$ at the open boundaries or in a similar process to $(ppp;ppp)$.
However, the tetrahedra from $X$ create $m$ excitations that cannot condense at the
open boundaries, and thus these operators are no longer logical. This reduces
the number of encoded qubits by a factor of $2$ compared to $(ppp;ppp)$.
Additionally, the periodic behavior described in Section \ref{sec:periodic} now
applies, resulting in additional logical operators when $3|L$. This leads to a
scaling
\begin{equation}
  k = \frac12 k_{(ppp;ppp)} + 2\tau(L;\;\infty)
\end{equation}
when $L_x=L_y=L$. Because the $e$ excitations can condense at both boundaries in
$z$, and the periodic behavior is unaffected by $L_z$, this scaling 
only depends on $L_x, L_y$. Note that the string operators on $(00\bar1)$ are
not confined by the $(001)$ face, since $e$ excitations are able to condense
there, hence the $\tau$ scaling is also unbounded by $L_z$.

In the case of $(ppm;ppe)$ boundaries, single $m$ or $e$ excitations are not
condensed at either boundary, and we have $(m_{ABC})$ and $(e_{ABC})$ on
$(001)$ and $(00\bar1)$ respectively. The logical operators here are described
by the discussion in Section \ref{sec:periodic}, and this process can occur on
both $(00\bar1)$ and $(001)$, giving twice as many. For $(pmm;pem)$, a similar
behavior occurs, with string operators on the $(e_{ABC})$ face of $(00\bar1)$
wrapping around the periodic-$x$ boundary. 

However, for $(ppe;ppm)$ there are no encoded logical operators. Neither
the $(001)$ $(e)$ boundary nor the $(00\bar1)$ $(m)$ boundary support string
operators. 

Finally, $(pem;pem)$ can be explained using the cascading procedure. As with
$(mem;eem)$ - a rotation of the \emph{tennis ball 1} configuration in Section
\ref{sec:superbdries} such that $z$ becomes $x$ - there are $k=2L_x$
encoded qubits. Indeed, the logical operators here are equivalent: excitations
are created at $(e_{ABC})$ or $(m_{ABC})$, and cascade towards $(e)$ or $(m)$
where they are condensed. These operators can be confined to single $yz$ planes,
leading to the linear dependence on $L_x$. The periodic boundaries in $x$ do not
impede or alter this process. However, when $3|L_x$, some of these operators can
be written as strings across the $(e_{ABC})$ and $(m_{ABC})$ boundaries. We
therefore get $4\tau(L_x;\;\infty) \leq 4L_x/3$ number of linear-weight
operators, leaving a minimum of $2L_x/3$ superlinear-weight operators. To avoid
this behavior or the need for a subsystem code, linear system size can be restricted to $3\!\nmid\!\!L_x$. In such cases, since the cascading operators form a complete logical algebra and no further linear or constant-weight operators are introduced, we, therefore, expect $(pem;pem)$ to be a stabilizer code with distance that is superlinear in $L_y, L_z$.

%% file: appendix/app_defects.tex
We now present the remaining results regarding the use of screw and edge
dislocations in the bulk, adding to the discussions in Section
\ref{sec:superdefects}.

\begin{figure}
  \centering
  \includegraphics[width=\linewidth]{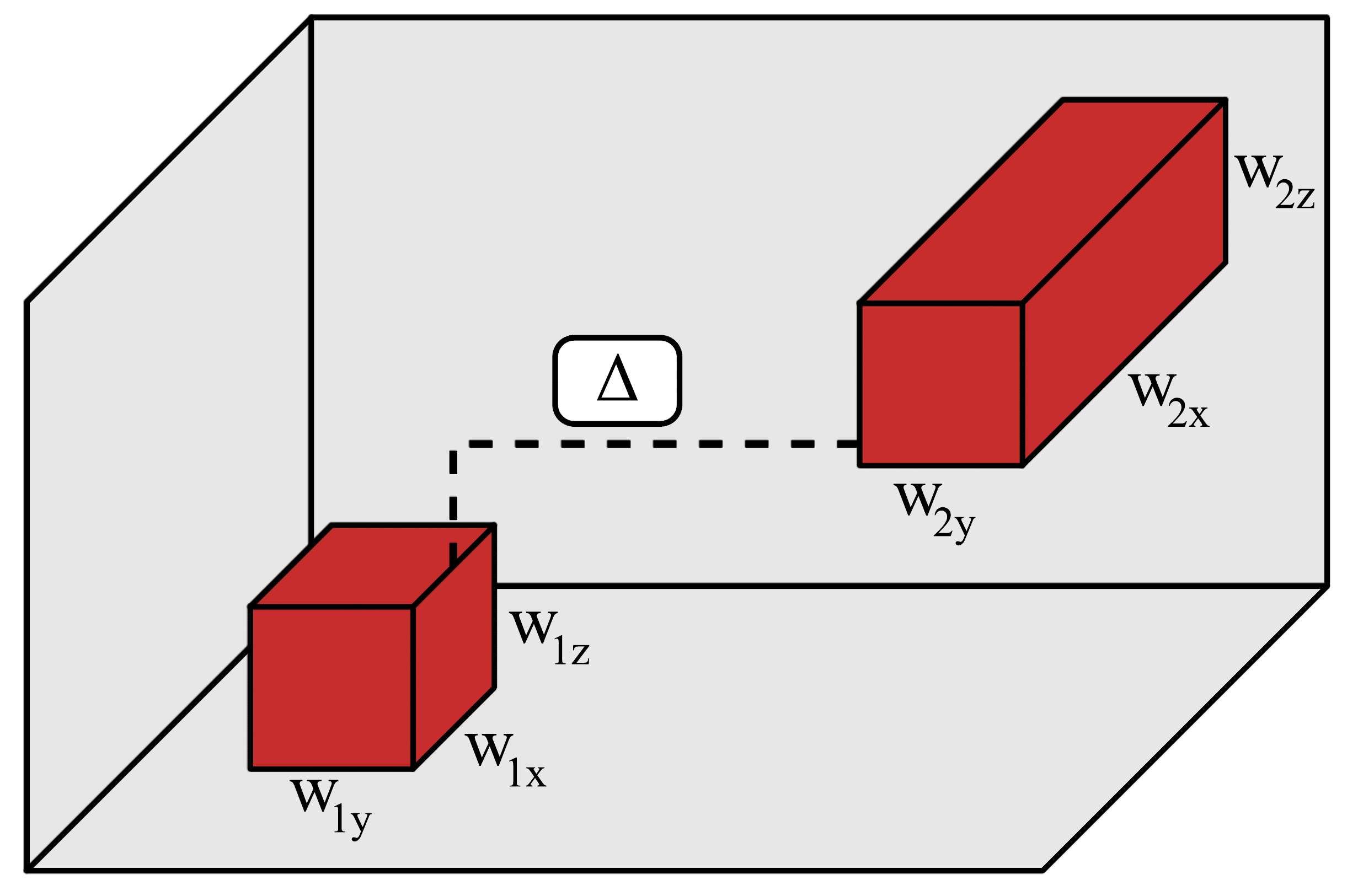}
  \caption{Two $\braket m$ vacancies of size $(w_{ix}, w_{iy}, w_{iz})$ for
  $i=1,2$ separated by a Manhattan distance $\Delta$ in the bulk.}
  \label{fig:twovacancies}
\end{figure}

\subsection{Additional Vacancy Encodings}

We now examine codes that incorporate vacancies into both a bulk model and those
involving boundaries. Consider a non-periodic lattice with
$L_\infty \gg \kappa$ in $x,y,z$, where $\kappa$ is the correlation length of
the system such that any interaction with boundary conditions can be deemed
negligible. One $\braket m$ or $\braket e$ vacancy of any size does not support
any logical operators, since any action that creates and condenses charges on
the same vacancy is trivial modulo the stabilizers. This is consistent with
results from the surface code \cite{kitaevQuantumErrorCorrection1997,
bombinTopologicalOrderTwist2010}. 

A similar result is found with one $\braket m$ and one $\braket e$ vacancy
separated in the bulk, since no excitation can condense at both vacancies. 

However, consider two $\braket m$ vacancies, each of width $(w_{ix}, w_{iy},
w_{iz})$ where $i=1,2$, separated by a Manhattan distance $\Delta$.
This is shown in Fig.~\ref{fig:twovacancies}. $m$ excitations can cascade in the
negative direction, moving between the two vacancies. Since cascading produces
additional excitations proportional to the distance traversed, we expect this
procedure to be a logical operator only if the negative vacancy is large enough
to condense these excitations. Without loss of generality, assume the two
vacancies are separated in $z$, with $i=1$ being in the negative direction. We
thus have the condition that $k > 0$ only if $\Delta \lesssim w_{1x} + w_{1y}$.

\begin{figure}
  \centering 
  \includegraphics[width=0.9\linewidth]{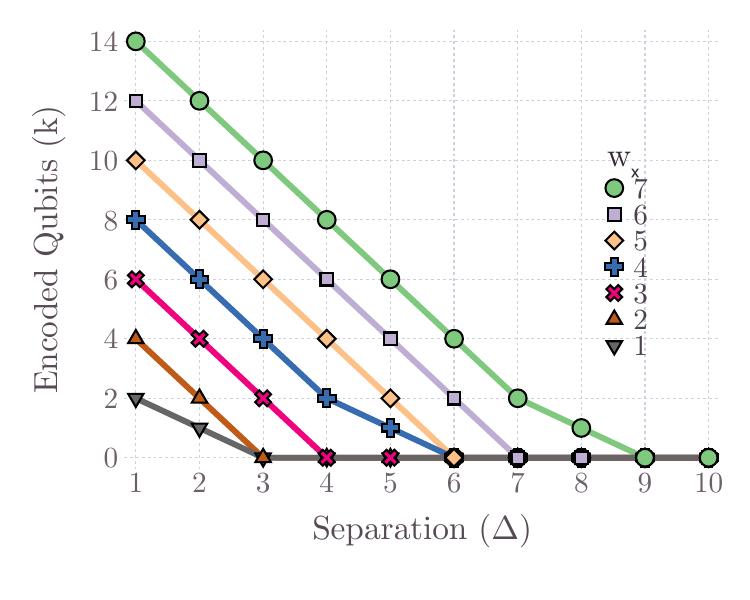}
  \vspace{-8mm}
  \caption{Numerical computation of the encoded qubits in a pair of $\braket m$
    vacancies oriented along the
  ${z}$ direction, separated by distance $\Delta$. The vacancy in the
  negative direction has
  size $(w_{1x},w_{1y},w_{1z}) = (w_x, 1, 1)$, where $w_x$ is varied in the
  figure, and the positive vacancy has $(1,1,1)$. We see the relationship in
  Eq.~(\ref{eq:k2vacancies}), with the slight deviations when $w_x =1,4,7$ due
  to the non-uniformity of composing $F_m$ operators.} 
\label{fig:k2vacancies}
\end{figure}

Numerically computing the ground state degeneracy confirms this result, and
provides a number of encoded qubits
\begin{equation}
  k = 2(w_{1x}+w_{1y} - \Delta) + c
  \label{eq:k2vacancies}
\end{equation}
where
\begin{equation}
  c =
  \begin{cases} 1 & \left[3\,|\,(1+w_{1x}+w_{1y})\right] \; \cap \; \left[w_{1x}
      + w_{1y}=\Delta
  \right] \\ 
    0 & \text{otherwise}
  \end{cases}
\end{equation}
as shown in Fig.~\ref{fig:k2vacancies}. Here, $c$ is a correction factor arising
when composing $F_m$ operators results in a deviation to the distance reached
for a given width. 

Note that the positive vacancy widths do not affect the scaling, since it only
needs one lattice point to condense the tip of the cascading operators. The
opposite result holds for $\braket e$ vacancies, where $w_{2}$ affects the
scaling, rather than $w_1$. 

These cascading operators give a form for the $\bar X$ logical operators. On the
other hand, $\bar Z$ logical operators are expected to be formed from moving $e$
charges around the $\braket m$ vacancy itself, similar to the periodic case.
Unlike those vacancies, however, many of these cage operators can be cleaned
onto a pyramidal region near the $\hat x + \hat y + \hat z$ vertex of the
vacancy and therefore will be local and constant-weight, independent of
$(w_{ix},w_{iy},w_{iz})$. This is problematic for the construction of a
superlinear-distance error correction code, and it is nontrivial to determine if
a subsystem approach is sufficient to create superlinear-weight dressed logical
operators. Crucially, the existing lemmas rely on bulk properties that no longer
hold in the vicinity of a defect, and so this remains an open question for
future research. It, therefore, appears that periodic vacancies provide a more
suitable method for improving $k$ while maintaining the partial self-correction
properties. 

\subsection{Additional Edge Dislocation Encodings}
As mentioned, a single edge dislocation (a pair of twists) in the bulk far from
any boundaries do not increase the ground state degeneracy.
On periodic boundary conditions (such that the dislocation line extends around
the periodic direction), the string-like operators seen in other configurations
emerge when $3|L$. 

However, there are additional linear-weight operators. Consider a $\braket{ee}$
dislocation extending in the $x$ direction. Acting on one trapezoidal prism in
the twist with the $T_X$ operator creates excitations in the neighboring $T_Z$
stabilizers by nature of their anti-commutation. By repeating $T_X$ on every
second trapezoidal prism, these excitations can move along the defect line.
Therefore, if $2|L_x$, these excitations can wrap around the periodic boundary
and annihilate. There are two such operators that do this, depending on the
starting position. If $L_x$ is not a multiple of $2$, these excitations can wrap
around the periodic boundary twice, annihilating again. In this case, there is
only one unique operator: the product of $T_X$ on all trapezoidal prisms.

Although the period-3 strings are independent when constructed on either of the
two $e$ twists, these new operators formed on either twist are equivalent modulo
stabilizers. Therefore, combining these two behaviors we get a scaling of
\begin{equation}
  k = 8\tau(L_x;\;L) + \begin{cases} 2 & 2|L_x \\ 1 &\text{otherwise}
  \end{cases}
\end{equation}
where $L$ is the appropriate linear system size to confine the string-like
operators. An equivalent result also holds for $\braket{mm}$. For mixed twists
$\braket{me}, \braket{em}$, the factor of ${2,1}$ does not appear since the
corresponding anti-commuting logical operator spans the separation between the
two twists, which must therefore both be of the same type. 

\subsection{\label{sec:enc_screw}Screw Dislocation Encodings}

A single $\braket R$ or $\braket L$ screw situated far from any boundary has no
logical operators. When extending around a periodic boundary, the period-3
strings are present. Since both $e$ and $m$ can condense on a screw dislocation
due to the lack of any stabilizer terms, we get twice as large a contribution to
the number of encoded qubits per screw. For one screw extending in the $z$
direction, we get
\begin{equation}
  k = 4\tau(L_z;\;L_\infty)
\end{equation}

For multiple screws, we first consider the case when the net Burgers vector is zero. That is, we have one $\braket L$ and one $\braket R$
separated by a Manhattan distance $\Delta$.
Computing the ground state degeneracy, we find the number of encoded qubits to
be
\begin{equation}
  k = 4\tau(L_z;\;L_\infty) +  2q_2(\Delta)
\end{equation} 
using the function $q$ from Eq.~(\ref{eq:qn}). Unlike edge dislocations, we thus
have a number of additional encoded qubits independent of the system size. Heuristically, this is linked to the idea that the action of the screw
dislocation introduces mobility along the defect. Spatially-translated
operators are now no longer necessarily independent. Since these additional
logical operators are spanning the bulk between the two screws, we expect this
configuration to have a superlinear distance, scaling with $\Delta$. However,
the constant $k$ value means that these encodings are not ideal candidates in
our search for improvements to the cubic code. Although $k$ could be increased
by potentially introducing additional screws, this runs into a similar problem
observed in the surface code: increasing the weight involves increasing the
separation between screws while increasing $k$ involves increasing the number
of screws. Achieving both simultaneously results in the system size increasing
significantly \cite{bravyiTradeoffsReliableQuantum2010}.

From a physical perspective, a nonzero Burgers vector would have a macroscopic
effect on the bulk of the material, rather than just on a localized region, and
is thus unlikely to be created in isolation
\cite{alloulDefectsCorrelatedMetals2009}. Nevertheless, we can still consider
the effect of two $\braket R$ or $\braket L$ screws in the bulk. In this case,
the formula is
\begin{equation}
  k = 4\left\lfloor \frac{L_z-1}{2} \right\rfloor + 
    4 \tau(L_z;\;L_\infty) + 2q_2(\Delta)
\end{equation}
gaining an additional set of logical operators that increases in number with
$L_z$.

%% file: appendix/appendix.tex
\subsection{\label{sec:periodicproof}Proof of Eq.~(\ref{eq:zmax})}
We first note the following lemma: 
\begin{lemma}
  Let $P$ be a local set of ($X$ or $Z$) Pauli operators with nontrivial excitations occurring on the dual lattice sites $S$, with $|S| > 1$. Choose an arbitrary site $s \in S$ to be the
  ``anchor'' of $P$. Define the new operator $P'$ by: for each $s_i \in S$, apply
  the operator $P$, translated so that the anchor point acts on $s_i$. Then, the
  pattern of excitations created by $P'$ is the pattern created by $P$, scaled by
  a factor of $2$ in all directions.
  \label{lemma:fractal}
\end{lemma}

We now present the proof of Eq.~(\ref{eq:zmax}):
\begin{proof}
  First, consider the case where $3\!\nmid\!\!L$. Here, the coloring of $A,B,C$ is
  not consistent across the periodic boundary. An operator that moves $e_A$
  across the boundary will transmute it into an $e_B$, for example. Repeating
  this will transmute $e_B$ into an $e_C$. Applying it a final time will return
  $e_C$ back into $e_A$, annihilating with the original charge. However,
  $e_A\times e_B\times e_C\sim 1$, and hence we have effectively moved a trivial
  charge around the boundary - therefore producing a trivial operator. This
  gives $z_\text{max}(L) = 0$ when $3\!\nmid\!\!L$ as required.

  We now demonstrate the arguments for $L=3,6,9$, before generalizing these
  results to arbitrary $L$. 

  Consider $L=3$. We employ the operator $G$, first shown in
  Fig.~\ref{fig:loop_path}, drawn into a 2D plane as in Fig.~\ref{fig:zmax_a} such that all excitations are only of type $e_A$. 

  \begin{figure}
    \centering 
    \includegraphics[width=.8\linewidth]{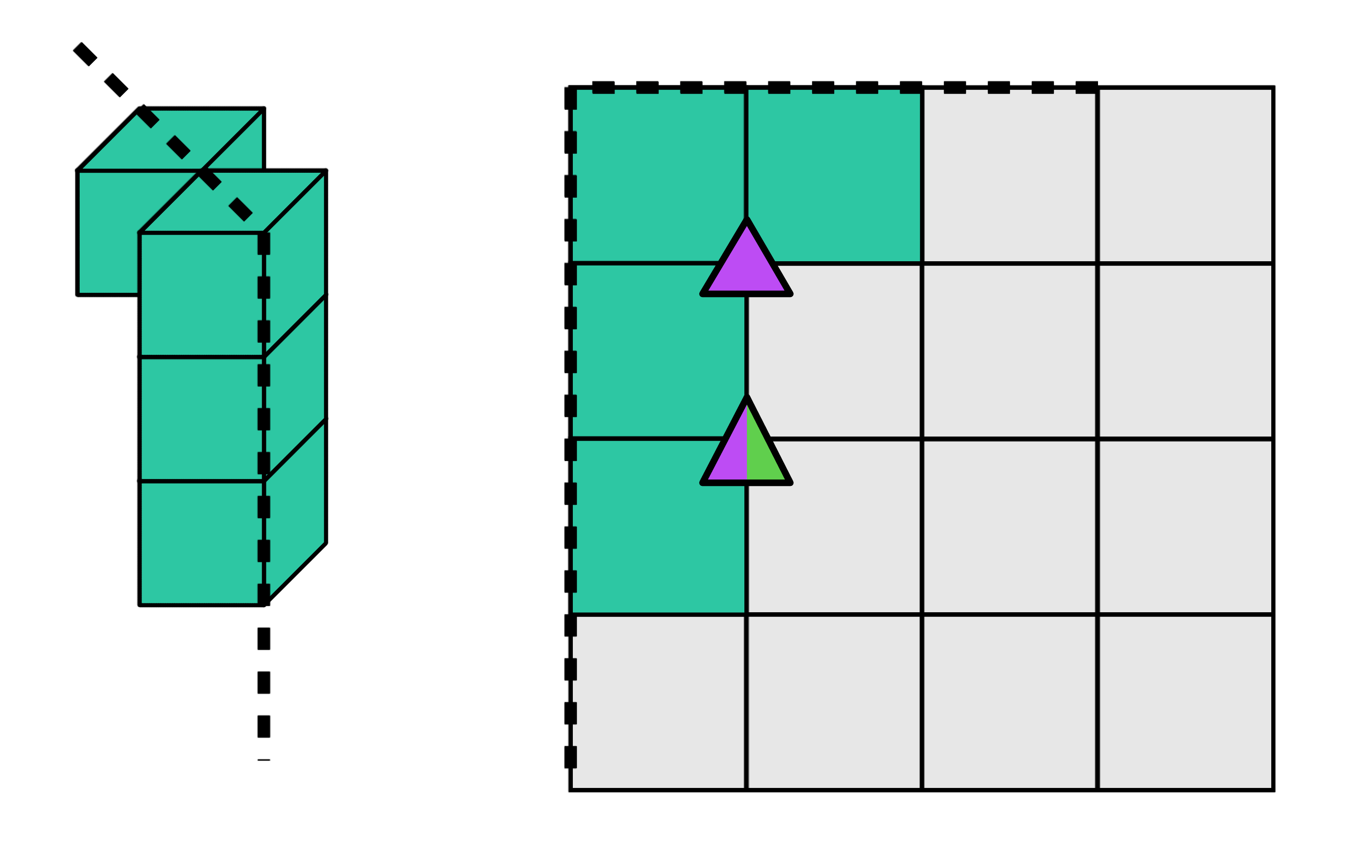}
    \caption{The operator $G$, constructed out of $IZ$ and $ZZ$ terms, which
    moves $e$ excitations along a line. In this representation, creating two
    $e_A$ in the top layer, for example, will also create an $e_A$ in the second
    and third layers. Considering the plane indicated by the dashed lines as a
    $2$D grid of just the $A$ diagonal gives the representation on the right,
    where each square corresponds to a location along that diagonal. This
    representation is used in Fig.~\ref{fig:zmax_b}.}
    \label{fig:zmax_a}
  \end{figure}

  Translating by $1$ to the right and reapplying $G$ two more times creates the pattern in Fig.~\ref{fig:zmax_b}a. Identifying the first and fourth columns
  via periodic boundary conditions produces an operator that commutes with all stabilizers in the topmost layer.
  In the next layer down, the residual charge is $e_A^3 \sim e_A$ and therefore
  nontrivial. This gives $z_\text{max}(3) = 1$, and we denote this operator
  $O_3$. 

  For $L=6$, we repeat $O_3$ twice, producing the excitation pattern in
  Fig.~\ref{fig:zmax_b}b. Notably, each excitation in Fig.~\ref{fig:zmax_b}a
  is now paired with another excitation, separated by a distance $\Delta = 3$
  horizontally. We thus apply $O_3$ onto each of the six excitation locations to
  the left of the dashed line, annihilating them. The residual charge created by
  this action produces the pattern in Fig.~\ref{fig:zmax_b}d. As predicted by
  Lemma \ref{lemma:fractal}, this is the same as that in
  Fig.~\ref{fig:zmax_b}a, but scaled by a factor of $2$. We now have
  a nontrivial residual charge in the third layer from the top, giving
  $z_\text{max}(6) = 2$. We denote this  operator as $O_6$. 

  For $L=9$, we cannot simply repeat $O_6$ twice. Instead, to annihilate all
  excitations in the topmost layer we must apply $O_3$ three times. Unlike with
  $O_6$, the residual net charge in the second layer is now nontrivial since
  the residual charge from $O_3$ is repeated thrice. We therefore again get that
  $z_\text{max}(9) = z_\text{max}(3) = 1$. 

  This process can be continued for arbitrary $L$. For each, $z_\text{max}$ is
  determined by the largest-support operator $O_j$ such that $j|L$, where
  $j=3,6,12,24,\ldots = 3\cdot 2^z$ for $z\in\mathbb Z$. Moreover, by Lemma
  \ref{lemma:fractal} each $O_j$ is equivalent to $O_{j/2}$ scaled by a factor
  of $2$, and therefore extends twice as far before reaching nontrivial residual
  charge. Since $O_3$ produces $z_\text{max} = 1$, we thus get that $O_j$
  produces $z_\text{max} = j/3$. 
  
  This gives the result
  \begin{equation}
    z_\text{max}(L) = \text{max}\left\{ 2^z \, : \, (3\cdot 2^z | L), \, z \in
    \mathbb Z \right \}
  \end{equation}
  for $3|L$, and $z_\text{max}(L) = 0$ otherwise, as required.
\end{proof}

Although this proof relied on a square lattice, it can be generalized to an
arbitrary $L_x, L_y$ by taking the minimum of $z_\text{max}(L_x)$ and
$z_\text{max}(L_y)$. Moreover, although we used here a diagonal $G$ operator,
equivalent fractal behavior emerges for operators confined to planes, such as the
$F$ operators used in the body of this paper. 

\begin{figure}[!t]
  \centering 
  \includegraphics[width=\linewidth]{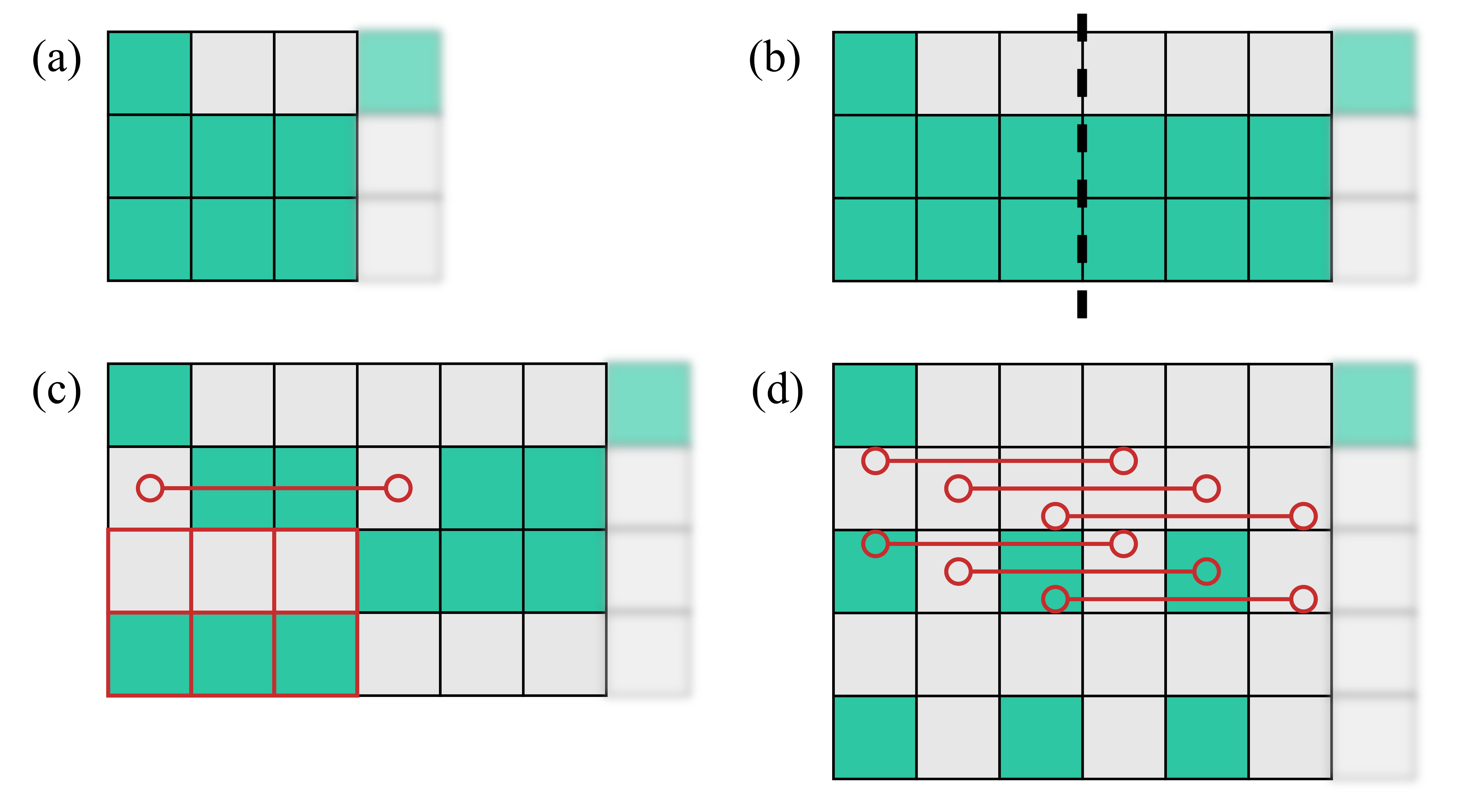}
  \caption{{(a)} Repeating the operator in Fig.~\ref{fig:zmax_a} three
  times produces this excitation pattern. The resulting operator is named
  $O_3$. If the first and fourth columns are identified via a periodic
  boundary, this commutes with all stabilizers in the first row.
  {(b)} Repeating the operator in (a) twice produces this excitation
  pattern, with a period of $6$. 
  {(c)} Applying $O_3$ onto the square indicated by the leftmost red
  circle annihilates two charges in the second layer, while flipping the squares
  indicated by the red outline.
  {(d)} Applying $O_3$ to each of the $6$ charges to the left of the
  dashed line in (b) produces an excitation pattern that is equivalent to (a)
  but scaled by a factor of $2$. This resulting operator is called $O_6$.}
  \label{fig:zmax_b}
\end{figure}

\begin{figure}[t]
  \centering 
  \vspace{12pt}
  \includegraphics[width=.7\linewidth]{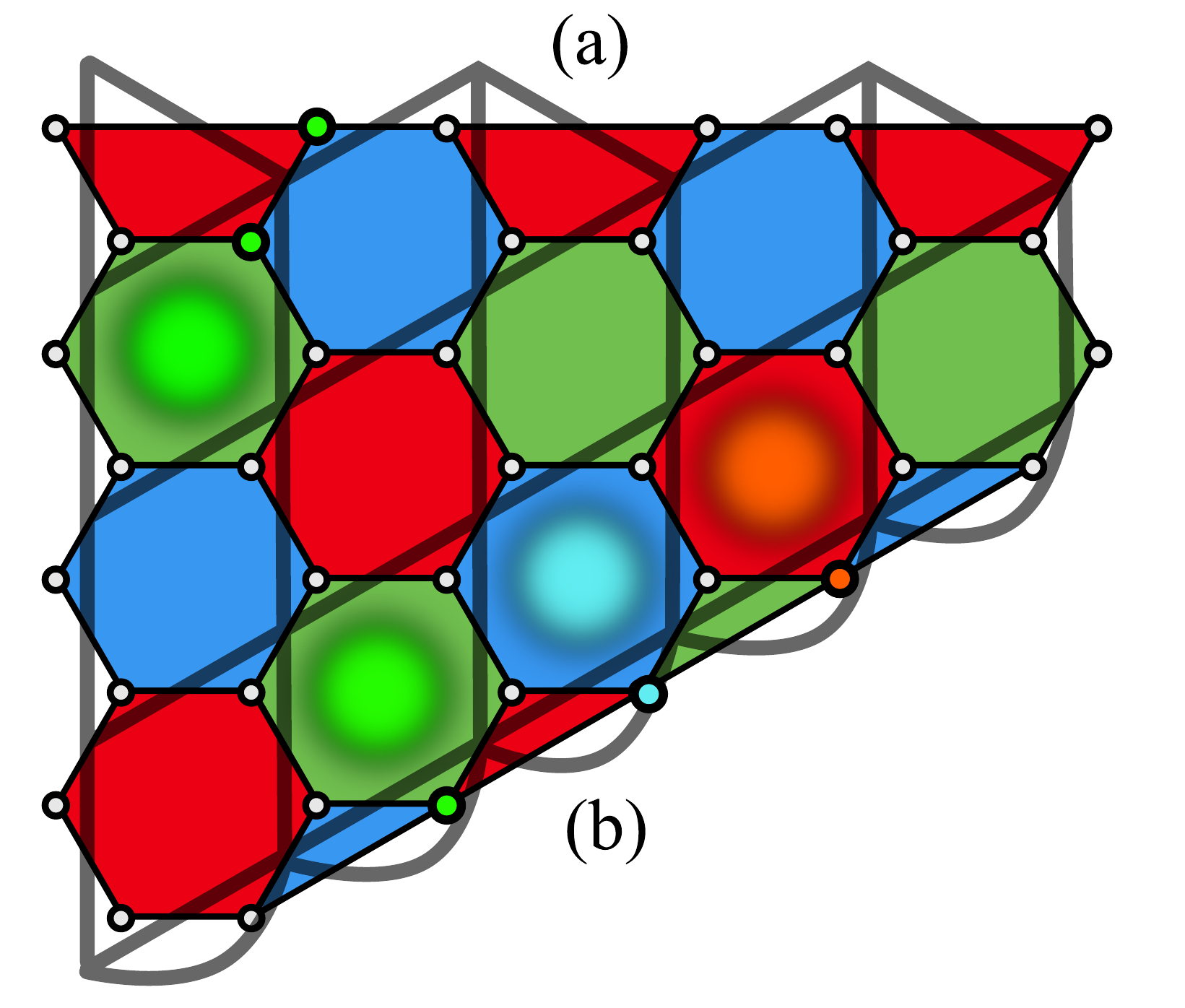}
  \caption{Two forms of the boundaries on the 6-6-6 color code: {(a)}
    Boundaries that condense a single color, in this case green. {(b)}
    Boundaries that condense a single Pauli charge, determined by which
    stabilizers are removed from the boundary group. Overlayed is the rhomboidal
    lattice used to construct the correspondence with the cubic code boundaries. Circles on the faces indicate charges, while colored circles on the vertices indicate acting with an $X$ (or equivalently $Z$) Pauli operator. } 
  \label{fig:colorcode2}
\end{figure}

\begin{figure}
  \centering 
  \includegraphics[width=\linewidth]{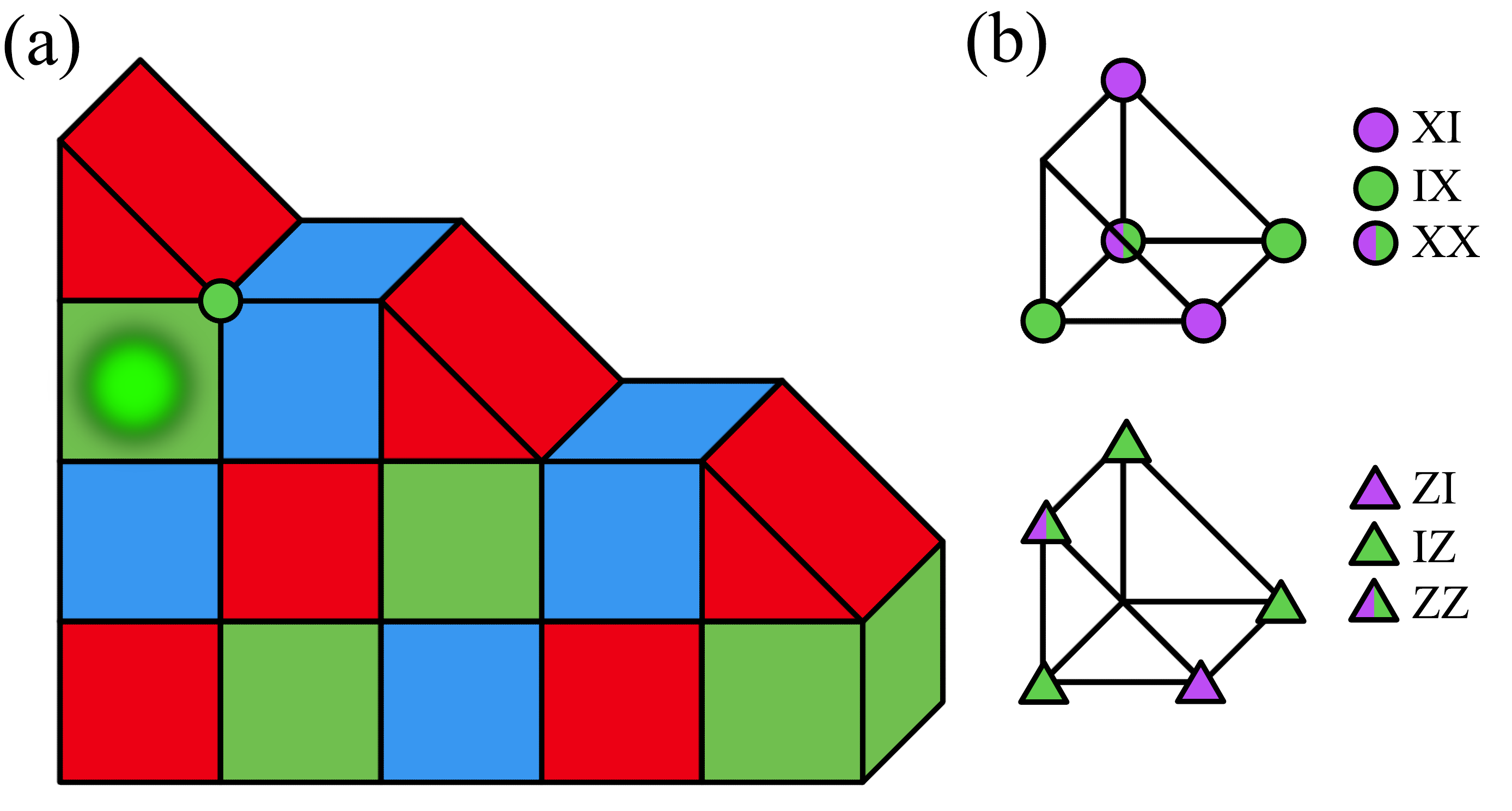}
  \caption{{(a)} A boundary on the $(100)$ face of the cubic code, which condenses
    only one charge type, such as $m_A$. The excitation and coloring shown are
    equivalent to that in Fig.~\ref{fig:colorcode2}. {(b)} The two forms of
    the stabilizers that are placed along the boundary. Note that the two terms
    anticommute when on the same location, but commute when neighboring in the
    $x$ direction.}
  \label{fig:colorcode3}
\end{figure}

\subsection{\label{sec:colorcode}Color Code Correspondence}
As discussed in Section \ref{sec:boundexc}, there is a direct correspondence
between the $(e_{ABC})$ and $(m_{ABC})$ boundaries, and the $(2+1)$D color code. 
Ref.~\cite{kesselringBoundariesTwistDefects2018} showed that these color codes
have $6$ distinct gapped boundaries, defined by their Lagrangian subgroups~\cite{levinProtectedEdgeModes2013}. We
should therefore be able to identify equivalent representations in the boundary
layers of the cubic code. Two of these boundaries directly correspond to the $e$
and $m$-condensing edges of the cubic code discussed in the body of this
manuscript. A third corresponds to taking products of $X$ and $Z$ stabilizers
such that a combined $em$ particle condenses. The remaining three, however,
represent distinct behavior not discussed previously. In the color code, these
three additional boundaries condense just red, green, or blue anyons, as in
Fig.~\ref{fig:colorcode2}. Correspondingly, in the cubic code, we should
observe the condensation of just $A$, $B$, or $C$ charges. These boundaries are
shown in Fig.~\ref{fig:colorcode3}. Notably, the new $X$ and $Z$ stabilizer
terms anticommute when at the same location, but commute when placed adjacent on
the $x$ direction. Therefore, we can construct a full $(3+1)$D model with this
boundary by alternating the $X$ and $Z$ stabilizers corresponding to the
triangular prisms. 

It remains an open question how these boundaries affect the encoding properties
of the cubic code model. Moreover, future works should investigate the
generalization of color code twist defects
\cite{kesselringBoundariesTwistDefects2018} to the cubic code model, and how
their results interplay with the analysis of twists conducted in this paper.

\vspace{18pt}

\subsection{\label{sec:appcode}Code Availability}

All numerical evaluations were performed using \verb`Julia`, the code for which
can be found at \url{https://github.com/corytaitchison/cubic-code}.